\documentclass[nofootinbib,reprint,amsmath,amssymb,showkeys, showpacs,aps]{revtex4-1}
\usepackage{graphicx}
\usepackage[T1]{fontenc}
\usepackage{dcolumn}
\usepackage{xcolor,colortbl}
\usepackage{multirow}
\usepackage{stackengine}
\usepackage{tikz-cd}
\usepackage{bm}
\usepackage{mathtools}
\usepackage{tikz}
\newcommand{\qm}[1]{``#1''}
\usepackage{hyperref}
\hypersetup{colorlinks, linkcolor={black},citecolor={blue},urlcolor={blue}}

\usepackage{accents}
\usepackage{scalerel}
\usepackage{bbm}
\usepackage{tabularx,siunitx}
\usepackage{wasysym}
\usepackage{ mathrsfs }
\newcommand*\fullcirc[1][0.3ex]{\tikz\fill (0,0) circle (#1);} 
\newcommand{\udt}[3]{#1^{#2}_{\phantom{#2}#3}}

\newcommand{\dut}[3]{#1_{#2}^{\phantom{#2}#3}}
\newcommand{\dudt}[4]{#1_{#2\phantom{#3}#4}^{\phantom{#2}#3}}

\newcommand{\lc}[1]{\accentset{\circ}{#1}}
\newcommand{\tg}[1]{\accentset{\land}{#1}}
\newcommand{\stg}[1]{\accentset{\scaleto{\diamond}{3.5pt}}{#1}}

\begin{document}
\title[Equivalent representations of gravity]{Comparing Equivalent  Gravities: common features and differences}
\author{Salvatore Capozziello$^{1,2,3}$}\email{capozziello@unina.it}
\author{Vittorio De Falco $^{2,3}$}\email{vittorio.defalco-ssm@unina.it}
\author{Carmen Ferrara$^1$}\email{ferraracarmen1@gmail.com}
\affiliation{$^1$ Dipartimento di Fisica "E. Pancini", Universit\'a di Napoli "Federico II", Complesso Universitario di Monte S. Angelo, Via Cinthia Edificio 6, I-80126 Napoli, Italy\\
$^2$ Scuola Superiore Meridionale, Largo San Marcellino 10, I-80138 Napoli, Italy,\\
$^3$ Istituto Nazionale di Fisica Nucleare, Sezione di Napoli, Complesso Universitario di Monte S. Angelo, Via Cinthia Edificio 6, I-80126 Napoli, Italy}

\date{\today}

\begin{abstract}
We discuss equivalent representations of gravity in the framework of metric-affine geometries pointing out basic concepts from where these theories stem out. In particular, we take into account   tetrads and spin connection to describe the so called  {\it Geometric  Trinity of Gravity}. Specifically, we consider General Relativity, constructed upon the metric tensor and based  on the curvature $R$; Teleparallel Equivalent of General Relativity, formulated  in terms of torsion $T$ and relying on tetrads and spin connection; Symmetric Teleparallel Equivalent of General Relativity, built  up on non-metricity $Q$, constructed from  metric tensor and affine connection. General Relativity is formulated as a geometric theory of gravity based on metric, whereas teleparallel approaches configure as gauge theories, where  auge choices permit not only to simplify calculations, but also to give  deep insight into the basic concepts of   gravitational field. Specifically, we point out how the foundation principles of General Relativity (i.e., the Equivalence Principle and the General Covariance) can be seen from the teleparallel point of view. These theories are dynamically equivalent and this feature can be demonstrated under three different standards: (1) the variational method; (2) the field equations; (3) the solutions. Regarding the second point, we provide a procedure  starting from the (generalised) second Bianchi identity and then deriving the field equations. Referring to the third point, we compare spherically symmetric solutions in  vacuum recovering the Schwarzschild metric and  the Birkhoff theorem in all the approaches. It is worth stressing that, in extending the formalisms to $f(R)$, $f(T)$, and $f(Q)$ gravities respectively, the dynamical equivalence is lost opening the discussion on  the different number of degrees of freedom intervening in the various representations of gravitational theories.
\end{abstract}
\maketitle

\tableofcontents{}

\section{Introduction}
\label{sec:intro}
In the 19$^{th}$ century, Newtonian mechanics was  considered as the best theory to describe gravity, since it was successfully exploited in everyday life and capable of describing the motion of planets and stars. However, in this period, there was a great cultural ferment around \emph{non-Euclidean geometries} starting from fundamental works by Gauss, Lobachevsky, Bolyai, Riemann, Bianchi, Ricci-Curbastro, and several others \cite{Bonola1955}. The Euclidean framework, the arena for classical Physics, was overtaken by the formulation of elliptical and hyperbolic geometries, stemming out from a rigorous axiomatic reformulation of the geometry foundations. Indeed, two approaches were more and more emerging from these studies: $(i)$ \emph{affine geometry}, introduced by Euler in 1748, deriving from the Latin word \emph{affinis}, meaning \qm{related}, and after promoted by M\"obius, Klein, and Weyl. It essentially focuses on the study of \emph{parallel lines}, based on the validity or redefinition of the fifth Euclid postulate, and on the \emph{affine transformations} \cite{Bennett2011}; $(ii)$ \emph{metric geometry}, introduced by Fr\'echet and Hausdorff, relies on a \emph{metric function} defining the concept of distance between any two points, members of a non-empty set \cite{Burago2001}.

Einstein, inspired by this line of nonconformist ideas, arrived, in 1915, to the formulation of General Relativity (GR)  \cite{Einstein1916}. This new vision of gravitational interaction, ruled by the   spacetime curvature, took time to be comprehended and accepted by the scientific community owed to the outcoming effects, retained to be too small to be measured and  observed at that time. The well-known subsequent astronomical confirmations constituted the success of GR \cite{Will1993}.

Although GR was not yet validated, some authors were however eager to advance proposals to extend it with the aim to fulfill more general purposes. In 1918, Weyl started to study the question on how to connect gravity and electromagnetism in a single and coherent geometric theory. To achieve this objective, he took into account an additional gauge field, which singles out a unique \emph{length connection}, whose four additional degrees of freedom (DoFs) are associated to the electromagnetic potentials. In the Weyl geometry, besides the GR connection, there is also an additional length connection, which is symmetric, metric incompatible, and gauge invariant. The consequence is that, during a parallel transport, both direction and length of vectors vary \cite{Weyl1922, Weyl1929}. However, Weyl's theory revealed to be in conflict not only with some experiences (for example, the frequency of spectral lines of atomic clocks depends on the location and past histories of the atoms), but even in a more fundamental way with Quantum Mechanics  (e.g., masses of particles rest on their past histories). 

In 1930, along the same line of thinking, Einstein himself proposed some modifications to his theory. Fascinated by teleparallelism and tetrad formalism, he initiated a prolific and extensive correspondence mainly with Cartan, Weitzenb\"ock, and Lanczos \cite{Unzicker2005,Goenner2004,Goenner2014}. Indeed, since the tetrad fields posses sixteen independent components, he associated ten of them to the metric tensor, whereas the other six were believed to be linked to a separate connection, entrusted to model  electromagnetic potentials. Unfortunately, he failed in his attempt, but his studies shed new light on the importance of   additional DoFs, which theoretically belong to the Lorentz group and physically are a consequence of the local Lorentz invariance.

In 1922, Cartan concentrated on a different direction, since he considered a natural extension of GR constituted not only by the Levi-Civita connection, but also by the \emph{torsion tensor} (essentially the antisymmetric part of a  metric compatible affine connection). Given these premises, he  developed all the ensuing  geometric formulation \cite{Cartan1952}, where he suggested that the torsion can be physically related to the intrinsic (quantum) angular momentum of matter and it vanishes as soon as vacuum regions are considered \cite{Cartan1922-1, Cartan1922-2, Cartan1922-3, Cartan1922-4, Cartan1922-5}. 

Around 1960, Kibble and Sciama revisited the theory formulating it within the gauge theory of the Poincar\'e group \cite{Kible1961, Sciama1962,Hehl1976}. This approach can be extended to the more general affine group, leading thus to the metric-affine gauge gravity \cite{Hehl1995}.

There have been other  proposals and experiments to probe  the fundamental nature  of gravitation, in particular, to establish  its geometric structure. In this vein, it was growing the awareness that affinity and metricity could be considered as two different and independent concepts, where the affine connection could not respect \emph{a priori}  the metric postulate. This perspective  is considered into the so-called   {\it Palatini approach}, where GR is   constituted by a metric tensor and an  affine connection, considered as two different geometric structures. Varying the Einstein-Hilbert action  with respect to the metric, the Einstein field equations are recovered; whereas, varying it with respect to the affine connection, the metric compatibility condition is naturally obtained and the Levi-Civita connection is restored \cite{Palatini}. This shows that GR structure entails metric compatibility, and the affine connection can be considered as a true dynamical field. As it is well-known, this coincidence does not work for extensions of GR as $f(R)$\cite{Allemandi}.

These considerations led  to the development of  theories of gravity  beyond the Einstein picture, where the field equations, besides  the scalar curvature,  can be formulated in terms of other geometric invariants. Furthermore, the affine connections were not considered anymore with an ancillary role with respect  to the metric tensor, but, contrarily, they assumed   a  dynamical fundamental role. These approaches  gave rise  to the current variegated realm of the \emph{Extended} and \emph{Alternative Theories of Gravity} (see e.g., \cite{Clifton2006,Capozziello2008,Sergei,Vasilis, Nojiri,Sotiriou2010,Faraoni2010,Capozziello2011R,Cai2016}).

In any case, GR revealed to be  extraordinarily successful   because passed several astrophysical and cosmological observational probes, like the Solar System tests \cite{Will1993,Ni2016,DeMarchi2020}, the  direct detection of gravitational waves  \cite{Abbott2016,Abbott2017-NS,Sathyaprakash2009,Bailes2021,LIGOScientific2021psn}, the recent black hole imaging \cite{EHC20191, EHC20192, EHC20193, EHC20194, EHC20195, EHC20196,Akiyama2022image,Ozel2016}, and
other robust confirmations   \cite{Stairs2003,Ciufolini2004,Abuter2020,Kramer2021}.
 
Despite these achievements, the theory exhibits various pathological issues, still matter of debate,  suggesting that approaches beyond Einstein gravity should be pursued \cite{Faraoni2010}. For example, from galaxies to cosmic evolution, the infrared behavior of  gravitational field presents several shortcomings mainly related to the  {\it Dark Matter} \cite{Mayet2016, Consiglio,Arun2017, Felix, Vesna}
and  \emph{ Dark Energy} problems \cite{Carroll2001, Planckcollaboration2020,Capozziello2008}, and the  tensions in cosmological parameters like $H_0$ \cite{Aloni2021,DiValentino2021}. At ultraviolet scales, the lack of 
renormalizability and unitarity of gravitational field points out that a final, self-consistent theory of  Quantum Gravity is not at hand  \cite{Hooft1974,Capozziello2011,Obukhov2017,Giulini2003,Rovelli2004,Gleiser2005,Thiemann2007}.
    
In general, the formulation of a new theory of gravity  to solve the above  issues is not a simple task. There are principles, constraints, mathematical consistencies, and the agreement with  observations that any novel approach must necessarily fulfill before  being accepted as a self-consistent picture. This is one of the thorny theoretical challenges of modern physics. 

In this perspective, we want to focus our attention on GR and its dynamically equivalent formulations,  in view to put in evidence similarities and differences towards a unified view of gravitational interaction. 

This paper is organized as follows: in Sec. \ref{sec:metric_affine_theories}, we describe the general framework, represented by the metric-affine theories of gravity, in which the so-called {\it Geometric Trinity of Gravity} \cite{BeltranJimenez:2019esp} can be formulated (Sec. \ref{sec:_geom_trin}). In Sec. \ref{sec:math_tools}, we provide the  mathematical tools necessary for the formulation of any theory of gravity. In Sec. \ref{sec:trinity_lagrangian}, we  discuss the Geometric Trinity of Gravity in terms of their Lagrangian equivalence. Section \ref{sec:dyn_field_equat} is devoted to the field equations derived from the second Bianchi identity. In Sec. \ref{sec:trinity_solutions}, we analyse the spherically symmetric solutions in the three equivalent  formulations, recovering, in all of them, the Schwarzschild metric and the Birkhoff theorem. Finally, Sec. \ref{sec:end} is devoted to the conclusions  and to the discussion  of some crucial issues necessary for any  self-consistent formulation of  gravity.

\emph{Notations.} We adopt the metric signature  $(-,+,+,+)$. Greek indices take values  $0,1,2,3$, while the lowercase Latin ones $1,2,3$. Capital Latin letters indicate tetrad indices. The flat metric is indicated by $\eta^{\alpha \beta }=\eta_{\alpha \beta }={\rm diag}(-1,1,1,1)$. The determinant of the metric $g_{\mu \nu}$ is denoted by $g$. Round (square) brackets around a pair of indices stands for  symmetrization (antisymmetrization) procedure, i.e., $A_{(ij)}=A_{ij}+A_{ji}$ ($A_{[ij]}=A_{ij}-A_{ji}$). 

\section{Metric-affine theories of gravity} 
\label{sec:metric_affine_theories}
A first extension of Einstein gravity starts by generalizing the affine connections which cannot be strictly Levi-Civita. A metric-affine theory is defined by the following triplet $\{\mathcal{M},g_{\mu\nu},\Gamma^{\rho}_{\ \mu\nu}\}$, where $\mathcal{M}$ is a four-dimensional spacetime manifold, $g_{\mu\nu}$ is a rank-two symmetric tensor (with 10 independent components), and $\Gamma^{\rho}_{\ \mu\nu}$ is the affine connection (endowed with 64 independent components). \emph{A priori} there is no relation between the metric and the affine connection, where the former is responsible to describe the \emph{casual structure}, whereas the latter deals with the \emph{geodesic scaffold}. As it is well-known, the structures coincide if the Equivalence Principle is the basic foundation of the theory, since it relates affine connection with the derivatives of the metric \cite{Will1993, Faraoni2010}.

Let us consider now a system of coordinates $\left\{x_0,x_1,x_2,x_3\right\}$ defined on $\mathcal{M}$, where $x_0$ labels the time and $\left\{x_1,x_2,x_3\right\}$ the space\footnote{It is important to note that, here, we are assuming that the spacetime can be split in space and time, as it normally occurs in several astrophysical and cosmological metrics.}. The metric $g_{\mu\nu}$ defines the line element ${\rm d}s^2=g_{\mu\nu}{\rm d}x^\mu {\rm d}x^\nu$. The notion of covariant derivative $\nabla$ acts on a generic $(1,1)$ tensor in the following way \cite{Vitagliano2011}
\begin{equation}
\nabla_\mu A^{\alpha}_{\ \beta}:=\partial_\mu A^{\alpha}_{\ \beta}- \Gamma^{\rho}_{\ \beta\mu}A^{\alpha}_{\ \rho}+ \Gamma^{\alpha}_{\ \rho\mu}A^{\rho}_{\ \beta}.
\end{equation}
The components of the general affine connection $\Gamma^\rho_{\ \mu\nu}$ can be uniquely decomposed as follows \cite{Jimenez2019,Bahamonde2021}:
\begin{equation} \label{eq:ACG}
\Gamma^\rho_{\ \mu\nu}:=\begin{Bmatrix} \rho\\ \mu \nu \end{Bmatrix}+K^\rho_{\ \mu\nu}+L^\rho_{\ \mu\nu},
\end{equation}
where $\begin{Bmatrix} \rho\\ \mu \nu \end{Bmatrix}$ is the Levi-Civita connection, $K^{\rho}_{\ \mu\nu}$ is the contortion tensor, and $L^{\rho}_{\ \mu\nu}$ is the disformation tensor, whose explicit expressions are \cite{Bahamonde2021}
\begin{subequations}
\begin{align}
\begin{Bmatrix} \rho\\ \mu \nu \end{Bmatrix}&:=\frac{1}{2}g^{\rho\lambda}(\partial_\mu g_{\lambda\nu}+\partial_\nu g_{\mu\lambda}-\partial_\lambda g_{\mu\nu}),\label{eq:LC}\\
K^\rho_{\ \mu\nu}&:=\frac{1}{2}(T_\mu^{\ \rho}{}_\nu+T_\nu^{\ \rho}{}_\mu-T^\rho_{\ \mu\nu}),\label{eq:contortion}\\
L^\rho_{\ \mu\nu}&:=\frac{1}{2}(Q^\rho_{\ \mu\nu}-Q_\mu^{\ \rho}{}_\nu-Q_\nu^{\ \rho}{}_{\mu})\label{eq:deformation}.
\end{align}
\end{subequations}
Notice that, while the Levi-Civita part is non-tensorial, the contortion and disformation terms are tensors  under changes of coordinates.
The three main geometric objects (related to the dynamics) are: the \emph{curvature tensor} $R^{\mu}_{\ \nu\alpha\beta}$, the \emph{torsion tensor} $T^{\rho}_{\ \mu\nu}$, and the \emph{non-metricity tensor}  $Q_{\rho\mu\nu}$. Their explicit expressions in terms of metric and connections are \cite{Bahamonde2021}:
\begin{subequations}
\begin{align}
R^{\mu}_{\ \nu\rho\sigma}&:=\partial_\rho \Gamma^{\mu}_{\ \nu\sigma}-\partial_\sigma \Gamma^{\mu}_{\ \nu\rho}+\Gamma^{\mu}_{\ \tau\rho}\Gamma^{\tau}_{\ \nu\sigma}-\Gamma^{\mu}_{\ \tau\sigma}\Gamma^{\tau}_{\ \nu\rho},\label{eq:curv_ten}\\
T^{\mu}_{\ \nu\rho}&:= 2\Gamma^{\mu}_{\ [\rho\nu]}\equiv\Gamma^{\mu}_{\ \rho\nu}-\Gamma^{\mu}_{\ \nu\rho},\label{eq:tor_ten}\\
Q_{\mu\nu\rho}&:=\nabla_\mu g_{\nu\rho}\equiv \partial_\mu g_{\nu\rho}-\Gamma^{\lambda}_{\ (\nu|\mu}g_{\rho)\lambda}\neq 0.\label{eq:nonm_ten}
\end{align}
\end{subequations}
These tensors show the following symmetries
\begin{subequations}
\begin{align}
    R^{\mu}_{\ \nu\rho\sigma}&=-R^{\mu}_{\ \nu\sigma\rho}, \label{eq:sym_Riemann}\\
    T^{\mu}_{\ \nu\rho}&=-T^{\mu}_{\ \rho\nu}, \label{eq:sym_tensor}\\
    Q_{\mu\nu\rho}&=Q_{\mu\rho\nu}. \label{eq:sym_nonmetricity}
\end{align}
\end{subequations}
The above geometric quantities, differently affect the parallel transport of a vector on a manifold. We have that:
\begin{itemize}
    \item \emph{curvature} manifests its presence when a vector is parallel transported along a closed curve on a non-flat background and come back to its starting point forming a non-null angle with its initial position; 
    \item \emph{torsion} entails a rotational geometry, where the parallel transport of two vectors is antysimmetric by exchanging the transported vectors and the direction of transport. This property results in the non-closure of parallelograms; 
    \item \emph{non-metricity} is responsible to alter the length of the vectors along the transport. 
\end{itemize}
In a generic metric-affine theory, all these effects can work together and could have also further meanings corresponding to physical quantities (e.g., the torsion tensor is linked to the spin in the Einstein-Cartan theory \cite{Hehl1976}). 

In general, the following \emph{Bianchi identities} hold \cite{Bahamonde2021}:
\begin{subequations} \label{eq:Bianchi_identity_general}
\begin{align}
R^{\mu}_{\ [\nu\rho\sigma]}&=\nabla_{[\nu}T^\mu_{\ \rho\sigma]}+T^\mu_{\ \alpha[\nu}T^\alpha_{\ \rho\sigma]},\\
\nabla_{[\alpha}R^{\mu}_{\ |\nu|\rho\sigma]}&=-R^\mu_{\ \nu\tau[\alpha}T^\tau_{\ \rho\sigma]},
\end{align}
\end{subequations}
which involve only curvature and torsion tensors.

\begin{figure}[!ht]
\centering
\includegraphics[trim=6cm 4cm 6cm 4cm,scale=0.5]{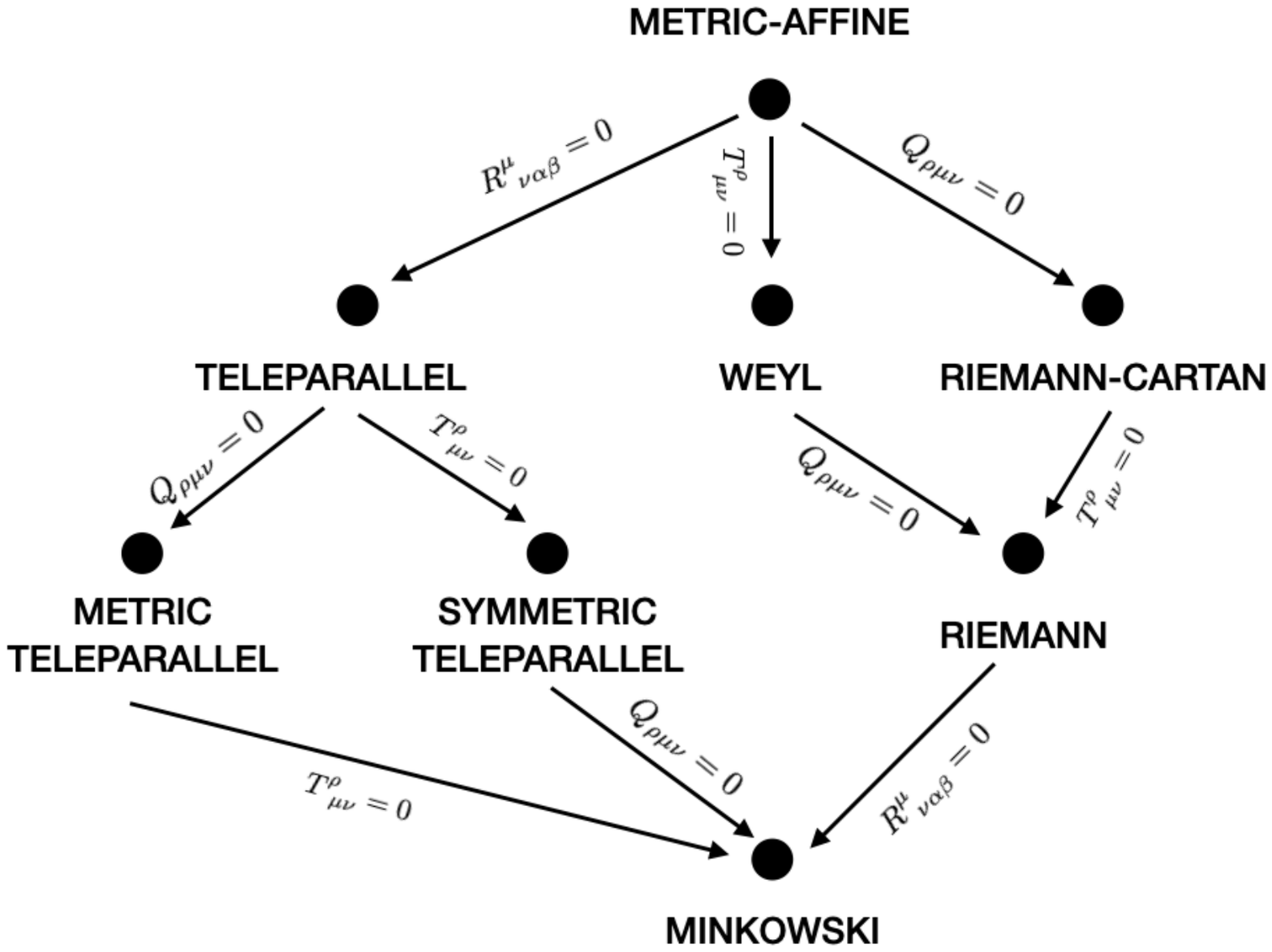}
\caption{A possible classification of gravity theories emerging within the arena of metric-affine geometry.}
\label{fig:Fig1}
\end{figure}

Metric-affine geometries are a broad class of theories, whose dynamics can be related to the tensors $R^{\mu}_{\ \nu\rho\sigma}$, $T^{\mu}_{\ \nu\rho}$, and $Q_{\mu\nu\rho}$ which can be generally classified as in Fig. \ref{fig:Fig1}. 
\begin{enumerate}
    \item[1)] The \emph{Riemann-Cartan geometry} is expressed in terms of metric compatible curvature and torsion tensors. It is also known in the literature as $U_4$ or Einstein-Cartan-Sciama-Kibble theory, where the role of the torsion is deputed to model the quantum spin effects present in the matter \cite{Hehl1971,Hehl1973b,Hehl1974b,Hehl1976}.
\item[2)] The \emph{Weyl geometry} is constructed by vanishing the torsion, where curvature and non-metricity are the only surviving geometric objects. This theory has interesting implications and moreover it represents also the origin of the $U(1)$ gauge theory \cite{Wheeler2018}.
\item[3)] \emph{Teleparallel geometries} are curvature-less and are based on the concept of \emph{Fernparallelismus} or \emph{parallelism at distance}, because two vectors can be immediately seen whether they are parallel or not, since the parallel transport of vectors becomes independent of the path \cite{Aldrovandi2013}. They admit two special subclasses, represented by
\begin{enumerate}
    \item[3.1)] \emph{metric teleparallel theories} expressed only in terms of the torsion tensor;
    \item[3.2)] \emph{symmetric teleparallel theories} described only by the non-metricity tensor. 
\end{enumerate}
\item[4)] The \emph{Riemannian geometry} represents the first arena within which Einstein framed his theory, constructed only upon the curvature tensor \cite{Misner1973,Romano2019}.
\item[5)] The \emph{Minkowski geometry} is obtained by setting curvature, torsion, and non-metricity to zero, where the flat metric $\eta_{\mu\nu}$, as well as zero affine connections, are adopted. This is the arena of  Special Relativity \cite{Misner1973,Romano2019}.
\end{enumerate}

\section{The Geometric Trinity of Gravity}
\label{sec:_geom_trin}
\begin{figure*}[ht!]
\includegraphics[trim=0cm 6cm 0cm 6cm,scale=0.6]{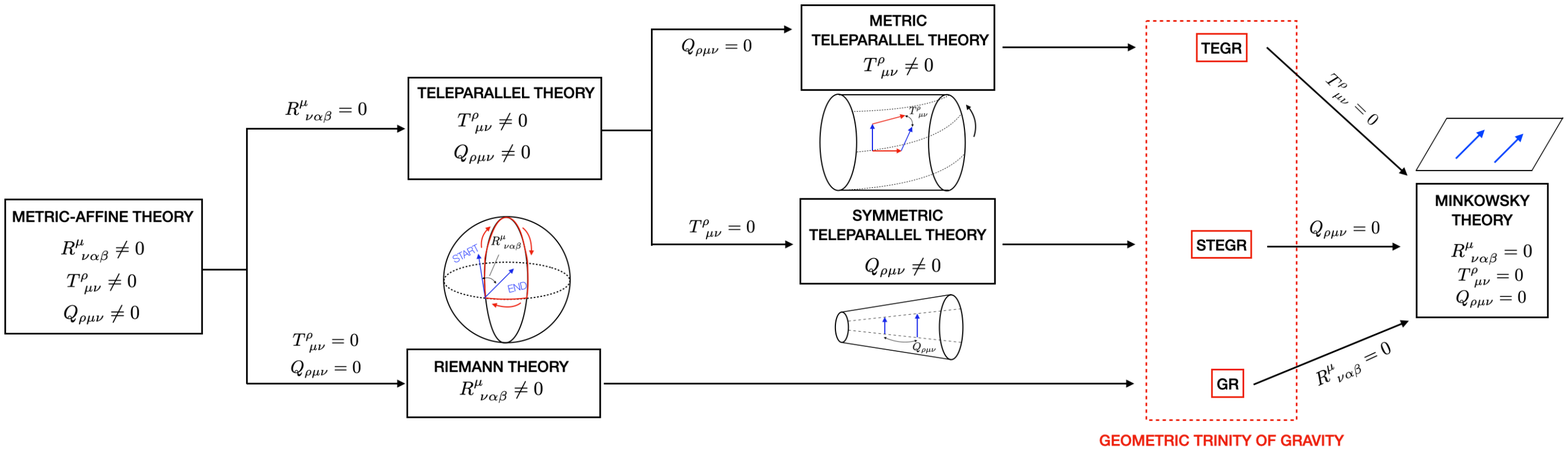}
\centering
\caption{The Geometric Trinity of Gravity framework and the dynamical role of tensor invariants. \emph{Curvature} rules how the tangent space rolls along a curve on a manifold; \emph{torsion}  how the tangent space twists around a curve when we parallel transport two vectors along each other; \emph{non-metricity} encodes the variation of vectors' length when they are moved along a curve \cite{Bahamonde2021}.}
\label{fig:Fig2}
\end{figure*}
Among the possible metric-affine gravity theories, Riemannian and teleparallel models are particularly interesting. GR is an example of Riemannian geometry, whereas the so-called metric teleparallel equivalent of GR (TEGR) and symmetric teleparallel equivalent of GR (STEGR) are examples of teleparallel geometries. See Fig. \ref{fig:Fig2}. These three theories constitute the so-called \emph{Geometric Trinity of Gravity} \cite{Jimenez2019}. 

A fundamental property of TEGR is that torsion replaces curvature for dynamics and it is able to provide the same descriptions of the gravitational interaction under a different perspective. In GR the geometric curvature is entrusted to model the gravitational force, whereas geodesics coincide with the free-falling test particle's trajectories. On the other hand, in TEGR, the gravitational interaction emerges through the torsion tensor and acts as a (gauge) force. This is the reason why, in the teleparallel framework, the concept of geodesics is replaced by force equations, analogously to what happens in electrodynamics where  the Lorentz force is present. STEGR shares several similar properties with TEGR. In this theory, one requires that curvature and torsion are both zero, and gravitational dynamics is attributed to the non-metricity tensor. 

GR is described in terms of the metric $g_{\mu\nu}$; TEGR in terms of the tetrads $e^A_{\ \mu}$ (accounting for the dynamical description of gravity) and spin connection $\omega^A_{\ B\mu}$ (flat connection outlining inertial effects); STEGR embodies the Palatini idea where metric $g_{\mu\nu}$ and affine connection $\Gamma^\mu_{\ \alpha\beta}$ are two separated  dynamical structures. Like other fundamental interactions in Nature, gravitation can be reformulated as a gauge theory through TEGR and STEGR. The most peculiar property of gravitation seems to be its \emph{universal character} that all objects, regardless of their internal structure, feel this force, which is encoded in the \emph{Equivalence Principle} of GR. In the teleparallel formulations, the Equivalence Principle is sometimes claimed to be not valid in the literature, instead we will underline how it can be recovered in such theories,  even if it does not lie at their foundation. This fact is extremely relevant because, if the Equivalence Principle  were shown to be violated at some fundamental level, the final theory of gravitation could be non-metric. 

In these equivalent pictures, we can  define alternative ways of representing the  gravitational field, accounting for the same DoFs, related to specific geometric   invariants: the Ricci curvature scalar $R$, the torsion scalar $T$, and the non-metricity scalar $Q$. In this sense, GR, TEGR, and STEGR give rise to the  the Geometric Trinity of Gravity. 

\begin{figure}
    \centering
    \includegraphics[trim=0cm 4cm 0cm 5cm,scale=0.29]{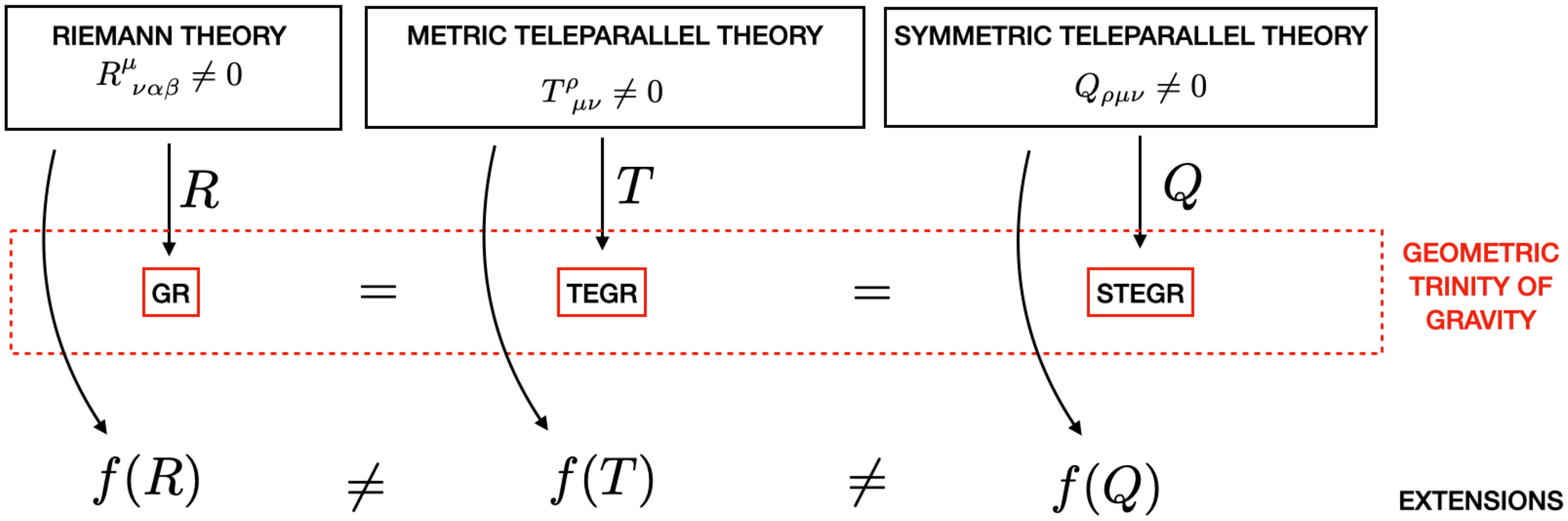}
    \caption{The Geometric Trinity of Gravity and  related extensions. The equivalence holds only for  theories linear in the scalar invariants. Extensions can involve further degrees of freedom which lead to the breaking of equivalence among different representaions of gravity. It can be restored identifying correct boundary terms.}
    \label{fig:Fig3}
\end{figure}

Similarly to GR where we can extend it to $f(R)$ gravity, $f(T)$ and $f(Q)$ gravity are the  extensions of TEGR and STEGR, respectively, where $f$ is a smooth function. It is worth  noticing  that, in general,  the equivalence among the three representations  is not  valid anymore among the  extensions, because they give rise to  dynamics with different DoFs (see Fig. \ref{fig:Fig3}). In particular, in $f(R)$ gravity we have field equations of fourth order, in metric representation, whereas $f(T)$ and $f(Q)$ still remains of second-order \cite{Faraoni2010,Pereira2012,Cai2016}. In addition, in $f(T)$ and $f(Q)$, we cannot choose, in general,  a  gauge to simplify the calculations, as in the cases of  TEGR and STEGR. On  the contrary,  we have to consider   field equations for the spin connection in $f(T)$ and affine field equations for $f(Q)$ \cite{Cai2016,Pereira2012,DAmbrosio2022}. In the next sections, we shall develop these points more in detail. 

\section{Tetrads and Spin Connection}
\label{sec:math_tools}
Before going into details of Trinity Gravity, some  considerations on the mathematical structure are in order.
To define a theory of gravity, we need to fix the underlying geometry,  the transformation properties, and the set of observables. GR is based on the metric tensor from which we can construct the Levi-Civita connection, and finally the curvature, which encodes the gravitational dynamics. The possibility to relate the metric and the geodesic structure, which essentially coincide, rely on the validity of the Equivalence Principle \cite{Faraoni2010}. However, GR can be reformulated also in terms of tetrad \cite{Aldrovandi2013,Pereira2012,Nakahara2003} and spin connection formalisms  \cite{Aldrovandi2013,Martin2019}, giving rise to the teleparallel equivalent theories of GR. In Sec. \ref{sec:tetrad}, we describe in detail the tetrad formalism, whereas, in Sec. \ref{sec:spin_connection}, we introduce the spin connection.

\subsection{ The tetrad formalism}
\label{sec:tetrad}
The geometric setting of any theory of gravity occurs in the tangent bundle, a natural construction always present in any smooth spacetime. In fact, at each point of the spacetime, it is possible to construct the tangent vector space attached to it. In the respective domains of definition, any vector or covector can be expressed in terms of a general linear orthonormal frame called tetrads or {\it vielbeine} (where \qm{viel} = \emph{many} and \qm{beine} = \emph{legs} in German, therefore dreibeine = three legs, vierbeine = four legs, etc.).

A tetrad field is a geometric construction, which permits to easily carry out the calculations on the tangent space, as it will be clearer in the following discussions. Physically, they represent the standard laboratory-apparatus of the observer for carrying out the measurements in space and time. Using a tetrad field means to adopt a \emph{Lagrangian point of view}, which entails to follow an individual fluid parcel as it moves through space and time. A tetrad field establishes a relationship between the manifold and its tangent spaces. This geometric structure is always present, independently of any prior gravity-model assumption. The theoretical framework intervenes to characterize the gravitational effects occurring in this frame. 

We first introduce the definition and properties of the tetrads (see Sec. \ref{sec:tetrads_definition}), and  then we present their anhonolonomy structure (see Sec. \ref{sec:anholonomy}) and its importance in the first Cartan structure equation (see Sec. \ref{sec:cartan_equation}). Finally we describe preferred frames represented by the inertial class and trivial tetrads (see Sec. \ref{sec:trivial_tetrads}).

\subsubsection{Tetrads: definition and properties}
\label{sec:tetrads_definition}
Let us assign a general metric spacetime ($\mathcal{M},g_{\mu\nu}$), being $\mathcal{M}$ a four-dimensional differential manifold of class $C^\infty$, whose tangent spaces $T_p\mathcal{M}$, at each point $p\in\mathcal{M}$, are Minkowski spacetimes with metric $\eta_{AB}$, and $g_{\mu\nu}$ the symmetric metric tensor. In these hypotheses, there exists a \emph{compatible atlas of charts} $\mathcal{A}$, being an open covering of $\mathcal{M}$. Therefore, for each $p\in\mathcal{M}$ there exists a chart $(\mathcal{U},\varphi)$ of domain $\mathcal{U}$, being an open neighbourhood of $p$, and a coordinate map $\varphi:\mathcal{U}\to\varphi(\mathcal{U})\subseteq \mathbb{R}^4$ (being an homeomorphism). In addition, for all $(\mathcal{U},\varphi),(\mathcal{V},\psi)\in \mathcal{A}$, the map $\psi \circ \varphi^{-1}:\varphi(\mathcal{U}\cap\mathcal{V})\to \psi(\mathcal{U}\cap\mathcal{V})$ is a $C^\infty$-diffeomorphism called \emph{coordinate transformation}. Therefore, to each point $p\in\mathcal{M}$, we can associate its \emph{coordinates} by $(x^0,x^1,x^2,x^3):=\varphi(p)\in\mathbb{R}^4$ \cite{Romano2019}. Defined the coordinate $x^\mu$-axes in $\mathbb{R}^4$, it is possible to construct the related coordinate curves $\gamma_{x^\mu}$ on $\mathcal{M}$ via the use of the charts. Therefore, all the parallel curves to coordinate axes in $\mathbb{R}^4$ forms the related grid on $\mathcal{M}$, which permits to uniquely identify the spacetime location of all points.
\begin{figure}[ht!]
    \centering
    \includegraphics[trim=0.7cm 0cm 0cm 3.4cm,scale=0.32]{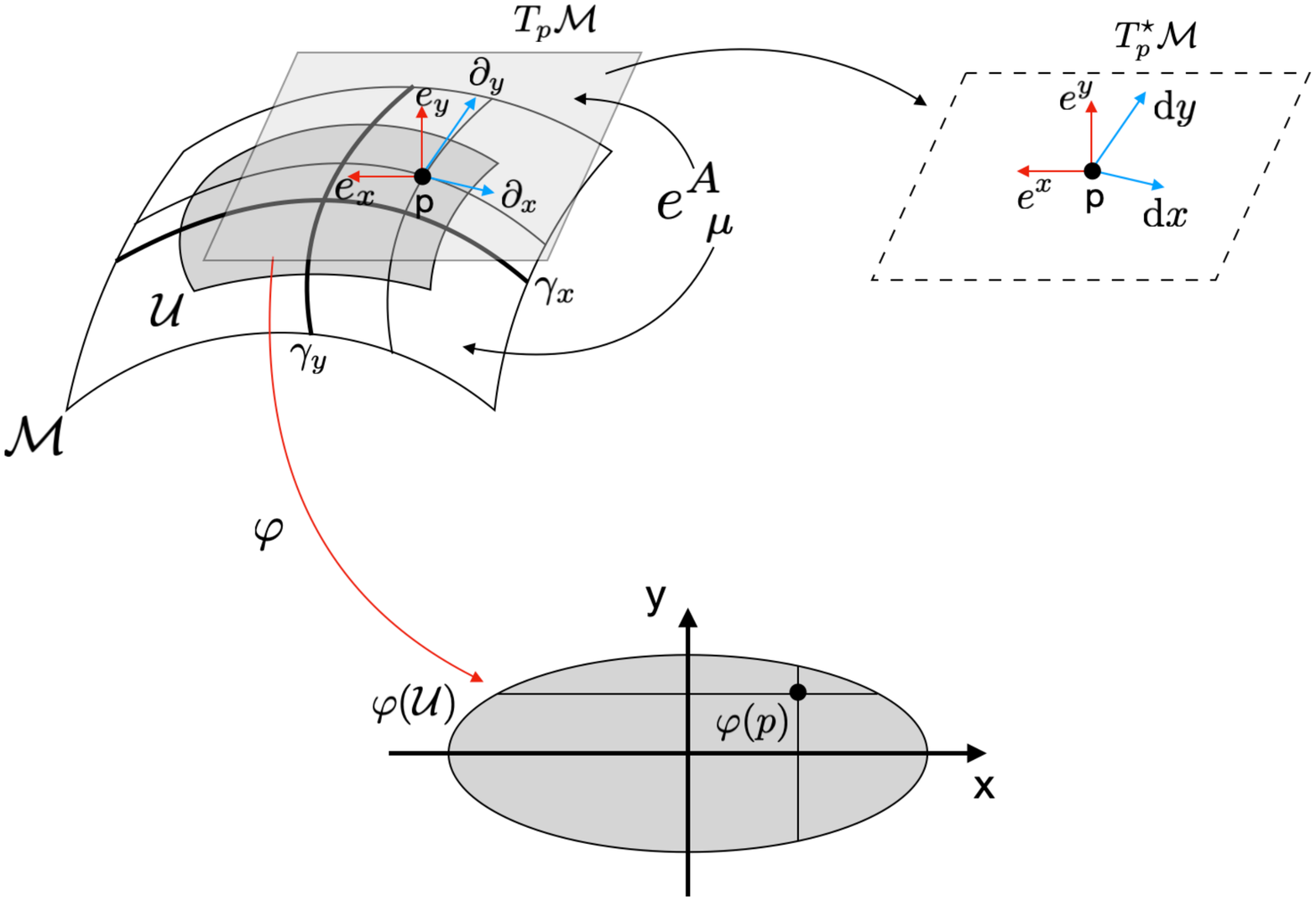}
    \caption{Two-dimensional picture to explain the tetrad formalism. Tetrads $e^A_{\ \mu}$ solder the coordinate chart $(\mathcal{U},\varphi)$ on the manifold $\mathcal{M}$ to the orthonormal basis $\left\{e_x,e_y\right\}$ in the tangent space $T_p\mathcal{M}$. They represent also the coefficients in the natural (holonomic) basis $\left\{\partial_x,\partial_y\right\}$. The coordinate map $\varphi$ assign at each point $p\in\mathcal{U}\subseteq\mathcal{M}$ the coordinates $\varphi(p)=(x,y)\in\varphi(\mathcal{U})\subseteq\mathbb{R}^4$. Passing from $T_p\mathcal{M}$ to the cotangent space $T_p^\star\mathcal{M}$ through $g_{\mu\nu}$ and $\eta_{AB}$, the natural basis $\left\{{\rm d}x,{\rm d}y\right\}$ is transformed into the orthonormal basis $\left\{e^x,e^y\right\}$ through the use of tetrads $e_A^{\ \mu}$.}
    \label{fig:Fig4}
\end{figure}

A natural differentiable basis or \emph{holonomic basis} of each tangent space $T_{p}\mathcal{M}$ is given by a sets of vectors tangent to the coordinate lines at each point $p$, i.e.,
\begin{equation}
\partial_{\mu} :=\left( \frac{\partial}{\partial x^{\mu}}\right)_p,
\end{equation}
as well as for covector fields defined on the cotangent space $T_p^\star\mathcal{M}$ (set of all linear maps $\alpha:T_p\mathcal{M}\to\mathbb{R}$) we have the following basis $\left\{{\rm d} x^{\mu}\right\}$ applied to the point $p\in\mathcal{M}$, which satisfies the orthonormality condition
\begin{equation}
    {\rm d}x^{\mu}\partial_{\nu} = \delta^{\mu}_{\nu}.
\end{equation}
The tangent $T_p\mathcal{M}$ and cotangent $T_p^\star\mathcal{M}$ spaces in $p\in\mathcal{M}$ are related through the metrics $g_{\mu\nu}$ and $\eta_{AB}$.

Every vector or covector applied to a point $p\in\mathcal{M}$ can be expressed in terms of the natural basis. Therefore, we can define a set of orthonormal vectors and covectors, which can be related to the natural basis through \cite{Aldrovandi2013}
\begin{equation} \label{eq:tetrad_basis}
    e_{A} := e^{\ \mu}_A\partial_{\mu}, \qquad e^{A} := e^A_{\ \mu} {\rm d}x^{\mu},
\end{equation}
where the set of coefficients $\left \{ e_{A}^{\ \mu} \right \}$ are called \emph{matrix of tetrad transformation} and belong to the linear group of all real $4\times4$ invertible matrices $GL(4,\mathbb{R})$. The tetrads act as a \emph{soldering agent} between the general manifold (Greek indices) and the Minkowski spacetime (capital Latin indices\footnote{Sometimes, capital Latin indices, referring to local coordinate indices, are also indicated by an over hat on the Greek indices, i.e., $e^A_{\ \mu}=e^{\hat \nu}_{\ \mu}$.}) as follows
\begin{equation}\label{eq:soldering_effect}
    g_{\mu\nu} = \eta_{AB}e^{A}_{\ \mu}e^{B}_{\ \nu},\qquad 
    \eta_{AB} = g_{\mu\nu}e^{\ \mu}_{A}e^{\ \nu}_{B}.
\end{equation}
Therefore, a tetrad field is a linear frame gluing together the coordinate charts on $\mathcal{M}$ to the preferred orthonormal basis $e_A$ on the tangent space, where calculations can be carried out in a considerably simplified manner. As $g_{\mu\nu}$ varies from point to point on the manifold $\mathcal{M}$, the vierbeine $e^{\ \mu}_A$ do the same. Calculating the determinant of (\ref{eq:soldering_effect}), we obtain $-g=e^2$, where $e$ denotes the determinant of $e^{\ \mu}_A$ and it is negative owed to the signature of $\eta_{AB}$. Generally speaking, we note that the vierbeine represent the square root of the metric. In Fig. \ref{fig:Fig4} the tetrads together with their properties are displayed. 

\subsubsection{Anholonomy of tetrad frames}
\label{sec:anholonomy}
Let us now analyse one of the  consequences in  using of the tetrad fields. A general tetrad basis $\left \{ e_A \right \}$ (cf. Eq. \eqref{eq:tetrad_basis}) satisfies the commutation relation \cite{Aldrovandi2013,Martin2019}
\begin{align}\label{eq:commutatotor_tetrad}
    [e_A,e_B] &:=e_Ae_B-e_Be_A\notag\\
    &=(e^{\ \mu}_A\partial_\mu)(e^{\ \nu}_B\partial_\nu)-(e^{\ \nu}_B\partial_\nu)(e^{\ \mu}_A\partial_\mu)\notag\\
    &=\left[e^{\ \mu}_A e^C_{\ \nu}(\partial_\mu e^{\ \nu}_B)-e^{\ \nu}_B e^C_{\ \mu}(\partial_\nu e^{\ \mu}_B)\right]e_C\notag\\
    &=e^{\ \mu}_A e^{\ \nu}_B \left[\partial_\nu e^C_{\ \mu}-\partial_\mu e^C_{\ \nu} \right]e_C\notag\\
    &=f^C_{\ AB} e_C,
\end{align}
where we have set 
\begin{equation} \label{eq:coeff_anhon}
f^C_{\ AB}:= e^{\ \mu}_A e^{\ \nu}_B \left[\partial_\nu e^C_{\ \mu}-\partial_\mu e^C_{\ \nu} \right],   
\end{equation}
which are known  as \emph{structure constants} or \emph{coefficients of anholonomy} related to the frame $\left\{e_A\right\}$. They quantify the failure of the parallelogram closure generated by the vectors $e_A$ and $e_B$. In general, when
$f^C_{\ AB}\neq0$, the tetrad basis is \emph{anholonomic} or \emph{non-trivial}, and the coefficients of anholonomy specify how much they depart from being holonomic. This approach reveals important properties of the underlying geometric framework on which we are working. In GR, they have been used in the  \emph{Bianchi classification}, which leads to eleven possible different spacetimes, useful to develop cosmological models \cite{Landau1975,Ryan1975,Stephani2009}. 

\subsubsection{The first Cartan structure equation}
\label{sec:cartan_equation}
Given a 1-form $\omega$ and defined ${\rm d}\omega$ as the exterior derivative, it can be written in components as
\begin{equation}
{\rm d}\omega=\partial_\mu \omega_\nu {\rm d}x^\mu\wedge{\rm d}x^\nu,     
\end{equation}
where $\wedge$ is the external product defined as
\begin{equation}
 {\rm d}x^\mu\wedge{\rm d}x^\nu={\rm d}x^\mu\otimes{\rm d}x^\nu-{\rm d}x^\nu\otimes{\rm d}x^\mu,   
\end{equation}
with $\otimes$ the tensorial product. Due to the antisymmetry of the exterior product and the Schwarz theorem, we have ${\rm d}^2\omega=0$ thanks to the \emph{Poincar\'e lemma} \cite{Nakahara2003}. 

We consider the 2-form ${\rm d}\omega$ applied to two vectors $u=u^\mu\partial_\mu,\ v=v^\nu\partial_\nu$, which can be written as \cite{Deruelle2018}
\begin{equation}
{\rm d}\omega(u,v)=u\omega(v)-v\omega(u)-\omega([u,v]_{\mathcal{L}}),
\end{equation}
where
\begin{subequations}
\begin{align}
{\rm d}\omega(u,v)&:=\partial_\mu \omega_\nu (u^\mu v^\nu-u^\nu v^\mu),\\
u\omega(v)&:=u^\mu v^\nu\partial_\mu \omega_\nu+u^\mu \omega_\nu \partial_\mu v^\nu,\\
\omega([u,v]_{\mathcal{L}})&:=\omega_\nu (u^\mu\partial_\mu v^\nu-v^\mu\partial_\mu u^\nu),
\end{align}
\end{subequations}
with $\omega=\omega_\mu {\rm d}x^\mu$ and $[u,v]_{\mathcal{L}}\equiv(\mathcal{L}_uv):=(u^\mu\partial_\mu v^\nu-v^\mu\partial_\mu u^\nu)\partial_\nu$. It is the Lie bracket or the Lie derivative of the vector field $v$ with respect to the vector field $u$. It is important to note that ${\rm d}\omega(u,v)$ produces a scalar. 

If we consider the tetrad basis $\left\{e_A\right\}$ and take $\omega=e^A$, then we have the following relation \cite{Deruelle2018}
\begin{align}
\left\{{\rm d}e^C(e_A,e_B)\right\}e_C&=\left\{e_A[e^C(e_B)]-e_B[e^C(e_A)]\right.\notag\\
&\left.-e^C([e_A,e_B]_{\mathcal{L}})\right\}e_C\notag\\
&=-e^C([e_A,e_B]_{\mathcal{L}}^Le_L)e_C\notag\\
&=-[e_A,e_B]_{\mathcal{L}}. \label{eq:relation_form}
\end{align}

Assigned a general metric-compatible affine connection $\Gamma^\lambda_{\ \alpha\beta}$, and the associated covariant derivative $\nabla$, we have
\begin{equation}
\nabla_{e_A}e_B=\gamma^C_{\ AB}e_C,    
\end{equation}
where $\gamma^C_{\ AB}$ are the \emph{Ricci rotation coefficients}, which measure the rotation of all frame tetrads when moved in various directions, encoding thus gravitational and non-inertial effects \cite{Nakahara2003,Deruelle2018}. When we use the natural basis, they reduce to the affine connection $\Gamma^C_{\ AB}$. It is important to note that such coeffiecients arise also in a flat spacetime when, generally, non-liner coordinates are exploited, since they give rise to non-inertial effects. In particular, in the considered tetrad basis, they assume the following expression and symmetries \cite{Nakahara2003}
\begin{align} \label{eq:simmetry_RRC}
\gamma_{\lambda\nu\mu}&:=e^A_{\ \mu}e^B_{\ \lambda}\nabla_A (e_\nu)_B\notag\\
&=-e^A_{\ \mu}(e_\nu)_B\nabla_A e^B_{\ \lambda}\notag\\
&=-e^A_{\mu}e^B_{\ \nu}\nabla_A (e_{ \lambda})_B=-\gamma_{\nu\lambda\mu},
\end{align}
where we have used the compatibility condition in the last equality. $\gamma^C_{\ AB}$ can be seen as the action of the \emph{connection 1-forms} $\omega^C_{\ B}$ on the tetrad basis $e_A$, i.e., \cite{Deruelle2018}
\begin{equation} \label{eq:connection_1form}
\gamma^C_{\ AB}=\omega^C_{\ B}(e_A)\quad \Leftrightarrow\quad \omega^C_{\ B}= \gamma^C_{\ AB}e^A.    
\end{equation}

Since we know that $\nabla_\mu \partial_\nu=\Gamma^\lambda_{\ \mu\nu}\partial_\lambda$, if we consider the commutator of $\nabla_\mu$ and $\partial_\nu$ we obtain\footnote{Here, we use the usual definition of torsion tensor, i.e., opposite to that defined in Eq. \eqref{eq:tor_ten}. This permits to easily follow the forthcoming discussions.}
\begin{align} \label{eq:commut_cov_der}
[\nabla_\mu,\partial_\nu]&=\nabla_\mu \partial_\nu-\nabla_\nu \partial_\mu\notag\\
&=\left(\Gamma^\lambda_{\ \mu\nu}-\Gamma^\lambda_{\ \nu\mu}\right)\partial_\lambda\notag\\
&=T^\lambda_{\ \mu\nu}\partial_\lambda,
\end{align}
where $T^\lambda_{\ \mu\nu}$ is the torsion tensor measuring the antisimmetry of the affine connections. In a coordinate-independent approach, the torsion $T$ (associated to the covariant derivative $\nabla$) is a $(1,2)$-type tensor, which acts on pairs of vectors $(v, u)$ to give another vector according to the following relation \cite{Misner1973,Deruelle2018}
\begin{align} \label{eq:torsion}
T(v,u):=\nabla_v u-\nabla_u v-[v,u]_{\mathcal{L}}.    
\end{align}
Applying Eq. \eqref{eq:torsion} to $\left\{e_A\right\}$, exploiting Eq. \eqref{eq:relation_form}, and considering $\omega^C_{\ B}(e_A)=(\omega^C_{\ D}\otimes e^D)(e_A,e_B)$, we obtain
\begin{align} \label{eq:torsion_onvectors}
T(e_A,e_B)&=\nabla_{e_A}e_B-\nabla_{e_B}e_A-[e_A,e_B]_{\mathcal{L}}\notag\\
&=\left[\omega^C_{\ B}(e_A)-\omega^C_{\ A}(e_B)+{\rm d}e^C(e_A,e_B)\right]e_C\notag\\
&=\left[(\omega^C_{\ D}\wedge e^D+{\rm d}e^C)(e_A,e_B)\right]e_C.
\end{align}
Defined $\Omega^C:=\omega^C_{\ D}\wedge e^D+{\rm d}e^C$ as the \emph{torsion differential 2-form}, Eq. \eqref{eq:torsion_onvectors} can be written as \cite{Nakahara2003,Deruelle2018}
\begin{equation} \label{eq:first_structure_equation}
T=\Omega^C\otimes e_C,    
\end{equation}
which is the \emph{first Cartan structure equation}. In the case of Riemman geometry, namely when the torsion vanishes, Eq. \eqref{eq:first_structure_equation} becomes \cite{Aldrovandi2013,Martin2019}
\begin{align} \label{eq:diff_form}
    de^C &:=-\omega^C_{\ A}\wedge e^A\notag\\
    &=-\frac{1}{2}\left(\gamma^C_{\ AB}-\gamma^C_{\ BA}\right)e^A \wedge e^B\notag\\
        &= -\frac{1}{2} e^{\ \mu}_A e^{\ \nu}_B (\partial_{\nu} e^C_{\ \mu}-\partial_{\mu}e^C_{\ \nu}) e^A \wedge e^B\notag\\
    &=- \frac{1}{2} f^C_{\ AB} e^A \wedge e^B,
\end{align}
where the anhonolonomy coefficients  emerge  as antisymmetric combination of the Ricci rotation coefficients. They are also related to the curls of the tetrad vector derivatives, as  it normally occurs to the components of a differential 2-form \cite{Misner1973,Nakahara2003}.

\subsubsection{Inertial frames and trivial tetrads}
\label{sec:trivial_tetrads}
Among the different frames,  a special class is represented by the \emph{inertial frames}, which can be denoted by $\left\{e'_A\right\}$, whose coefficients of anhonolonomy $f'^C_{\ AB}$ locally satisfy the condition $f'^C_{\ AB} = 0$. For Eq. \eqref{eq:diff_form} we have ${\rm d}e'^A=0$, which is locally exact and can be written as $e'^A={\rm d}x'^A$ and therefore it is holonomic. Therefore, all coordinate bases belong to this family. It is worth noting that \emph{this is not a local property, but it holds everywhere for all frames being part of this inertial class} \cite{Aldrovandi2013}.   

\begin{figure}[ht!]
    \centering
    \includegraphics[scale=0.33]{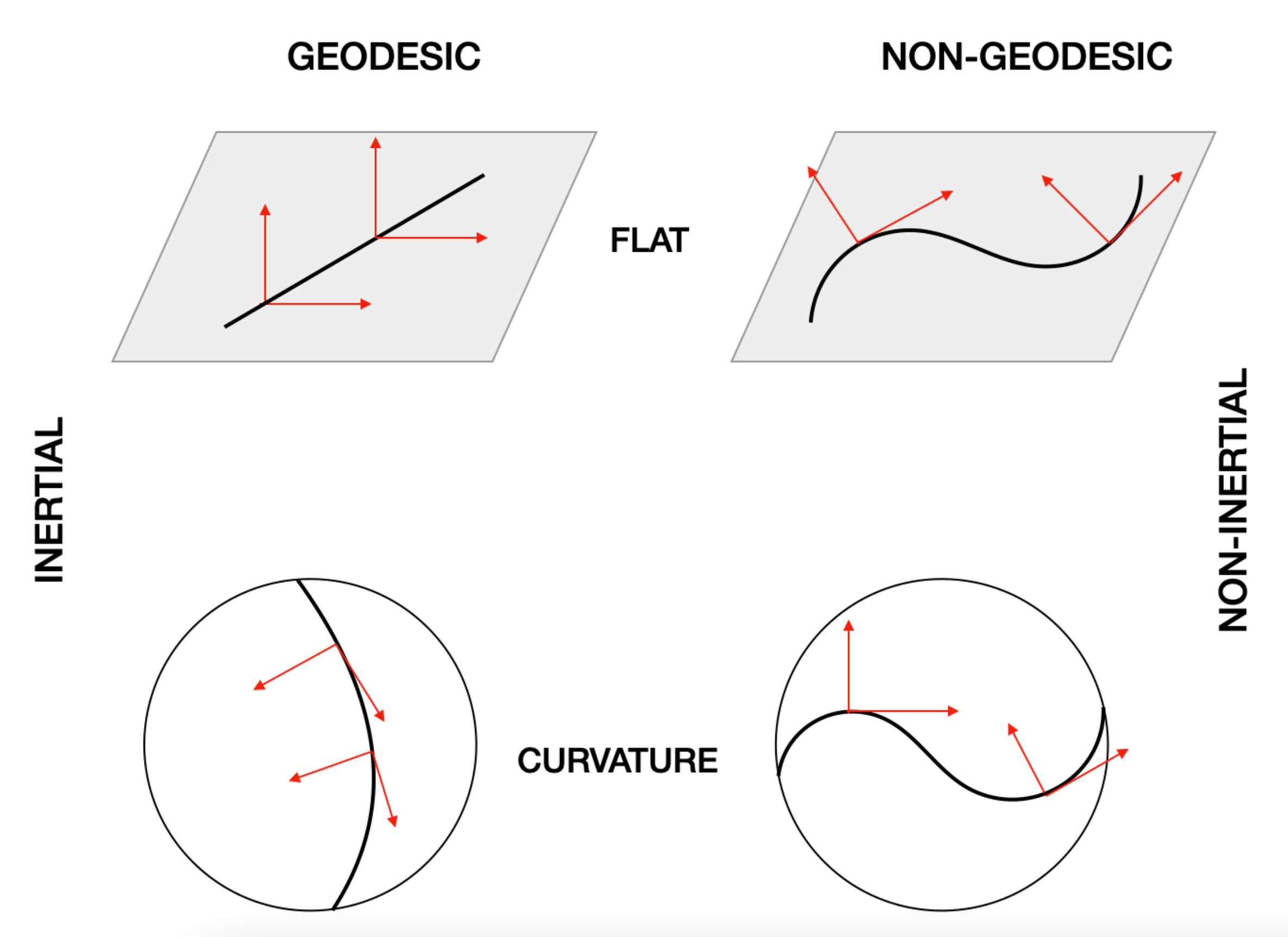}
    \caption{This figure shows how tetrads behave in terms of inertial and gravitational effects. When no gravity is present, and we consider inertial effects only (i.e., we move along geodesics), we obtain trivial (holonomic) tetrads, whereas when non-inertial contributions take place (i.e., following non-geodesic orbits) the tetrads become anholomic. The situation is analogue when gravitation is switched on. Along geodesic we obtain inertial frames, whereas along non-geodesic trajectories we have the most general anholomic frames.}
    \label{fig:Fig6}
\end{figure}

In absence of gravitation, the anholonomy is only caused by inertial forces present in these frames. The metric $g_{\mu\nu}$ reduces to the Minkowski metric $\eta_{\mu\nu}$. In all coordinate systems, $\eta_{\mu\nu}$ is a function of the spacetime point, and independently of whether $\left\{e_A\right\}$ is holonomic (inertial) or not. In this case, tedrads always relate the tangent Minkowski space to a Minkowski spacetime
\begin{equation}\label{eq:trivial_tetrads}
    \eta_{AB} = \eta_{\mu\nu} e^{\ \mu}_{A} e^{\ \nu}_{B}.
\end{equation}
These are the frames appearing in Special Relativity, which are usually called \emph{trivial frames} or \emph{trivial tetrads}. They are  very useful when we deal with  spaces involving torsion \cite{Martin2019}. Of course, in absence of inertial forces, the class of inertial frames is, consequently, represented by vanishing structure coefficients. These concepts are  sketched in Fig. \ref{fig:Fig6}.

\subsection{The spin connection}
\label{sec:spin_connection}
The spin connection plays a fundamental role when we deal with tetrads, because it encodes the inertial effects occurring in the considered frame. Let us briefly recall the fundamental properties of the Lorentz group (see Sec. \ref{sec:Lorentz_group}),  then we discuss the associated Lorentz algebra as well as its properties (see Sec. \ref{sec:Lorentz_algebra}). Lorentz connections will be first introduced under a mathematical point of view (see Sec. \ref{sec:Lorentz_connection_mathematical}) together with the fundamental tetrad postulate (see Sec. \ref{sec:tetrad_postulate}), and then the same subject will be considered under a physical perspective (see Sec. \ref{sec:Lorentz_connection_physical}). 

\subsubsection{The Lorentz group}
\label{sec:Lorentz_group}
Electromagnetism is framed under the standard of Special Relativity by postulating \cite{Romano2019}:
\begin{enumerate}
    \item[1)] \emph{the optical isotropy principle:} all inertial frames are optically isotropic, i.e., the light propagates in these frames with velocity $c=1/\sqrt{\epsilon_0\mu_0}$ in any direction;
    \item[2)] \emph{the principle of relativity:} the laws of physics assume the same form in all inertial reference frames.
\end{enumerate}
Given two inertial frames and assuming that one is moving with respect to the other with uniform velocity $\boldsymbol{v}:=(v^1,v^2,v^3)$, the \emph{Lorentz transformation} is a linear (affine) map relating the temporal and spatial coordinates of the two inertial observers \cite{Romano2019}
\begin{equation} \label{eq:coordinate_transform}
\Lambda^\mu_{\ \nu}:x^\mu\longrightarrow x'^\mu=\Lambda^\mu_{\ \nu}(x)x^\nu,
\end{equation}
which leaves invariant the following quadratic form
\begin{equation} \label{eq:quadratic_form}
\eta_{\mu\nu}x^\mu x^\nu=-t^2+x^2+y^2+z^2.    
\end{equation}
A general Lorentz transformation is given by \cite{Weinberg1995}
\begin{equation} \label{eq:general_Lorentz}
\Lambda^\alpha_{\ \beta}=\mathcal{G}\cdot\begin{bmatrix}
\gamma & -\gamma \mathcal{R}^i_{\ j}\frac{v^j}{c}\\
 -\gamma \mathcal{R}^i_{\ j}\frac{v^j}{c} & \mathcal{R}^i_{\ j}\left(\delta^i_j+(\gamma-1)\frac{v^iv^j}{v^2}\right)
\end{bmatrix},
\end{equation}
where $v\equiv|\boldsymbol{v}|:=\sqrt{(v^1)^2+(v^2)^2+(v^3)^2}$ is the modulus of the spatial velocity $\boldsymbol{v}$, $\gamma:=(1-\frac{v^2}{c^2})^{-1/2}$ is the Lorentz factor, $\mathcal{R}^i_{\ j}$ is a rotation matrix, and $\mathcal{G}$ is one of the following operators $\left\{\mathbbm{1},\mathbbm{P},\mathbb{T},\mathbb{P}\cdot\mathbb{T}\right\}$ with
\begin{subequations}
\begin{align}
\mathbbm{1}&:={\rm diag}(1,1,1,1),\\
\mathbbm{P}&:={\rm diag}(1,-1,-1,-1),\\
\mathbbm{T}&:={\rm diag}(-1,1,1,1),
\end{align}
\end{subequations}
being the unitary, parity, and time reversal operators, respectively. The expression of $\Lambda^\alpha_{\ \beta}$ shows that a Lorentz transformation is defined in terms of six parameters: three related to the rotation angles and the other three to the components of the spatial velocity $\boldsymbol{v}$.

The set of all Lorentz transformations of Minkowski spacetime forms the \emph{(homogeneous) Lorentz orthogonal group} $O(1,3)$. The requirement \eqref{eq:quadratic_form}, together with \eqref{eq:coordinate_transform}, entails, in matrix notation, that $\eta=\Lambda^T\eta\Lambda$. This gives rise to ${\rm det}^2\Lambda=1$, namely \emph{proper} (${\rm det}\Lambda=1$) and \emph{improper} (${\rm det}\Lambda=-1$) Lorentz transformations, which can be further subdivided (cf. Eq. \eqref{eq:general_Lorentz}) in \emph{orthochronous} ($\Lambda^0_{\ 0}\ge1$) and \emph{non-orthochronous} ($\Lambda^0_{\ 0}\le-1$) \cite{Weinberg1995,Maggiore2005}. The proper orthochronous Lorentz transformations form the \emph{restricted Lorentz special orthogonal group} $SO^+(1,3)$. Therefore, \emph{the Lorentz group is a six-dimensional, non-compact, non-Abelian, and real Lie group endowed with four connected components} \cite{Weinberg1995,Maggiore2005}.
The Lorentz group is closely involved in all known fundamental laws of Nature describing the related symmetries of space and time. In particular, in GR, we  consider the  \emph{local Lorentz invariance}, because in every small enough regions of spacetime, thanks to the Equivalence Principle,  the gravitational effects can be neglected, i.e., this occurs in the \emph{local inertial frame} (LIF), which permits to recover the Special Relativity physics.

At each point of a Riemannian spacetime, the metric $g_{\mu\nu}$ determines a tetrad up to the local Lorentz transformations in the tangent space. In other words, a tetrad vector (covector) base $\left \{ e_A \right \}$ ($\left \{ e^A \right \}$) is not unique, because it is always possible to find another base $\left \{ \bar{e}_A \right \}$ ($\left \{ \bar{e}^A \right \}$) by performing a local Lorentz transformation, namely
\begin{equation}\label{eq:change_tetrad}
    \bar{e}{}^{A}_{\ \mu} = \Lambda^{A}_{\ B}e^{B}_{\ \mu},
\end{equation}
such that
\begin{equation}\label{eq:constraint_tetrad}
    g_{\mu\nu} = \eta_{AB}\bar{e}^A_{\ \mu}\bar{e}^B_{\ \nu}\qquad \eta_{AB} =g_{\mu\nu} \bar{e}_A^{\ \mu}\bar{e}_B^{\ \nu}.
\end{equation}

\subsubsection{The Lorentz algebra}
\label{sec:Lorentz_algebra}
Another important feature of the Lorentz group is that it admits a \emph{Lorentz algebra} $\mathfrak{L}$ \cite{Weinberg1995,Maggiore2005}. If we consider an infinitesimal transformation in $SO^+(1,3)$ we have
\begin{equation}
\Lambda^\alpha_{\ \beta}=\delta^\alpha_\beta+\omega^\alpha_{\ \beta}+\mathcal{O}[(\omega^\alpha_{\ \beta})^2].    
\end{equation}
Applying $\eta=\Lambda^T\eta\Lambda$, at linear order in $\omega^\alpha_{\ \beta}$,  $\omega_{\mu\nu}=-\omega_{\nu\mu}$ is an antisymmetric $4\times4$ matrix with six independent indices. Therefore, we can associate six generators to the Lorentz algebra labeled by $J_{AB}$, with $J_{AB}=-J_{BA}$ \cite{Maggiore2005}, where each of them can be expressed in the \emph{four-vector representation} by a $4\times4$ matrix as follows
\begin{equation} \label{eq:generators}
\left(J_{AB}\right)^C_{\ D}:=2i\eta_{[B|D}\delta^C_{A]}=i(\eta_{BD}\delta^C_A-\eta_{AD}\delta^C_B).
\end{equation}
Each element of the Lorentz group can be written as \cite{Maggiore2005}
\begin{equation}
\Lambda=e^{\frac{i}{2}\omega_{AB}J^{AB}}.    
\end{equation}

\subsubsection{The derivation of Lorentz connection}
\label{sec:Lorentz_connection_mathematical}
Some geometric objects with an established behaviour may lose the covariant character under point-dependent transformations, e.g., ordinary derivative of covariant objects. In order to supply for this defective behaviour, it is fundamental to introduce \emph{connections} $\omega_\mu$ fulfilling the following properties: $(i)$ they behave like vectors in the spacetime indices; $(ii)$ they act as non-tensor in the algebraic indices to compensate this effect and to reestablish the correct tensorial trend. The \emph{linear connections} fulfilling these requirements belong to the subgroup $SO^+(1,3)$ of $GL(4,\mathbb{R})$, and they are dubbed \emph{Lorentz connections}. It is worth  noticing that all Lorentz connections exhibit  the presence of torsion (see Ref. \cite{Aldrovandi2013}, and discussions therein).

A Lorentz connection,  also known as \emph{spin connection}, $\omega_\mu$ is a 1-form acting in the Lorentz algebra, namely
\begin{equation}\label{eq:spin_connection}
\omega_\mu: J_{AB}\in \mathfrak{L}\longrightarrow  \omega_\mu:=\frac{1}{2}\omega^{AB}{}_{\mu}J_{AB},    
\end{equation}
where $\omega^{AB}{}_{\mu}$ are the spin connection coefficients, which are antysymmetric in the $AB$ indices owed to the antisymmetry of $J_{AB}$, i.e., $\omega^{AB}{}_{ \mu}=-\omega^{BA}{}_{\mu}$. This permits to introduce the \emph{Fock-Ivanenko covariant derivative} \cite{Capozziello:2009dz,Aldrovandi2013}
\begin{equation}\label{eq:Fock_Ivanenko_CD}
    \mathcal{D}_{\mu}:=\partial_{\mu}-\omega_\mu=  \partial_{\mu} - \frac{i}{2} \omega^{AB}{}_{\mu} J_{AB}.
\end{equation}
where $J_{AB}$ is the generator in the appropriate representation of the Lorentz group. The right member of Eq. \eqref{eq:Fock_Ivanenko_CD} acts only on tangent (algebraic) space indices. If we apply Eq. \eqref{eq:generators} to the field $e^C$ we obtain
\begin{align} \label{eq:FIderivative}
\mathcal{D}_\mu e^C&=\partial_\mu e^C-\frac{i}{2} \omega^{AB}{}_{\mu}\left[i(\eta_{BD}\delta^C_A-\eta_{AD}\delta^C_B)\right]e^D\notag\\
&=\partial_\mu e^C+\frac{1}{2}\left[\omega^A{}_{D\mu}\delta^C_A+\omega^B{}_{D\mu}\delta^C_B\right]e^D\notag\\
&=\partial_\mu e^C+\omega^C{}_{D\mu} e^D.
\end{align}
Considering Eq. \eqref{eq:FIderivative} and splitting $e^A$ by Eq. \eqref{eq:tetrad_basis}, we obtain the following expressions
\begin{subequations}
\begin{align}
\mathcal{D}_\mu(e^C_{\ \lambda}dx^\lambda)&=\mathcal{D}_\mu(e^C_{\ \lambda}){\rm d}x^\lambda+e^C_{\ \lambda}\mathcal{D}_\mu({\rm d}x^\lambda)\notag\\
&=\mathcal{D}_\mu(e^C_{\ \lambda}){\rm d}x^\lambda+e^C_{\ \lambda}(\delta^\lambda_\mu+e_E^{\ \lambda}e^D_{\ \mu}\omega^E{}_{D\rho}{\rm d}x^\rho)\notag\\
&=\mathcal{D}_\mu(e^C_{\ \lambda}){\rm d}x^\lambda+e^C_{\ \mu},\label{eq:FIRST_DER}\\
\mathcal{D}_\mu(e^C_{\ \lambda}dx^\lambda)&=\partial_\mu(e^C_{\ \lambda}dx^\lambda)+\omega^C{}_{D\mu}e^D_{\ \lambda}{\rm d}x^\lambda\notag\\
&=\partial_\mu(e^C_{\ \lambda})dx^\lambda+e^C_{\ \mu}+\omega^C{}_{D\mu}e^D_{\ \lambda}{\rm d}x^\lambda. \label{eq:SECOND_DER}
\end{align}
\end{subequations}
Equating Eq. \eqref{eq:FIRST_DER} with \eqref{eq:SECOND_DER} we obtain
\begin{equation} \label{eq:deriv_tetr}
\mathcal{D}_\mu(e^C_{\ \lambda})=\partial_\mu(e^C_{\ \lambda})+\omega^C{}_{D\mu}e^D_{\ \lambda}.
\end{equation}

\subsubsection{The tetrad postulate}
\label{sec:tetrad_postulate}
In non-coordinate bases $\left\{ e_A \right\}$, the covariant derivative $\tilde{\nabla}$ of an algebraic (1,1) tensor $X^A_{\ B}$ can be written in terms of the spin connection as
\begin{equation}
\tilde{\nabla}_\mu X^A_{\ B}:=\partial_\mu +\omega^A{}_{C\mu}X^C_{\ B}-\omega^C{}_{B\mu}X^A_{\ C}.   
\end{equation}
Instead, the covariant derivative of a vector $V$, considered in the coordinate bases $\left\{\partial_\mu\right\}$, is
\begin{align} \label{eq:connection_norm}
\nabla V&=(\nabla_\mu V^\nu) {\rm d}x^\mu\otimes\partial_\nu\notag\\
&=(\partial_\mu V+\Gamma^\nu_{\ \mu\lambda}V^\lambda){\rm d}x^\mu\otimes\partial_\nu.
\end{align}
If we consider now the same vector $V$ written in a mixed basis, tetrad and coordinate basis, gives
\begin{align} \label{eq:connection_tilde}
 \tilde{\nabla} V&=(\tilde{\nabla}_\mu V^A){\rm d}x^\mu\otimes
 e_A\notag\\
 &=(\partial_\mu V^A+\omega^A{}_{B\mu}V^B){\rm d}x^\mu\otimes e_A\notag\\
 &=[\partial_\mu(e^A_{\ \lambda}V^\lambda)+\omega^A{}_{B\mu}e^B_{\ \lambda}V^\lambda]{\rm d}x^\mu\otimes (e^{\ \nu}_A\partial_\nu)\notag\\
 &=[\partial_\mu V^\nu+(e^{\ \nu}_A \partial_\mu e^A_{\ \lambda}+\omega^A{}_{B\mu}e^{\ \nu}_A e^B_{\ \lambda})V^\lambda]{\rm d}x^\mu\otimes\partial_\nu\notag\\
 &=[\partial_\mu V^\nu+(e^{\ \nu}_A\mathcal{D}_\mu e^A_{\ \lambda})V^\lambda]{\rm d}x^\mu\otimes\partial_\nu.
\end{align}
This is a crucial point, because the operations \eqref{eq:connection_norm} and \eqref{eq:connection_tilde} are in principle distinct. However, it is reasonable to assume $\nabla\equiv \tilde{\nabla}$, because the same covariant derivative of a vector cannot change in terms of which type of basis one chooses. This is the so-called \emph{tetrad postulate}, which is valid for any affine connection, defined on a smooth manifold $\mathcal{M}$, and no metric is involved. 

Therefore, it implies (cf. Eqs. \eqref{eq:connection_norm} and \eqref{eq:connection_tilde}) 
\begin{equation}\label{eq:gamma_lorentz}
    \Gamma^{\lambda}_{\ \mu \nu} \equiv e^{\ \lambda}_A \mathcal{D}_{\mu} e^A_{\ \nu}.
\end{equation}
This identity entails several significant implications on the spin connections: $(i)$ since it does not possess a tensorial character, it acquires a non-homogeneous term under the Fock-Ivanenko covariant derivative owed to the affine connection \cite{Aldrovandi2013}; $(ii)$ a spin connection is naturally induced by the affine connection; $(iii)$ it can be also regarded as the gauge field generated by local Lorentz transformations; $(iv)$ inverting Eq. \eqref{eq:gamma_lorentz} with respect to the spin connection, we obtain \cite{Aldrovandi2013}
\begin{equation}\label{eq:spin_connect_der}
    \omega^A_{\ B\mu} = e^A_{\ \lambda} e^{\ \nu}_{B} \Gamma^{\lambda}_{\ \mu\nu} + e^{A}_{\ \sigma}\partial_{\mu}e^{\ \sigma}_{B} \equiv e^{A}_{\ \nu} \nabla_{\mu} e^{\ \nu}_{B};
\end{equation} 
$(v)$ according to Eq. \eqref{eq:spin_connect_der}, the connection 1-form $\omega^C_{\ B}$ (cf. Eqs. \eqref{eq:connection_1form}, \eqref{eq:simmetry_RRC}) can be written as
\begin{equation}
\omega^{AB}=\omega^{AB}{}_\mu {\rm d}x^\mu,    
\end{equation}
and \emph{the Ricci rotation coefficients are the spacetime indices of the spin connection components}; $(vi)$ the covariant derivative of the tetrad, expressed in terms of the affine and spin connections, vanishes identically (cf. Eq. \eqref{eq:spin_connect_der}), namely
\begin{equation}\label{eq:tetrad_postulate}
    \nabla_{\mu}e^{A}_{\nu} = \partial_{\mu}e^{A}_{\ \nu}-\Gamma^{\lambda}_{\ \mu\nu}e^{A}_{\ \lambda}+\omega^{A}_{\ B\mu}e^{B}_{\ \nu} = 0;
\end{equation}
$(vii)$ we note that $\nabla_{\mu}$ is the covariant derivative linked to the connection $\Gamma^{\lambda}_{\mu\nu}$ when acts on external indices and can be defined for tensorial fields, whereas the Fock-Ivanenko derivative $\mathcal{D}_{\mu}$ acts on internal indices and can be defined for all tensorial and spinorial fields \cite{Aldrovandi2013}; $(viii)$ from the metric compatibility condition, we obtain a sort of consistency check given by (cf. Eqs. \eqref{eq:deriv_tetr} and \eqref{eq:gamma_lorentz})
\begin{align} \label{eq:compatibility}
0&=\nabla_\lambda g_{\mu\nu}=\partial_\lambda g_{\mu\nu}-\Gamma^\sigma_{\ \lambda\mu}g_{\sigma\nu}-\Gamma^\sigma_{\ \lambda\nu}g_{\mu\sigma}\notag\\
&=\partial_\lambda(e_{\ \mu}^A e_{\ \nu}^B \eta_{AB})-e_A^{\ \sigma}g_{\sigma\nu}\mathcal{D}_\lambda e^A_{\ \mu} -e_A^{\ \sigma}g_{\mu\sigma}\mathcal{D}_\lambda e^A_{\ \nu}\notag\\
&=-e^A_{\ \nu}e^D_{\ \mu}(\omega_{AD\lambda}-\omega_{DA\lambda}),
\end{align}
which implies $\omega_{AB\mu} = -\omega_{BA\mu}$, i.e., $\omega_{AB\mu}$ is Lorentzian. If the metric postulate (\ref{eq:compatibility}) is not valid, the corresponding spin connection cannot assume values in the  Lorentz algebra, because it is not a Lorentz connection \cite{Aldrovandi2013}. Therefore, we have this equivalence: \emph{metric compatibility holds if and only if we choose a Lorentz connection}.

\subsubsection{Physical considerations on the Lorentz connection}
\label{sec:Lorentz_connection_physical}
We have seen how the tetrads transform under local (point-dependent) Lorentz transformations $\Lambda^{A}_{\ B}(x)$ (cf. Eq. \eqref{eq:change_tetrad}), and now let us apply the same transformations to the spin connections. Let us first consider the inertial frames (see Sec. \ref{sec:trivial_tetrads}) $\left\{e'{}^A_{\ \mu}\right\}$, which, in general coordinates $\left\{x'{}^\mu\right\}$, can be written in the holonomic form $e'{}^A_{\ \mu}=\partial_\mu x'{}^A$, where $x'{}^A=x'{}^A(x^\mu)$ is a point-dependent vector. Under a local transformation $x^A=\Lambda^A_{\ B}(x)x'{}^B$,  we have $e^A_{\ \mu}=\Lambda^A_{\ B}(x)e'{}^B_{\ \mu}$ by transforming the vectors $x^A$ and $x'{}^A$ in the coordinate base $\left\{\partial_\mu\right\}$. 

Let us evaluate $\partial_\mu x'{}^A$, which gives ($\partial'_A\equiv \partial/\partial  x'{}^A$) 
\begin{align}
\partial_\mu x'{}^A&=\partial_\mu(\Lambda^A_{\ B}(x) x^B)\notag\\
&=(\partial_\mu x^B)\Lambda^A_{\ B}(x)+x^B(\partial_\mu \Lambda^A_{\ B}(x)),\\
\partial_\mu x'{}^A&=e'{}^C_{\ \mu}\partial'_C x'{}^A=e'{}^A_{\ \mu}=e^C_{\ \mu}\Lambda^A_{\ C}(x).
\end{align}
Therefore, gathering together the above results, we have (using Eq. \eqref{eq:FIderivative} and $\mathcal{D}_\mu x^A=e^A_{\ \mu}$)
\begin{equation}\label{eq:tetrad_and_spincon}
    e^{A}_{\ \mu} = \partial_{\mu}x^{A} + \overset{\fullcirc}{\omega}^{A}{}_{B\mu}x^{B} \equiv \mathcal{D}_{\mu}x^{A},
\end{equation}
where
\begin{equation}\label{eq:Lorentz_conn_and_spn_con}
    \overset{\fullcirc}{\omega}^A{}_{B\mu}:= \Lambda^{A}_{\ C}(x)\partial_{\mu}\Lambda^{C}_{\ B}(x)
\end{equation}
is defined as a \emph{purely inertial spin connection}, because it physically manifests the inertial effects occurring in the Lorentz rotated frame $e^A_{\ \mu}$. From Eq. \eqref{eq:Lorentz_conn_and_spn_con}, we learn that \emph{the Lorentz connections physically represent the inertial effects present in a given frame. In the inertial frames (i.e., $e'{}^A_{\ \mu}=\partial_\mu x'{}^A$), these effects are absent since the Lorentz connections vanish, $\omega'^{AB}{}_\mu=0$ for Eq. \eqref{eq:tetrad_and_spincon}} \cite{Martin2019}. 

\begin{figure} [ht!]
    \centering
    \includegraphics[trim=0.4cm 2cm 0cm 3cm,scale=0.33]{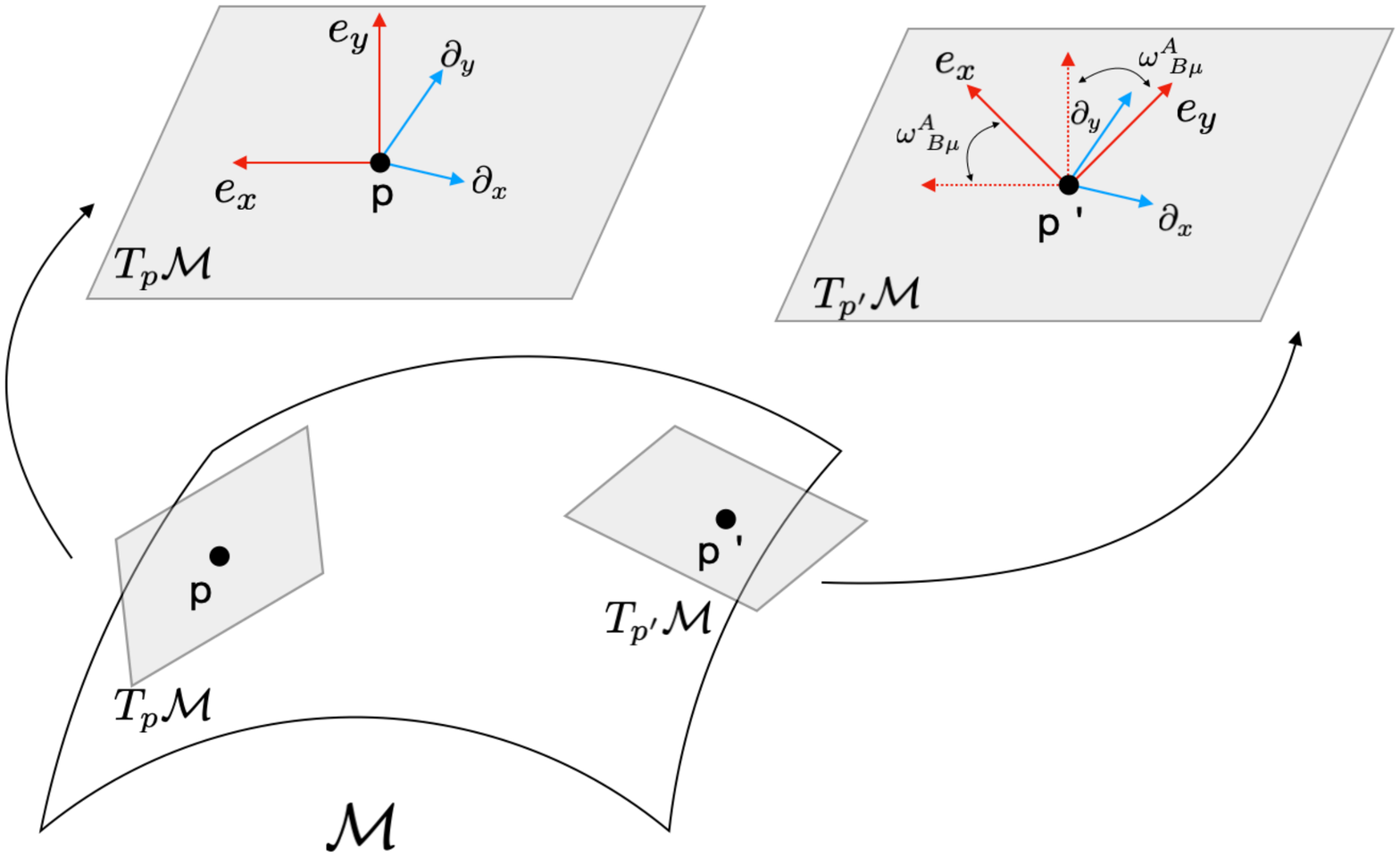}
    \caption{Two-dimensional picture displaying  the role of the spin connection $\omega^A_{\ B\mu}$. It translates the inertial effects present in the tetrad anhonolomic frame $\left\{ e_x,e_y \right\}$. When we pass from $p\in\mathcal{M}$ to $p'\in\mathcal{M}$, the related tetrads in $T_p\mathcal{M}$ and $T_{p'}\mathcal{M}$ exhibit a rotation, modeled by the spin connection. Instead, the inertial holonomic frame $\left\{\partial_x,\partial_y\right\}$ does not undergo any rotation, because it admits vanishing spin connection.}
    \label{fig:Fig5}
\end{figure}

To better understand these results, let us consider the transformation of the spin connection under local Lorentz transformations, which leads to \cite{Aldrovandi2013,Martin2019}
\begin{equation}\label{eq:spin_conn_via_LLT}
    \omega^A{}_{B\mu} =\underbrace{\Lambda^{A}_{\ C}(x)\omega'^{C}{}_{D\mu}\Lambda^{D}_{\ C}}_\text{non inertial}+\underbrace{\Lambda^{A}_{\ C}\partial_{\mu}\Lambda^{C}_{\ B}(x)}_\text{inertial}.
\end{equation}
When we pass from a frame to another one, there are two distinct contributions: (1) \emph{non-inertial effects} connected with the new frame; (2) \emph{inertial contributions} due to the rotation of the new frame with respect to the previous one. Therefore, starting from inertial frames ($\omega'^{AB}{}_\mu=0$), it is possible to obtain a class of non-inertial frames (cf. Eq. \eqref{eq:spin_conn_via_LLT}) via local Lorentz transformations. It is important to note that all these infinite frames are related through global (point-independent) Lorentz transformations $\Lambda^A_{\ B}={\rm const}$ \cite{Martin2019}. In Fig. \ref{fig:Fig5} we display the spin connection mechanism. 

From Eqs. \eqref{eq:spin_connect_der} and \eqref{eq:Lorentz_conn_and_spn_con}, the coefficients of anholonomy \eqref{eq:commutatotor_tetrad} can be written as ($\overset{\fullcirc}{\omega}^{A}_{\ BC}=\overset{\fullcirc}{\omega}^{A}_{\ B\mu}e^{\ \mu}_C$) \cite{Aldrovandi2013,Martin2019}
\begin{equation} \label{eq:f_omega}
    f^{C}_{\ AB} =\overset{\fullcirc}{\omega}^{C}_{\ BA}-\overset{\fullcirc}{\omega}^{C}_{\ AB}.
\end{equation}
From this relation we can define the spin connection in term of the structure constants as
\begin{equation}
    \overset{\fullcirc}{\omega}^{A}_{\ BC} = \frac{1}{2} (f_{B}{}^A{}_{C} +f_{C}{}^A{}_{B} -f^A_{\ BC}). 
\end{equation}
Let us  show now other two important implications of the purely inertial connection. Inserting its expression \eqref{eq:Lorentz_conn_and_spn_con} into the definitions of curvature and torsion tensors (cf. Eqs. \eqref{eq:curv_ten} and \eqref{eq:tor_ten}), we obtain the following relations \cite{Aldrovandi2013,Martin2019}
\begin{subequations}
\begin{align}
R^A_{\ B\mu\nu}&=\partial_\nu\overset{\fullcirc}{\omega}^A_{\ B\mu}-\partial_\mu\overset{\fullcirc}{\omega}^A_{\ B\nu}+\overset{\fullcirc}{\omega}^A_{\ E\nu}\overset{\fullcirc}{\omega}^E_{\ B\mu}\notag\\
&\hspace{3.1cm}-\overset{\fullcirc}{\omega}^A_{\ E\mu}\overset{\fullcirc}{\omega}^E_{\ B\nu}\equiv0,\label{eq:Riemann_SC}\\
T^A_{\ \nu\mu}&=\partial_\nu e^A_{\ \mu}-\partial_\mu e^A_{\ \nu}+\overset{\fullcirc}{\omega}^A_{\ E\nu} e^E_{\ \mu}-\overset{\fullcirc}{\omega}^A_{\ E\mu} e^E_{\ \nu}. \label{eq:Torsion_SC}
\end{align}
\end{subequations}
To prove that Eq. \eqref{eq:Riemann_SC} is identically vanishing, we have used the property $\Lambda^{E}_{\ C}\partial_{\mu}\Lambda^{C}_{\ E} = -\Lambda^{C}_{\ E}\partial_{\mu}\Lambda^{E}_{\ C}$. This  result, physically tells that inertial effects cannot generate \qm{curvature effects}, but it is possible  to produce only non-null torsional effects, see Eq. \eqref{eq:Torsion_SC}. However, if we consider trivial tetrads (i.e., $e^A_{\ \mu}=\partial_\mu x^a$ and $\overset{\fullcirc}{\omega}^A_{\ B\mu}=0$), we can further nullify also the torsion tensor.   

\section{Equivalent representations of gravity: The Lagrangian level}
\label{sec:trinity_lagrangian}
Let us consider now  the Geometric Trinity of Gravity, taking  into account  its mathematical and physical aspects. We discuss first the   formulation of gravity according to GR in Sec. \ref{sec:GR}. Gravity under the standard of  gauge description is considered in  Sec. \ref{sec:Teleparallel}).   In Sec. \ref{sec:summary_results}, the basic concepts of GR, TEGR and STEGR are  compared and discussed. 

The notations we are going to use are the following: \emph{over-circles} refer to  quantities built up on the Levi-Civita connection (i.e., $\lc{A}^\mu_{\ \ \nu}$), \emph{over-hats} denote  quantities related to the teleparallel connection (i.e., $\tg{A}^\mu_{\ \ \nu}$), and \emph{over-diamonds} denote  quantities  involving non-metricity (i.e., $\stg{A}^\mu_{\ \ \nu}$).

\subsection{Metric formulation of gravity:\\ The case of General Relativity}
\label{sec:GR}
The GR  is the first geometric formulation of gravity in curved spacetimes. We first recall its basic principles  (Sec. \ref{sec:principlesGR}), and  implications related to the geodesic equations (see Sec. \ref{sec:geodesic}). The fundamental geometric object is the metric tensor, which allows to define uniquely the Levi-Civita connection, which, in turn, determines the Riemann curvature tensor (see Sec. \ref{sec:Riemann}, for the description of its properties and symmetries). Then, Lagrangian and field equations of GR are presented in Sec. \ref{sec:Lagrangian_FEs}. Finally, we discuss the tetrad formalism in GR (see Sec. \ref{sec:tetrad_GR}).

\subsubsection{Principles of General Relativity}
\label{sec:principlesGR}
The Einstein theory is essentially  based on the following  pillar ideas, which can be stated as follows \cite{Misner1973,Romano2019,Faraoni2010}:
\begin{itemize}
    \item[(1)]\emph{Relativity Principle}: there is no preferred inertial frames, i.e. all frames are good  for Physics;
    \item[(2)] \emph{ General Covariance Principle:} the basic laws of Physics can be formulated in tensor form in any smooth four-dimensional manifold $\mathcal{M}$. This means that field equations must be ” covariant” in form, i.e. they must be invariant  under the action of spacetime diffeomorphisms;
    \item[(3)] \emph{ Equivalence Principle:} in any smooth four-dimensional manifold $\mathcal{M}$, it is possible to consider a small spacetime region $\mathcal{W}$ where spatial and temporal gravitational changes are negligible. Therefore, there always exists a LIF where  gravitational effects can be nullified. In other words,  inertial effects are locally indistinguishable from gravitational effects (which means the equivalence between the inertial and the gravitational masses). 
    \item[(4)] \emph{Causality Principle:} each point of spacetime has to admit a universally valid notion of past, present and future.
    \end{itemize}

The first two principles are strictly related. They configures the extension of   Relativity Principle of Special Relativity to any reference frame independently of the acceleration state. In other words, they figure out  a sort of \emph{democracy principle for all observers}, i.e., all observers have the same right to describe the physical reality  \cite{Romano2019, Boskoff}.

Regarding the third principle, it permits to locally recover the Physics of Special Relativity. Geometrically, it translates in determining the tangent plane in every point of a smooth manifold. Furthermore, gravity is the only interaction that cannot be switched off in  \emph{absolute}, as instead it occurs for electromagnetic  and other fields. Therefore, the gravitational field can be defined as what remains when we have deactivated the other interactions in an absolute way and independently from the observer. It can be only locally nullified in the LIFs, physically coinciding with local free-falling frames. Due to the underlying Riemannian geometric description, LIF is defined by the \emph{Riemann theorem} for every $p\in\mathcal{M}$ in a local chart $(\mathcal{U},\varphi)$ of $p$ as \cite{Romano2019,Boskoff}
\begin{align} \label{eq:LIF}
g_{\mu\nu}(\varphi(p))=\eta_{\mu\nu},\qquad \partial_\lambda g_{\mu\nu}(\varphi(p))=0.
\end{align}
This holds if we assume that inertial and gravitational mass coincides (see Refs. \cite{Romano2019, Boskoff}, for more details). This is the \emph{(weak) equivalence principle} or also known as \emph{universality of free fall}, stating that the trajectory of a point mass in a gravitational field depends only on its initial position and velocity and it is independent of its composition and structure. Therefore, the inertial effects may be globally eliminated by an appropriate choice of the reference frame (see Sec. \ref{sec:Lorentz_connection_physical}), whereas the gravitational field can be only locally disregarded not eliminated \cite{Romano2019,Boskoff}. 

The fourth principle is needed to ensure the uniqueness of the  time notion despite of spacetime deformations and singularities. As it is well known, several issues of modern physics are questioning the Causality Principle but we will not go into this discussion in this paper.

\subsubsection{Geodesic equations}
\label{sec:geodesic}
Starting from the universality of free fall postulate in LIF via the coordinates $\left\{\xi^\mu\right\}$, a test particle will draw a straight line, whose equation of motion is given by
\begin{equation}\label{eq:geodesic_LIF}
    \frac{{\rm d}^{2}\xi^{{\alpha}}}{{\rm d}s^{2}}=0,
\end{equation}
where ${\rm d}s^{2}=\eta_{\alpha\beta}d\xi^{\alpha}d\xi^{\beta}$ is the line element. Since in such a frame it is not possible to experience the existence of gravitational effects, we perform a change of coordinates $\xi^{\alpha}=\xi^{\alpha}(x^{\mu})$, with $\left\{x^{\mu}\right\}$ the new coordinates. Applying this transformation to Eq. \eqref{eq:geodesic_LIF}, we obtain
\begin{equation}\label{eq:geodesic_general}
    \frac{{\rm d}^{2}x^{\lambda}}{{\rm d}s^{2}} +\lc{\Gamma}^\lambda_{\ \mu\nu}\frac{{\rm d}x^{\mu}}{{\rm d}s}\frac{{\rm d}x^{\nu}}{{\rm d}s}=0,
\end{equation}
where $\lc{\Gamma}^\lambda_{\ \mu\nu}$ is the affine connection responsabile of the geodesic spacetime structure, which arises from the gravitational force acting on the test particle and being responsible of the departure from the straight trend. Its expression is now given by
\begin{equation}\label{eq:LC_coordinates}
    \lc{\Gamma}^\lambda_{\ \mu\nu} := \frac{\partial x^{\lambda}}{\partial\xi^{\sigma}} \frac{\partial^{2} \xi^{\sigma}}{\partial x^{\mu}\partial x^{\nu}},
\end{equation}
which explicitly shows that it is not a tensor. Physically they are the apparent forces acting on the body due to the curved geometric background induced by gravity. 

Therefore, assigned the metric tensor ${\rm d}s^{2}=g_{\mu\nu}{\rm d}x^{\mu}{\rm d}x^{\nu}$, in a generic coordinate system $\left\{x^\mu\right\}$, the \emph{geodesic equation} is described by Eq. \eqref{eq:geodesic_general}. In a metric compatible and torsion-free spacetime, we have that the unique affine symmetric connection is the \emph{Levi-Civita one} via the \emph{Levi-Civita theorem} \cite{Misner1973,Romano2019}. The condition $\lc{\nabla}_\lambda g_{\mu\nu}=0\;$ gives $\;\lc{\Gamma}^\lambda_{\ \mu\nu}\equiv\begin{Bmatrix} \lambda\\ \mu \nu \end{Bmatrix}$ (see Eq. \eqref{eq:LC}).

\subsubsection{The Riemann curvature tensor}
\label{sec:Riemann}
We have seen the effect of  geometric curvature in the geodesic equation, but to quantify it as a field we have to introduce the \emph{Riemann curvature tensor} $\lc{R}^\alpha_{\ \beta\mu\nu}$ (see Eq. \eqref{eq:curv_ten} with $\Gamma^\lambda_{\ \mu\nu}=\lc{\Gamma}^\lambda_{\ \mu\nu}$) arising from the commutation of covariant derivatives on a generic vector $v^\alpha$, that is
\begin{align}\label{eq:Riemann_tensor}
 [\lc{\nabla}_{\mu}, \lc{\nabla}_{\nu}] v^{\alpha}& = \lc{R}^{\alpha}_{\ \beta\mu\nu}v^{\beta}.
\end{align}
The above equation is telling us that the Schwarz theorem, applied to covariant derivatives, does not hold, otherwise we have a flat spacetime (i.e., $\lc{R}^{\alpha}_{\ \beta\mu\nu}=0$). The gravitational field is fully encoded in this tensor. 

The Riemann tensor maintains the symmetry \eqref{eq:sym_Riemann} in a generic metric-affine theory. However in GR (due to the symmetries of the Levi-Civita connection) it acquires the following further symmetries \cite{Misner1973}
\begin{subequations}
\begin{align} \label{eq:symmetry_Riemmann_GR}
\lc{R}_{\mu\nu\alpha\beta}&= -\lc{R}_{\nu\mu\alpha\beta}, \\ \lc{R}_{\mu\nu\alpha\beta} &= \lc{R}_{\alpha\beta\mu\nu}.
\end{align}
\end{subequations}
The two Bianchi identities \eqref{eq:Bianchi_identity_general} have both the right members equal to zero, since GR is torsion-free. Due to the symmetries \eqref{eq:symmetry_Riemmann_GR}, we can define the symmetric Ricci tesor $\lc{R}_{\alpha\beta}=\lc{R}^\mu_{\ \alpha\mu\beta}$ and the scalar curvature $\lc{R}=\lc{R}^\alpha_{\ \alpha}$.

Let us consider now a one-parameter family of geodesics $\gamma_s(t)$, where $t$ is the affine parameter along the geodesic, and $s\in[a,b]\subset\mathbb{R}$ labels the curves. We assume that the collection of these curves defines a smooth two-dimensional surface $x^\mu(t,s)$ embedded in $\mathcal{M}$. Provided that this family of geodesics forms a congruence, the parameters $t$ and $s$ are the coordinates on this surface. 

A natural vector basis adapted to the coordinate system is given by $\left\{T^\mu,S^\mu\right\}$, whose expressions are \cite{Romano2019}
\begin{align}
T^\mu=\frac{\partial x^\mu}{\partial t},\qquad S^\mu=\frac{\partial x^\mu}{\partial s}.    
\end{align}
Then, we define the relative velocity $V^\mu$ and acceleration $A^\mu$ along the geodesics as follows
\begin{subequations}
\begin{align}
V^\mu&=T^\nu \lc{\nabla}_\nu T^\mu,\\ 
A^\mu&=T^\nu \lc{\nabla}_\nu V^\mu.
\end{align}
\end{subequations}
We then obtain the \emph{geodesic deviation equation} \cite{Romano2019}
\begin{equation}
A^\mu=\lc{R}^\mu_{\ \lambda\alpha\beta}T^\lambda T^\alpha S^\beta,    
\end{equation}
where the relative acceleration between two close geodesics is proportional to the Riemann curvature tensor, which characterizes the behaviour of a one-parameter family of neighbouring geodesics. 

\subsubsection{Lagrangian formalism and field equations}
\label{sec:Lagrangian_FEs}
The GR dynamics is derived from the \emph{Hilbert-Einstein action}, whose expression is given by \cite{Corry1973}
\begin{equation}\label{eq:GR_action}
    S_{\rm GR}:= \frac{c^4}{16 \pi G} \int {\rm d}^{4}x \sqrt{-g}\left(\mathscr{L}_{\rm GR}+\mathscr{L}_{\rm m}\right),
\end{equation}
where $\mathscr{L}_{\rm GR}:=\lc{R}(g)$ is the Einstein-Hilbert Lagrangian, coinciding with the Ricci curvature scalar, and $\mathscr{L}_{\rm m}$ is the matter Lagrangian. In this case, the fundamental object is the metric, as underlined in the curvature scalar $\lc{R}(g)$. The total DoFs are represented by the ten independent components of the metric tensor, from which we must subtract the four-parameter diffeomorphisms underlying the invariance (gauge symmetries' freedom) and other four by a suitable choice of the coordinates (gauge fixing) \cite{Misner1973,Zakharov2006,Jimenez2019}. Therefore, the gravitational dynamical DoFs becomes two, corresponding thus to the \emph{graviton}, massless spin-2 particle, related to the $X$ and $+$ polarizations of gravitational waves \cite{Faraoni2010, Boskoff}.

Applying the principle of least action to Eq. \eqref{eq:GR_action}, we derive the GR field equations in presence of matter
\begin{equation} \label{eq:GR_equations}
\lc{G}_{\mu\nu}:=\lc{R}_{\mu\nu} - \frac{1}{2}g_{\mu\nu}\lc{R} = \frac{8 \pi G}{c^{4}}\mathfrak{T}_{\mu\nu},
\end{equation}
where $\lc{G}_{\mu\nu}$ is the Einstein tensor and
\begin{equation} \label{eq:energy_momentum_tensor}
    \mathfrak{T}^{\mu\nu} = -\frac{1}{2\sqrt{-g}} \frac{\delta\mathscr{L}_{m}}{\delta g_{\mu\nu}}
\end{equation}
is the (second-order) energy-momentum tensor, which is symmetric, satisfies the conservation equations $\lc{\nabla}_\mu \mathfrak{T}^{\mu\nu}=0$, and physically represents the source of gravitational field. 

Particular consideration has to be devoted to matter fields and gravity, because some subtleties can arise. For example, (1) ambiguity in the matter coupling; (2) treatment of bosonic and fermionic fields. In GR,  it is clear that a point particle  follows the geodesic equations according to the Levi-Civita part of the connection. More problematic issues are linked to bosons (coupling only to the metric) and fermions (coupling with metric and connection). Therefore, when matter fields are taken into account, one must either consider minimally coupled fields or formulate  consistent theories in  metric-affine formalism. For example in GR, the presence of fermions requires the introduction of tetrads and spin connection  \cite{Jimenez2019}.    

\subsubsection{Tetrad formalism in General Relativity}
\label{sec:tetrad_GR}
GR conceives the gravitational interaction as a change in the geometry of spacetime itself, where we pass from the Minkowski $\eta_{\mu\nu}$ to the Riemannian metric $g_{\mu\nu}$, and from partial $\partial$ to covariant derivatives $\nabla$. The metric plays the role of the fundamental field, which is defined everywhere. In order  to study how gravitation couples with others fields, we have to introduce the tetrads to deal with spinors in curved spacetimes. In addition, tetrads encode the  Equivalence Principle since they are locally defined, as gravitation is locally equivalent to an accelerated frame. Therefore, to obtain the effects of gravitation on general sources (particles or fields), we need to: $(i)$ write all the related equations in the Minkowski spacetime in general coordinates, represented by trivial tetrads;
$(ii)$ replace the holonomic tetrads with the anholonomic tetrads, keeping the same formulae. The resulting equations hold in GR. \emph{Einstein’s vierbein theory becomes thus a gauge field theory for gravity}.

Once we assign a general (anhonolomic) tetrad $\left\{e^A_{\ \mu}\right\}$, we can rewrite the Riemann tensor according to the Cartan structure equations (see Sec. \ref{sec:cartan_equation}) as \cite{Deruelle2018}
\begin{subequations}
\begin{align}
&{\rm d}e^C+\lc{\omega}^A_{\ B}\wedge e^B=0,\\    
&\lc{\omega}_{AB}+\lc{\omega}_{BA}={\rm d}g_{AB},\\
&{\rm d}\lc{\omega}^A_{\ B}+\lc{\omega}^A_{\ C}\wedge\lc{\omega}^C_{\ B}=\frac{1}{2}\lc{r}^A_{\ BCD}e^C\wedge e^D,
\end{align}
\end{subequations}
where $\lc{r}^A_{\ BCD}$ is the Riemann curvature tensor in the tetrad frame, with
\begin{subequations}
\begin{align}
&\lc{\omega}^A_{\ B\mu}:=e^A_{\ \nu}\lc{\nabla}_\mu e_B^{\ \nu},\label{eq:omegaLC}\\ 
&\lc{f}^A_{\ BC}:=\lc{\gamma}^A_{\ BC}-\lc{\gamma}^A_{\ CB},\\
&{\rm d}g_{AB}=\partial_C\ g_{AB}e^C,\\
&\lc{\gamma}^A_{\ BC}=\frac{1}{2}(\lc{f}^A_{\ BC}-g_{CL}g^{AM}\lc{f}^L_{\ BM}-g_{BL}g^{AM}\lc{f}^L_{\ CM})\notag\\
&\hspace{1cm} +\lc{\Gamma}^A_{\ BC}.
\end{align}
\end{subequations}
It is important to note that we can uniquely associate the Lorentz connection to the Levi-Civita connection via Eq. \eqref{eq:spin_connect_der}. In addition, if we consider the natural basis, then we have $\lc{\omega}^A_{\ BC}=0$ and therefore $\lc{\gamma}^A_{\ BC}\equiv\lc{\Gamma}^A_{\ BC}$. Using the above cited equations, it is possible to extract the components of $\lc{r}^A_{\ BCD}$, which are \cite{Deruelle2018} 
\begin{align}
\lc{r}^A_{\ BCD}&=\partial_D\lc{\gamma}^A_{\ BC}-\partial_C\lc{\gamma}^A_{\ BD}+\lc{\gamma}^A_{\ CM}\lc{\gamma}^M_{\ DB}\notag\\
&-\lc{\gamma}^A_{\ DM}\lc{\gamma}^M_{\ CB}-\lc{\gamma}^A_{\ MB}\lc{\gamma}^M_{\ CD}.    
\end{align}
Also in this case, in the natural basis, we re-obtain the standard definition of the Riemann curvature tensor \eqref{eq:Riemann_tensor}.

\subsection{Gauge formulation of gravity:\\
The case of Teleparallel Gravity}
\label{sec:Teleparallel}
A gauge formulation of gravity is possible in the Teleparallel Gravity Theory. We first show that this general theory can be seen as a translation gauge theory (see Sec. \ref{sec:trans_gauge_theo}), then we analyse the concepts of geodesics and autoparallel curves in this new  framework (see Sec. \ref{sec:autoparallels}). We finally concentrate on two important teleparallel subtheories: the metric teleparallel gravity,  (in Sec. \ref{sec:TEGR}) and the  symmetric teleparallel gravity,  (see Sec. \ref{sec:STEGR}). Two important realizations of these approaches are TEGR and STEGR, respectively. 

\subsubsection{Translation gauge theory}
\label{sec:trans_gauge_theo}
In a modern vision of physics, it is very important to settle  theories in a gauge framework \cite{Blagojevic2001}. In Sec. \ref{sec:tetrad_GR} we have seen that also GR can be converted in a gauge theory. Let us now sketch  how GR can be formulated  as a \emph{gauge theory of translations} \cite{Blagojevic2001,Aldrovandi2013}. 

This picture of GR can be achieved by both invoking the N\"other theorem and recalling that the source of the gravitational field is given by the energy and momentum. Indeed, provided that gravitational Lagrangian is invariant under spacetime translations, the energy-momentum current is covariantly conserved. We will see that a metric teleparallel theory is more suitable to express gravity in this context, because it entails more benefits, and the introduction of tetrads reveals to be more natural. 


This approach was first proposed by Lasenby, Doran, and Gull in 1998 \cite{Lasenby1998}. Its geometric setting is the tangent bundle, where the gauge transformations take place. Let us first introduce $\left \{ x^{\mu}\right \}$ and $\left \{ x^{A} \right \}$ as the coordinates on $\mathcal{M}$ and $T_{p}\mathcal{M}$, respectively. Now, let us consider the following infinitesimal local translation
\begin{equation} \label{eq:transofrmation}
    x^A\longrightarrow \bar{x}^{A}=x^{A}+\varepsilon^{A}(x^{\mu}),
\end{equation}
where $\varepsilon^{A}(x^{\mu})$ are the infinitesimal parameters of the transformation. The set of translations forms the \emph{translation Lie group} $O(1,3)$, whose generators are
\begin{equation}
    P_{A}:= \partial_{A}.
\end{equation}
They generate the \emph{Abelian translation algebra}, because they satisfy the following trivial commutation rules
\begin{equation}
    [P_{A},P_{B}]\equiv[\partial_{A}, \partial_{B}]=0.
\end{equation}
The infinitesimal transformation, written in terms of the generators, has the following expression
\begin{equation} 
    \delta \bar{x}^{A}=\varepsilon(x^\mu)^{B} \partial_{B} x^{A}=\varepsilon(x^\mu)^{A}.
\end{equation}
A general source field $\Psi=\Psi(\bar{x}^A(x^\mu))$ transforms under the map \eqref{eq:transofrmation} as follows \cite{Aldrovandi2013,Martin2019}
\begin{equation}
    \delta_\varepsilon \Psi= \epsilon^{A}(x^\mu)\partial_{A} \Psi.
\end{equation}
Let $\varepsilon^A={\rm constant}$ be a global translation, then the ordinary derivative $\partial_\mu \Psi$ transforms covariantly, because
\begin{equation} \label{eq:non_covariant}
\partial_\varepsilon(\partial_\mu\Psi)=\varepsilon^A\partial_A(\partial_\mu\Psi).    
\end{equation} 
For a local translational transformation $\varepsilon^A(x^\mu)$, $\partial_\mu \Psi$ does not transform covariantly, because \cite{Aldrovandi2013,Martin2019}
\begin{equation}
\partial_\varepsilon(\partial_\mu\Psi)=\underbrace{\varepsilon^A(x^\mu)\partial_A(\partial_\mu\Psi)}_\text{correct}+\underbrace{(\partial_\mu\varepsilon^A(x^\mu))\partial_A\Psi}_\text{spurious},    
\end{equation}
where the spurious term spoil the translational gauge covariance. However, in order to save this gauge covariance, we follow the praxis exploited in all other gauge theories \cite{Maggiore2005}. Like in the electromagnetic case, where we include the gauge potential field $A_{\mu}$ to guarantee the covariance of the theory, also here we have to set forth the \emph{translational gauge potential 1-form} $B_{\mu}$, assuming values in the Lie algebra of the translation group, to guarantee the covariance of the gravity theory. Therefore, we introduce the following \emph{gauge covariant derivative} (see Sec. \ref{sec:trivial_tetrads})
\begin{equation} \label{eq:trans_trivial_tetrads}
e'_\mu\Psi\equiv\partial_\mu\Psi=\partial_\mu+B^A_{\ \mu}\partial_A\Psi, 
\end{equation}
which holds in the class of Lorentz inertial frames (see Sec. \ref{sec:trivial_tetrads}). To recover the gauge covariance, we require that the gauge potential $B_\mu$ transforms according to
\begin{equation} \label{eq:gauge_potential}
    \delta_\varepsilon B^{A}_{\ \mu} = -\partial_{\mu} \varepsilon^{A}(x^\mu).
\end{equation}
Indeed, now $e_\mu\Psi$ transforms covariantly 
\begin{equation}
\partial_\varepsilon(e'_\mu\Psi)=\underbrace{\varepsilon^A(x^\mu)\partial_A(\partial_\mu\Psi)}_\text{correct},        
\end{equation}
since the potential \eqref{eq:gauge_potential} equals the spurious term in Eq. \eqref{eq:non_covariant}, cancelling it out. The above construction is based on trivial tetrads. However, for a general non-trivial tetrad field, it has the following expression  
\begin{equation}
e_\mu\Psi=e^A_{\ \mu}\partial_A\Psi, \qquad e^A_{\ \mu}=\partial_\mu x^A+B^A_{\ \mu},     
\end{equation}
where $B^A_{\ \mu}\neq -\partial_\mu \varepsilon^A(x^\mu)$ and $e'^A_{\ \mu}\neq\partial_\mu x^A$. Now let us consider a Lorentz transformation \eqref{eq:coordinate_transform}, and let us assume that the gauge potential $B^A_{\ \mu}$ transforms as a Lorentz vector in the algebraic index, namely it satisfies
\begin{equation}
B^A_{\ \mu}\longrightarrow \Lambda^A_{\ B}(x)B^B_{\ \mu}.   
\end{equation}
Therefore, the generalization of Eq. \eqref{eq:trans_trivial_tetrads} becomes
\begin{equation} \label{eq:trans_general_tetrads}
e_\mu\Psi=\partial_\mu+\overset{\fullcirc}{\omega}^A_{\ B\mu}x^B\partial_A\Psi+B^A_{\ \mu}\partial_A\Psi, 
\end{equation}
where
\begin{equation}
e^A_{\ \mu}=\partial_\mu x^A +\overset{\fullcirc}{\omega}^A_{\ B\mu}x^B+B^A_{\ \mu}=\overset{\fullcirc}{\mathcal{D}}_\mu x^A+B^A_{\ \mu}.  
\end{equation}
For general non-trivial tetrads, we need to upgrade the gauge potential \eqref{eq:gauge_potential} as follows
\begin{equation} \label{eq:gauge_potential_general}
    \delta_\varepsilon B^{A}_{\ \mu} = -\overset{\fullcirc}{\mathcal{D}}_{\mu} \varepsilon^{A}(x^\mu).
\end{equation}
In the context of teleparallel gravity, we have applied the following \emph{translation coupling prescription}
\begin{equation} \label{eq:tra_coup_pre}
e'^A_{\ \mu}\longrightarrow e^A_{\ \mu},   
\end{equation}
from which,  the \emph{gravitational coupling prescription}, assumed in GR, naturally emerges
\begin{equation}
\eta_{\mu\nu}\longrightarrow g_{\mu\nu}.   
\end{equation}
It is important to stress that the local Lorentz invariance is a fundamental symmetry respected by all physical laws in Nature, therefore, we must impose that our new theory be locally Lorentz invariant. Such a requirement entails the additional \emph{Lorentz gravitational coupling prescription}, which is a direct consequence of the strong Equivalence Principle \cite{Aldrovandi2013,Martin2019}. Indeed, this prescription is based on the General Covariance Principle, which can be seen as an active version of the strong Equivalence Principle, namely given an equation valid in presence of gravitation, the corresponding special relativistic equation is locally recovered (at a point or along a trajectory), i.e.,
\begin{align}
\partial_\mu\Psi\to&\mathcal{D}'_\mu\Psi=\partial_\mu\Psi\notag\\
&+\frac{1}{2}e'{}^A_{\ \mu}\left(f'_B{}^C_{\ A}+f'_A{}^C_{\ B}-f'^C_{\ BA}\right)S^B_{\ C}\Psi.    
\end{align}
where $\Psi$ is a general field, and $S^B_{\ C}$ are the generators of the Lorentz group in the same representation to which $\Psi$ belongs. However, in presence of gravitation, we obtain 
\begin{align} \label{eq:fullgptg}
\partial_\mu\Psi\to&\mathcal{D}_\mu\Psi=\partial_\mu\Psi\notag\\
&+\frac{1}{2}e^A_{\ \mu}\left(f_B{}^C_{\ A}+f_A{}^C_{\ B}-f^C_{\ BA}\right)S^B_{\ C}\Psi,    
\end{align}
which represents the \emph{full (Lorentz plus translational) gravitational coupling prescription in teleparallel gravity}. We have therefore the following scheme
\begin{equation}
\underbrace{\begin{Bmatrix}
e'^A_{\ \mu}\longrightarrow e^A_{\ \mu}\\
\partial_\mu\longrightarrow \mathcal{D}_\mu
\end{Bmatrix}}_\text{grav. coupling prescription in TG}\ \Leftrightarrow\ \underbrace{\eta_{\mu\nu}\longrightarrow g_{\mu\nu}}_\text{grav. coupling prescription in GR}.   
\end{equation}

\subsubsection{Autoparallels and geodesics}
\label{sec:autoparallels}
Let us consider the equation of motion of a free test particle first described in the inertial frames $e'^A_{\ \mu}$, i.e., \cite{Martin2019}
\begin{equation}\label{eq:eom_inertial}
    \frac{{\rm d}u'^{A}}{{\rm d}\sigma} = 0,
\end{equation}
where $u'^{A}$ is the anholonomic  four–velocity of the test particle and ${\rm d}\sigma$ is the Minkowskian line element ${\rm d}\sigma^{2} = \eta_{\mu\nu}{\rm d}x^{\mu}{\rm d}x^{\nu}$. We note that Eq. \eqref{eq:eom_inertial} is written in a particular class of reference frames, and under a local Lorentz transformation \eqref{eq:coordinate_transform}, it is non-covariant since
\begin{equation}
\frac{{\rm d}u'^{A}}{{\rm d}\sigma}=\underbrace{\Lambda^A_{\ B}(x)\frac{{\rm d}u^{B}}{{\rm d}\sigma}}_\text{correct}+\underbrace{\frac{{\rm d}\Lambda^A_{\ B}(x)}{{\rm d}\sigma}
u^{B}}_\text{spurious}\ . 
\end{equation}
This is an apparent failure of the covariance, because if we consider the anhonolomic frame $e^A_{\ \mu}$, associated to $e'^A_{\ \mu}$ through local Lorentz transformation (cf. Eq.\eqref{eq:change_tetrad}), we immediately recover the covariance, because
\begin{equation}
\frac{{\rm d}u'^{A}}{{\rm d}\sigma}=0\ \longrightarrow\ \frac{{\rm d}u^{B}}{{\rm d}\sigma}+\overset{\fullcirc}{\omega}^A_{\ B\mu}u^B u^\mu =0.    
\end{equation}

In Sec. \ref{sec:geodesic}, we have defined the geodesic equation \eqref{eq:geodesic_general} in GR. This notion must be revised in the parallel framework. Let us consider a chart $(\mathcal{U},\varphi)$ on the manifold $\mathcal{M}$ and let $\gamma^\mu(\tau)$ be the parametric equation of a curve $\gamma$ contained in $\mathcal{U}$, where $\tau$ is the affine parameter along $\gamma$. The tangent vector $\dot{\gamma}$ to $\gamma$, in the natural basis $\left\{\partial_\mu\right\}$ along $\gamma$, is given by the following expression \cite{Romano2019}
\begin{equation}
\dot{\gamma}(\tau):=\frac{{\rm d}\gamma^\mu}{{\rm d}\tau}\partial_\mu.    
\end{equation}
A vector $Y^\mu(\tau)$ is defined to be \emph{parallel transported along $\gamma$} if it fulfills the following request
\begin{equation} \label{eq:autoparallel}
\frac{{\rm d}Y^\mu}{{\rm d}\tau}:=\nabla_\gamma Y^\mu\equiv\frac{{\rm d}Y^\mu}{{\rm d}\tau}+\Gamma^\mu_{\ \alpha\beta}Y^\alpha\frac{{\rm d}\gamma^\beta}{{\rm d}\tau}=0,    
\end{equation}
where, for the moment, we do not specify $\Gamma^\mu_{\ \alpha\beta}$. Eq. \eqref{eq:autoparallel} represents a system of first order differential equations in the unknown $Y^\mu(\tau)$, which admits a unique solution once the initial condition $Y_0^\mu:=Y^\mu(\tau_0)$ has been provided. It is important to note that $Y^\mu(\gamma(\tau))$ depends on the curve $\gamma$. Therefore, a curve $\gamma(\tau)$ is said to be \emph{autoparallel} if its tangent vector $\dot{\gamma}(\tau)$ satisfies \cite{Romano2019}
\begin{equation} \label{eq:eqgeo}
\nabla_\gamma \dot{\gamma}\equiv \frac{{\rm d}^2 x^\mu}{{\rm d}\tau^2}+\Gamma^\mu_{\ \alpha\beta}\frac{{\rm d}x^\alpha}{{\rm d}\tau}\frac{{\rm d}x^\beta}{{\rm d}\tau}=0,   
\end{equation}
or, in other words, if it remains parallel to itself along $\gamma(\tau)$, where $x^\mu$ are the coordinates of $\gamma(\tau)$ in the chart $(\mathcal{U},\varphi)$. Eq. \eqref{eq:eqgeo} is a system of second order differential equations, which admit a unique solution once initial position and velocity have been assigned. It is worth noticing that, in GR, autoparallels and geodesic equations coincide, whereas, in teleparallel gravity, they give rise to two different structures, because the autoparallels are related to the affine connection, whereas the geodesic to the concept of metric, since it measures the minimal lengths between two or more points. In the teleparallel framework,  Eq. \eqref{eq:eqgeo} becomes (cf. Eq. \eqref{eq:ACG})
\begin{subequations} \label{eq:eqgeo_teleparallel}
\begin{align}
&\frac{{\rm d}^2 x^\mu}{{\rm d}\tau^2}+\lc{\Gamma}^\mu_{\ \alpha\beta}\frac{{\rm d}x^\alpha}{{\rm d}\tau}\frac{{\rm d}x^\beta}{{\rm d}\tau}=-K^\mu_{\ \alpha\beta}\frac{{\rm d}x^\alpha}{{\rm d}\tau}\frac{{\rm d}x^\beta}{{\rm d}\tau},\label{eq:eqgeo_teleparallel_TEGR}\\
&\frac{{\rm d}^2 x^\mu}{{\rm d}\tau^2}+\lc{\Gamma}^\mu_{\ \alpha\beta}\frac{{\rm d}x^\alpha}{{\rm d}\tau}\frac{{\rm d}x^\beta}{{\rm d}\tau}=-L^\mu_{\ \alpha\beta}\frac{{\rm d}x^\alpha}{{\rm d}\tau}\frac{{\rm d}x^\beta}{{\rm d}\tau}.\label{eq:eqgeo_teleparallel_STEGR}
\end{align}
\end{subequations}
Therefore, Eqs. \eqref{eq:eqgeo_teleparallel} recover a new aspect of GR, seen not anymore geometrically as a minimal distance path, but in the gauge paradigm as a sort of \emph{Lorentz force-like interaction} for the contortion tensor and \emph{kinetic energy-like interaction} regarding the disformation tensor, acting on the test particle \cite{Bahamonde2021}. It is important to note that if we impose the Weitzenb\"ock gauge in TEGR and the coincident gauge in STEGR, we reduce to $\frac{{\rm d}^2 x^\mu}{{\rm d}\tau^2}=0$, meaning that we are in the LIF. 

Another fundamental implication of autoparallels in teleparallel gravity is that they are sensitive to parameter changes, because it is possible to obtain another curve, although we do not alter the locus of its points. Therefore, if $\gamma(\lambda)$ is autoparallel, then $\mu(\tau)\equiv\gamma(\lambda(\tau))$ might be not autoparallel. This change of parameterization $\lambda=\lambda(\tau)$ entails that Eq. \eqref{eq:eqgeo} becomes \cite{Romano2019} 
\begin{equation} \label{eq:eqgeo_rep}
\frac{{\rm d}^2 x^\mu}{{\rm d}\tau^2}+\Gamma^\mu_{\ \alpha\beta}\frac{{\rm d}x^\alpha}{{\rm d}\tau}\frac{{\rm d}x^\beta}{{\rm d}\tau}=-\left(\frac{{\rm d}\lambda}{{\rm d}\tau}\right)^2\frac{{\rm d}^2\tau}{{\rm d}\lambda^2}\frac{{\rm d}\gamma^\mu}{{\rm d}\tau}.  
\end{equation}
We immediately see that the autoparallel character of the curve $\gamma(\lambda)$ is conserved under the parameter change $\lambda=\lambda(\tau)$ if and only if $\tau=a\lambda+b$, with $a,b$ being real arbitrary constants. Here $\lambda,\mu$ are called \emph{canonical parameters}. 

\subsubsection{Metric teleparallel gravity}
\label{sec:TEGR}
Metric (or torsional) teleparallel gravity (TG), known also as simply teleparallel gravity, is obtained by assuming the metric compatibility. The theory is geometrically described only by the torsion tensor. In Sec. \ref{sec:trans_gauge_theo}, we have already seen that tetrads $e^A_{\ \mu}$ and spin connection $\omega^A_{\ B\mu}$ play a fundamental role in describing gravity. Indeed, GR can be recast as a translational gauge theory, where the related gravitational field strength arises from the commutation relation of the covariant derivatives, see Eqs. \eqref{eq:commut_cov_der} and \eqref{eq:trans_general_tetrads}, namely\footnote{We define the torsion tensor as minus of that defined in Eq. \eqref{eq:commut_cov_der}, for having the signs in agreement when compared to those of GR.} 
\begin{equation}
[e_\mu,e_\nu]=\tg{T}^A_{\ \nu\mu}\partial_A,
\end{equation}
where the torsion (antisymmetric in the indices  $\mu\nu$)
\begin{equation} \label{eq:torsion_gen}
\tg{T}^A_{\ \mu\nu}=\partial_\nu B^A_{\ \mu}-\partial_\mu B^A_{\ \nu}+\overset{\fullcirc}{\omega}^A_{\ B\nu}B^B_{\ \mu}-\overset{\fullcirc}{\omega}^A_{\ B\mu}B^B_{\ \nu}
\end{equation}
represents the field strength. Adding the vanishing term
\begin{equation}
\overset{\fullcirc}{\mathcal{D}}_\mu(\overset{\fullcirc}{\mathcal{D}}_\nu x^A)-\overset{\fullcirc}{\mathcal{D}}_\nu(\overset{\fullcirc}{\mathcal{D}}_\mu x^A)\equiv0    
\end{equation}
to Eq. \eqref{eq:torsion_gen}, it becomes
\begin{equation}\label{eq:TORSION_TETRAD}
\tg{T}^A_{\ \mu\nu}= \partial_\nu e^A_{\ \mu}-\partial_\mu e^A_{\ \nu}+\overset{\fullcirc}{\omega}^A_{\ B\nu}e^B_{\ \mu}-\overset{\fullcirc}{\omega}^A_{\ B\mu}e^B_{\ \nu}.   
\end{equation}
Exploiting Eqs. \eqref{eq:gamma_lorentz} and \eqref{eq:TORSION_TETRAD}, we have that
\begin{equation}
\tg{T}^\lambda_{\ \mu\nu}=e_A^{\ \lambda}\tg{T}^\lambda_{\ \mu\nu}:=\Gamma^\lambda_{\ \nu\mu}-\Gamma^\lambda_{\ \mu\nu}.    
\end{equation}
The spin connection is linked to the inertial effects present in the tetrad frame, it is covariant under both diffeomorphisms and local Lorentz transformations (see Sec. \ref{sec:spin_connection}), assuring  the same properties also for the torsion tensor. It is important to \emph{associate at each tetrad the related spin connection}, therefore in TG we have always to provide the couple $\{e^A_{\ \mu},\overset{\fullcirc}{\omega}^A_{\ B\mu}\}$ \cite{Martin2019}. There exist frames in TG where the related spin connection vanishes, which are called \emph{proper frames} $\{e^A_{\ \mu},0\}$. This definition leads to the \emph{Weitzenb\"ock gauge}, which produces the \emph{Weitzenb\"ock connection} $\tg{\Gamma}^{\lambda}_{\ \nu\mu}=e_{A}^{\ \lambda}\partial_{\mu}e^{A}_{\ \nu}$, being the \emph{distant parallelism condition} from where TG takes its name.

A natural question spontaneously arises: \emph{given a tetrad frame, how do we operatively associate the related spin connection?} The simplest solution is tho choose proper frames, but, \emph{a priori}, we do not know which are the related tetrads. Therefore, we have to find a strategy to answer  this question. As one can verify, determining them from the field equations is, in general, not a simple task (see Ref. \cite{Martin2019}, for details). The method we propose relies on first determining the inertial effects in the trivial tetrad frame and then associating the related spin connection (see Ref. \cite{Hohmann2019}, for another method). In this approach, let us first introduce the concept of \emph{reference tetrad} $e^A_{\rm(r)\mu}$, in which gravity is switched off, that is
\begin{equation} \label{eq:reference_tetrad}
e^A_{\rm(r)\mu}:=\lim_{G\to0}e^A_{\ \mu}.    
\end{equation}
Through this process we are basically exploiting the Equivalence Principle or the inverse translational coupling prescription \eqref{eq:tra_coup_pre}. This has the effect to consider a trivial tetrad, where the  anhonomaly coefficients are zero (see Sec. \eqref{sec:trivial_tetrads}) and therefore the torsion tensor vanishes. In formulae, this can be written as (cf. Eq. \eqref{eq:f_omega})
\begin{equation}
\tg{T}^A_{\ BC}(e^A_{\ \mu},\overset{\fullcirc}{\omega}^A_{\ B\mu})=\overset{\fullcirc}{\omega}^A_{\ BC}-\overset{\fullcirc}{\omega}^A_{\ BC}-f^A_{\ BC}(e_{\rm (r)})=0,    
\end{equation}
from which we have
\begin{equation}
\overset{\fullcirc}{\omega}^A_{\ BC}=\frac{1}{2}e^C_{\rm(r)\mu}\left[f_B{}^A_{\ C}(e_{\rm (r)})+f_C{}^A_{\ B}(e_{\rm (r)})-f^A_{\ BC}(e_{\rm (r)})\ \right].    
\end{equation}
Since they differ only by the gravitational content, they represent the gravitational effects inside the tetrad frame. This approach can be schematized as follows
\begin{equation}
\begin{tikzcd}[row sep=large, column sep = large]
\overbrace{e^A_{\ \mu}}^\text{general tetrad} \arrow[r,"\text{gravity off}"]
&[5em] \overbrace{e^A_{\rm(r)\mu}}^\text{reference tetrad} \arrow[d, "\rotatebox{270}{\footnotesize{\text{reduce to}}}"]  \\[5ex]
                 & \underbrace{e'^A_{\ \mu}}_\text{trivial tetrad} \arrow[lu, "\overset{\fullcirc}{\omega}^A_{\ B\mu}"]
\end{tikzcd}
\end{equation}
The coefficients of anhonolonomy \eqref{eq:coeff_anhon}, in presence of  torsion, read as (cf. Eq. \eqref{eq:torsion_onvectors})
\begin{equation}
\overset{\fullcirc}{\omega}^C_{\ AB}-\overset{\fullcirc}{\omega}^C_{\ BA}=f^C_{\ AB}+T^C_{\ AB}.    
\end{equation}
This expression can be recombined as follows
\begin{equation} \label{eq:fund_ident}
\frac{1}{2}\left(f_B{}^C_{\ A}+f_A{}^C_{\ B}-f^C_{\ BA}\right)=\overset{\fullcirc}{\omega}^C_{\ BA}-\tg{K}^C_{\ BA},    
\end{equation}
where the \emph{contortion tensor}   
\begin{equation} \label{eq:contortion}
\tg{K}^C_{\ BA}=\frac{1}{2}\left(\tg{T}_B{}^C_{\ A}+\tg{T}_A{}^C_{\ B}-\tg{T}^C_{\ BA}\right)\,,    
\end{equation}
has been introduced.
Eq. \eqref{eq:fund_ident} is a further development of Eq. \eqref{eq:fullgptg}. Using the fundamental identity of the theory of Lorentz connections, we obtain (cf. Eq. \eqref{eq:omegaLC}) \cite{Kobayashi1996,Martin2019}\footnote{Equation \eqref{eq:identity_LorentzC} is important, but its derivation is also not trivial at all. Here, we provide an intuitive proof, although a more rigorous demonstration can be found in Sec. II.6 of Ref. \cite{Kobayashi1996}. Let us suppose to have the tetrads $\lc{e}^A_{\ \mu}$ in GR and $\tg{e}^A_{\ \mu}$ in TG such that they have the same coefficients of anholonomy $\lc{f}^A_{\ BC}=\tg{f}^A_{\ BC}$, guaranteed by the fact that there exists an isomorphism assuring this property. This implies $\lc{\mathcal{D}}_\mu=\tg{\mathcal{D}}_\mu$, which then gives Eq. \eqref{eq:identity_LorentzC}.}
\begin{equation} \label{eq:identity_LorentzC}
\overset{\fullcirc}{\omega}^C_{\ B\mu}-\tg{K}^C_{\ B\mu}=\overset{\circ}{\omega}^C_{\ B\mu},  
\end{equation}
which joins together GR and TG in a single compact expression. We remark that this \qm{combined} coupling prescription has been obtained from the General Covariance Principle, and it is thus consistent with the strong Equivalence Principle. In Eq. \eqref{eq:identity_LorentzC}, there is $\overset{\circ}{\omega}^C_{\ B\mu}$ in GR, enclosing  both gravitation and inertial effects in an indistinct form, whereas in TG, $\overset{\fullcirc}{\omega}^C_{\ B\mu}$ describes the inertial effects and $\tg{K}^C_{\ B\mu}$ represents only the gravitation. \emph{This is a new and elegant perspective to see the strong Equivalence Principle in TG}. Therefore, in a local frame where the GR spin connection vanishes, we obtain the identity $\overset{\fullcirc}{\omega}{}^C_{\ B\mu}=\tg{K}^C_{\ B\mu}$, where inertial effects compensate gravitation \cite{Martin2019}, resembling the free-falling cabin' situation. 

Another fundamental ingredient of TG theory is represented by the \emph{superpotential}, whose expression is \cite{Capozziello2021}
\begin{equation} \label{eq:superpotential}
\tg{S}_{A}^{\ \mu\nu} := \tg{K}^{\mu\nu}{}_{A}-e_{A}^{\ \nu} \tg{T}^\mu+e_{A}^{\ \mu}\tg{T}^\nu,
\end{equation}
where $\tg{T}^{\alpha\mu}{}_{ \alpha}:=\tg{T}^\mu$ is dubbed \emph{torsion vector}. This permits then to introduce the \emph{torsion scalar}
\begin{align} \label{eq:torsion_tensor}
    \tg{T}&:= \frac{1}{2} \tg{S}_{A}^{\ \mu\nu}\tg{T}^{A}_{\ \mu\nu}\notag\\
    &=\frac{1}{4}\tg{T}^\rho_{\ \mu\nu}\tg{T}_\rho^{\ \mu\nu}+\frac{1}{2}\tg{T}^\rho_{\ \mu\nu}\tg{T}^{\nu\mu}{}_\rho-\tg{T}_\mu \tg{T}^\mu,    
\end{align}
which is quadratic in the all possible torsion tensor combinations. In particular, in the last equality, the first term resembles that of the usual Lagrangian of internal gauge theories, whereas the other two stem out from the tetrad soldered character allowing thus to set at the same level internal and external indices \cite{Martin2019}. 

Since TG is curvatureless we have that  
\begin{equation}\label{eq:LCR_scalar}
    \tg{R}=\lc{R} + \tg{T}+\frac{2}{e} \partial_{\mu}\left(e \tg{T}^{\mu}\right)=0,
\end{equation}    
from which we immediately derive    
\begin{equation}\label{eq:TGR_scalar}    
    \lc{R} =-\tg{T}-\underbrace{\frac{2}{e} \partial_{\mu}\left(e \tg{T}^{\mu}\right)}_\text{boundary term}.
\end{equation}
In Sec. \ref{sec:dyn_field_equat}, the above calculations will be derived  in details. Therefore, a particular TG Lagrangian is 
\begin{equation}\label{eq:STEGR_lagrangian}
    S_{\rm TEGR}= -\frac{c^4}{16\pi G}\int {\rm d}^{4}x\ e \underbrace{\mathscr{L}_{\rm TEGR}}_{-\tg{T}}+\int {\rm d}^{4}x e\ \mathscr{L}_m,
\end{equation}
up to a boundary term, which gives no contributions, because the boundary is fixed and the variation of the tetrads over there is vanishing.
Eq.\eqref{eq:LCR_scalar} is dynamically equivalent to that of GR (cf. Eq.\eqref{eq:GR_action}), namely $S_{\rm TEGR}=S_{\rm GR}$. This specific TG theory is called TEGR. 

The related field equations are \cite{Aldrovandi2013}
\begin{equation} \label{eq:TG_FEs}
\tg{G}_{\mu\nu}:=\frac{1}{e}\partial_\lambda (e\ \tg{S}_{\mu\nu}{}^\lambda)-\frac{4\pi G}{c^4}\mathfrak{t}_{\mu\nu}=\frac{4\pi G}{c^4} \mathfrak{T}_{\mu\nu},    
\end{equation}
where $\tg{G}_{\mu\nu}$ is the TG Einstein tensor and
\begin{equation}
 \mathfrak{t}_{\mu\nu}=\frac{c^4}{4\pi G}\tg{S}_{\lambda\nu}{}^\rho\Gamma^\lambda_{\ \rho\mu}-g_{\mu\nu}\frac{c^4}{16\pi G}\tg{T}   
\end{equation}
is the energy-momentum (pseudo) tensor of the gravitational field. This equation shows that Eq. \eqref{eq:superpotential} is linked to the gauge representation of the gravitational energy-momentum tensor, namely \cite{Aldrovandi2013}
\begin{equation}
\tg{S}_A{}^{\mu\nu}=-\frac{8\pi G}{c^4 e}\frac{\partial \mathscr{L}_{\rm TEGR}}{\partial(\partial_\nu e^A_{\ \mu})}.    
\end{equation}
The field equations \eqref{eq:TG_FEs} can be also equivalently written in a more explicit form as \cite{Capozziello2021}
\begin{align} \label{eq:TEGR_FE_final}
\tg{G}_{\mu\nu}&:=\frac{1}{e}e^A_{\ \mu} g_{\nu\rho} \partial_\sigma(e\tg{S}_A^{\ \rho\sigma})-\tg{S}_B^{\ \sigma}{}_{\nu}\tg{T}^B_{\ \sigma\mu}\notag\\
&+\frac{1}{2}\tg{T}g_{\mu\nu}-e^A_{\ \mu}\omega^B_{\ A\sigma}\tg{S}_{B\nu}{}^\sigma=\frac{8\pi G}{c^4}\mathfrak{T}_{\mu\nu}.  
\end{align}
In Sec. \ref{sec:dyn_field_equat} we will explicitly show that these field equations coincide with those of GR. An important issue is related to the matter couplings, because the presence of torsion introduces some difficulties when dealing with fermions and bosons. Indeed, they are very sensitive to the appearance of distortions in the connections, and the unique resolution of this problem consists in resorting to the Weitzenb\"ock gauge (see Refs. \cite{Obukhov2004,Aldrovandi2013,Jimenez2019}, for more details).

Looking at the torsion scalar expression \eqref{eq:torsion_tensor}, we see that it is possible to obtain new theories by considering the following general definition of torsion scalar 
\begin{equation}\label{eq:general_torsion}
    \tg{T}_{\rm gen}:=-\frac{c_{1}}{4} \tg{T}_{\alpha \mu \nu} \tg{T}^{\alpha \mu \nu}-\frac{c_{2}}{2} \tg{T}_{\alpha \mu \nu} \tg{T}^{\mu \alpha \nu}+c_{3} \tg{T}_{\alpha} \tg{T}^{\alpha},
\end{equation}
where $c_{1}, c_{2}, c_{3}$ are some free real constants, whose explicit values characterize the gravity model known under the name of \emph{three-parameter
Hayashi-Shirafuji theory} \cite{Hayashi1979}. The general torsion scalar \eqref{eq:general_torsion} is invariant under both general coordinates and local Lorentz transformations, independently of the numerical values of the coefficients, because it relies only on the properties of the torsion tensor. On the contrary, the   
equivalence with GR, and then TEGR,  is achieved only for $c_1=c_2=c_3=1$, which is naturally obtained within the TG gauge paradigm, without resorting to hypotheses related to GR \cite{Aldrovandi2013,Martin2019}. This crucial aspect makes TG a self-consistent theory. 

The N\"other energy-momentum pseudotensor $\mathfrak{t}_\mu^{\ \rho}$ entails $\partial_\mu\mathfrak{t}_\mu^{\ \rho}=0$ \cite{Maggiore2005}. In addition, considering the $\partial_\mu$ derivative of Eq.\eqref{eq:TG_FEs}, we obtain $\partial_\mu \mathfrak{T}^{\mu\nu}=0$, which shows that the energy-momentum tensor is conserved under ordinary derivative, which implies that the spacetime charges $\mathcal{Q}^\mu:=\int e{\rm d}^3x \mathfrak{T}^{0\mu}$ are conserved. In addition, being the TG field equations symmetric, it is thus very easy to be compared with the GR ones \cite{Aldrovandi2013,Martin2019}. Therefore, the antisymmetric part of the energy-momentum tensor \eqref{eq:energy_momentum_tensor} is vanishing, namely
\begin{equation}
\mathfrak{T}_{[\mu\nu]}=e^A_{\ [\mu}g_{\nu]\rho}\tg{T}_A^{\ \rho}=0.   
\end{equation}
Another way to see this identity is through the 
invariance of the action under local Lorentz transformations \cite{Aldrovandi2013,Martin2019}. 
In TEGR, the covariance eliminates six of the sixteen equations, which means that we are able to determine the tetrads up to a local Lorentz transformation, which is equivalent to determine the metric tensor.  

The role of  spin connection is not dynamical in TEGR and we will show that it trivially satisfies 
the field equations. The same result is also confirmed by exploiting the constrained variational principle via the Lagrange multipliers (see Ref. \cite{Martin2019} and references therein, for details). 

Let us consider the following TG Lagrangians
\begin{equation}
\mathscr{L}_{\rm TEGR}(e^A_{\ \mu},0),\qquad \mathscr{L}_{\rm TEGR}(e^A_{\ \mu},\overset{\fullcirc}{\omega}^A_{\ B\mu}),    
\end{equation}
which are both dynamically equivalent to the Hilbert-Einsten action. Therefore,  the following identity holds
\begin{align}
&\mathscr{L}_{\rm TEGR}(e^A_{\ \mu},\overset{\fullcirc}{\omega}^A_{\ B\mu})+\partial_\mu\left[\frac{ec^4}{8\pi G}\tg{T}^\mu(e^A_{\ \mu},\overset{\fullcirc}{\omega}^A_{\ B\mu})\right]\notag\\
&=\mathscr{L}_{\rm TEGR}(e^A_{\ \mu},0)+\partial_\mu\left[\frac{ec^4}{8\pi G}\tg{T}^\mu(e^A_{\ \mu},0)\right],    
\end{align}
which explicitly reads as \cite{Martin2019}
\begin{equation}
\tg{T}^\mu(e^A_{\ \mu},\overset{\fullcirc}{\omega}^A_{\ B\mu})=\tg{T}^\mu(e^A_{\ \mu},0)-\overset{\fullcirc}{\omega}^\mu.    
\end{equation}
Therefore, we arrive to the conclusion that
\begin{equation} \label{eq:TGLag_SC}
 \mathscr{L}_{\rm TEGR}(e^A_{\ \mu},\overset{\fullcirc}{\omega}^A_{\ B\mu})= \mathscr{L}_{\rm TEGR}(e^A_{\ \mu},0)+\partial_\mu\left[\frac{ec^4}{8\pi G}\overset{\fullcirc}{\omega}^\mu\right].  
\end{equation}
This proves that the spin connection enters in the Lagrangian as a total derivative, justifying also the possibility to reduce the calculations in TEGR by  adopting the Weitzenb\"ock gauge in any case. In addition, if we vary the Lagrangian in Eq. \eqref{eq:TGLag_SC} with respect to the spin connection, we obtain an identically vanishing equation. Therefore, the spin connection does not contribute to the TG field equations, representing non-dynamical DoFs. This fact shows also that TEGR can be considered as a \emph{pure tetrad teleparallel gravity} \cite{Maluf2013}, which assumes that for whatever tetrad one chooses, the spin connection is zero, treating these two objects as independent structures (see Ref. \cite{Martin2019}, for details and for its implications). 

We have understood that the spin connection is not relevant, if we are interested in searching for the solutions of  TEGR field equations. However, formally, its presence fulfills a paramount role, because: $(i)$ it guarantees the covariance of the action under local Lorentz transformations and diffeomorphisms; $(ii)$ it is endowed with a regularizing power, because it removes the divergent inertial effects from the Lagrangian, dubbed thus \emph{renormalized action} (see Ref. \cite{Martin2019}, for details); $(iii)$ it permits to obtain a regular field theory and naturally produces, in its action, a Gibbons-Hawking-York term, which permits to be coherently related to the formulation of a quantum gravity theory (see Refs. \cite{Aldrovandi2006a,Aldrovandi2006b,Oshita2017}, for details).   

Finally,  analysing the DoFs of TEGR, we start from the vierbeine $e^A_{\ \mu}$ with 16 components.  We have to subtract 6 DoFs related to the inertial effects due to the spin connection and other 8 non-dynamical DoFs due to diffeomorphisms (the same as in GR). The result is  2 DoFs as in the case of GR \cite{Jimenez2019}. Also for this feature, TEGR is dynamically equivalent to GR.

\subsubsection{Symmetric teleparallel gravity}
\label{sec:STEGR}
Symmetric teleparallel gravity (STG) is a formulation of gravitational interaction described only in terms of  non-metricity \eqref{eq:nonm_ten}. While TG theories have been extensively discussed, STG patterns have only recently received a growing attention, and there are still some crucial points to be disclosed and better understood. This theory can be either formulated in terms of metric tensor or tetrads, although the former is the most common presentation followed in the literature \cite{Nester1999}. In STG, the symmetric affine connection (cf. Eq. \eqref{eq:ACG}) assumes a fundamental dynamical role and represents an independent structure. This hypothesis is not trivial at all, because it requires considerable efforts in determining all the affine components already in the simplest cases both at astrophysical and cosmological levels (see e.g., Refs. \cite{Capozziello:2022wgl,Hohmann2021,DAmbrosio2022}, for more details).  

The presence of non-metricity entails particular geometric effects, which gives rise to counterintuitive implications from those analysed in the previous theories. They can be summarised in the following points:
\begin{itemize}
    \item 
raising up or  lowering down indices of vectors or tensors under the covariant derivative $\stg{\nabla}$, is not straightforward like in metric case, namely given a vector $v^\mu$, we have
\begin{equation}
g_{\nu\lambda}\stg{\nabla}_\mu v^\lambda=\stg{\nabla}_\mu v_\nu - v^\lambda\stg{Q}_{\mu\nu\lambda};    
\end{equation}
\item non-metricity does not preserve the length of  vectors; indeed given two vectors $v=v^\mu \partial_\mu$ and $w=w^\mu\partial_\mu$ parallel along a curve $\gamma$, their tangent vectors are $T=T^\mu\partial_\mu$ with $T^\mu\equiv \dot{\gamma}^\mu$, namely $T^\lambda\stg{\nabla}_\lambda v^\mu=0$ and $T^\lambda\stg{\nabla}_\lambda w^\mu=0$. Let us calculate the evolution of the scalar product of the vectors
\begin{equation} \label{eq:non_conservation_length}
T^\lambda\stg{\nabla}_\lambda v\cdot w=T^\lambda v^\mu w^\nu \stg{Q}_{\lambda\mu\nu},    
\end{equation}
where $v\cdot w:=g_{\mu\nu}v^\mu w^\nu$, which is not conserved, as well as the norm of a vector $|v|:=\sqrt{v\cdot v}$, and therefore it is not possible to normalize it. It follows also that the angles, between two vectors, do not in general conserve, namely
\begin{equation}
T^\lambda\stg{\nabla}_{\lambda}\left(\frac{v\cdot w}{\vert v\vert \vert w\vert}\right)\neq0.    
\end{equation}
STG is, in general, not a \emph{conformal theory}, but it is possible to reduce it to a conformal one (see Ref. \cite{Gakis2020}, for details). The above results imply also the impossibility to define a proper time along a curve as in GR;
\item Given a four-velocity $u^\mu$, we define
\begin{subequations}
\begin{align} 
a^\mu&:=u^\lambda\stg{\nabla}_\lambda u^\mu,\label{eq:acc}\\
\tilde{a}_\mu&:=u^\lambda\stg{\nabla}_\lambda u_\mu=a_\mu+\stg{Q}_{\lambda\nu\mu}u^\lambda u^\nu,\label{eq:an_acc}
\end{align}
\end{subequations}
where $a^\mu$ is the \emph{acceleration}, whereas $\tilde{a}_\mu$ is the \emph{anomalous acceleration}. In particular, this implies that the four-velocity is not anymore orthogonal to the four-acceleration, because 
\begin{align}
u_\mu a^\mu&=u_\mu u^\lambda \stg{\nabla}_\lambda u^\mu\notag\\
&=  u^\lambda \stg{\nabla}_\lambda(u_\mu u^\mu)- u^\mu u^\lambda \stg{\nabla}_\lambda u_\mu\notag\\
&=\stg{Q}_{\lambda\mu\nu}u^\lambda u^\mu u^\nu +2u_\mu a^\mu - \tilde{a}_\mu u^\mu,
\end{align}
from which we obtain 
\begin{equation}
a^\mu u_\mu=\tilde{a}_\mu u^\mu -\stg{Q}_{\lambda\mu\nu}u^\lambda u^\mu u^\nu.   
\end{equation}
From Eq. \eqref{eq:an_acc} we get
\begin{equation}
 \left(\tilde{a}_\mu-a_\mu\right)u^\mu=\stg{Q}_{\lambda\mu\nu}u^\lambda u^\mu u^\nu.   
\end{equation}
Therefore, the non-metricity tensor expresses how much the anomalous acceleration deviates from the standard acceleration, and it is also responsible to depart the acceleration from the spatial hypersurface orthogonal to the four-velocity;
\item the acceleration of autoparallels (cf. Eq. \eqref{eq:autoparallel}) in STG becomes 
\begin{equation} \label{eq:STG_autoparallel}
a^\mu=0,\qquad \tilde{a}_\mu=\stg{Q}_{\lambda\nu\mu}u^\lambda u^\nu;    
\end{equation}
\item in order to recover the length conservation (cf. Eq. \eqref{eq:non_conservation_length}) and the autoparallel definition (cf. Eq. \eqref{eq:STG_autoparallel}), we have to impose
\begin{equation}
\stg{Q}_{(\lambda\mu\nu)}=0,\qquad \stg{Q}_{(\lambda\mu)\nu}=0,    
\end{equation}
but these two conditions are too strict constraints. These issues can be solved by resorting to the \emph{Weyl conformal transformations} (see Ref. \cite{Wheeler2018}, for details). 
\end{itemize}

Let us consider now the STG action, constituted by the most general quadratic Lagrangian \cite{Jimenez2019,DAmbrosio2022}:
\begin{equation}\label{eq:STG_Lagrangian_general}
S_{\rm STEGR}:=\int {\rm d}^{4}x\ \sqrt{-g} \left[\frac{c^4}{16\pi G}\underbrace{\mathscr{L}_{\rm STEGR}}_{\stg{Q}}+\mathscr{L}_m\right],
\end{equation}
where $\stg{Q}$ is the so-called \emph{non-metricity scalar}, whose expression is given by \cite{Capozziello2021,DAmbrosio2022}
\begin{align}\label{eq:nonmetricity_scalar}
\stg{Q} &:= g^{\mu\nu}\left(\stg{L}^\alpha_{\ \beta\mu}\stg{L}^\beta_{\ \nu\alpha}-\stg{L}^\alpha_{\ \beta\alpha}\stg{L}^\beta_{\ \mu\nu}\right)\notag\\
&=\frac{1}{4}\left( \stg{Q}_{\alpha} \stg{Q}^{\alpha}-\stg{Q}_{\alpha \beta \gamma} \stg{Q}^{\alpha \beta \gamma}\right)\notag\\
&+\frac{1}{2}\left( \stg{Q}_{\alpha \beta \gamma} \stg{Q}^{\beta \alpha \gamma}-\stg{Q}_{\alpha} \stg{\bar{Q}}^{\alpha}\right),    
\end{align}
where $\stg{Q}_\alpha:=\stg{Q}_{\alpha\lambda}{}^\lambda$ and $\stg{\bar{Q}}_\alpha:=\stg{Q}^\lambda_{\ \lambda\alpha}$ represent two independent traces of the non-metricity tensor. This gives rise to the STEGR theory, where it is possible to show the validity of the following formula (see Sec. \ref{sec:STEGR_field_equations}) \cite{DAmbrosio2022}
\begin{equation} \label{eq:scalr_Q}
\stg{Q}=\lc{R}+\lc{\nabla}_\mu(\stg{Q}^\mu-\stg{\bar{Q}}^\mu),    
\end{equation}
and using the following GR identity \cite{Misner1973}
\begin{equation}
 \lc{\nabla}_\mu(\stg{Q}^\mu-\stg{\bar{Q}}^\mu)\equiv\frac{1}{\sqrt{-g}}\partial_\mu\biggr{[}\sqrt{-g}(\stg{Q}^{\mu}-\stg{\bar{Q}}^\mu)\biggr{]},   
\end{equation}
we see that the STEGR action is dynamically equivalent to  GR up to a boundary term, which is vanishing because the boundary is fixed and the variation of the metric over there is vanishing.
\begin{equation}\begin{aligned}\label{eq:STG_LAGRANGIAN_GENERAL}
\stg{Q}_{\rm gen} &:= c_1 \stg{Q}_{\alpha \beta \gamma} \stg{Q}^{\alpha \beta \gamma}+c_2 \stg{Q}_{\alpha \beta \gamma} \stg{Q}^{\beta \alpha \gamma}+c_3 \stg{Q}_{\alpha} \stg{Q}^{\alpha}\\
&+c_4 \stg{\bar{Q}}_{\alpha} \stg{\bar{Q}}^{\alpha}+c_5 \stg{Q}_{\alpha} \stg{\bar{Q}}^{\alpha}.
\end{aligned}
\end{equation}
where $c_1,\ c_2,\ c_3,\ c_4,\ c_5$ are real free constant parameters, and this gives rise to the \emph{five-parameter family of quadratic theories} or the so-called \emph{New GR} (see Ref. \cite{Blixt} and references therein).

We can introduce a \emph{superpotential} or the \emph{non-metricity conjugate} as \cite{Gakis2020,Capozziello2021,DAmbrosio2022}
\begin{align} \label{eq:Pamn}
\stg{P}^\alpha_{\ \mu\nu}&:=\frac{1}{2\sqrt{-g}}\frac{\partial(\sqrt{-g}\stg{Q})}{\partial \stg{Q}_\alpha^{\ \mu\nu}}\notag\\
&=\frac{1}{4}\stg{Q}^\alpha_{\ \mu\nu}-\frac{1}{4}\stg{Q}_{(\mu}{}^\alpha_{\ \nu)}-\frac{1}{4}g_{\mu\nu}\stg{Q}^{\alpha\beta}{}_\beta\notag\\
&+\frac{1}{4}\left[\stg{Q}_\beta^{\ \beta\alpha}g_{\mu\nu}+\frac{1}{2}\delta^\alpha_{(\mu}\stg{Q}_{\nu)}{}^\beta_{\ \beta}\right].
\end{align}
Through this definition, we can describe the non-metricity scalar \eqref{eq:nonmetricity_scalar} equivalently as
\begin{equation}
\stg{Q}:=\stg{Q}_{\alpha\mu\nu}\stg{P}^{\alpha\mu\nu}.    
\end{equation}
We can introduce also the further quantity \cite{Gakis2020,Capozziello2021}
\begin{align} \label{eq:qmn}
\frac{1}{\sqrt{-g}}\stg{q}_{\mu\nu}&:=\frac{1}{\sqrt{g}}\frac{\partial(\sqrt{-g}\stg{Q})}{\partial g^{\mu\nu}}+\frac{1}{2}g_{\mu\nu}\stg{Q}\notag\\
&=\frac{1}{4}\left(2\stg{Q}_{\alpha\beta\mu}\udt{\stg{Q}}{\alpha\beta}{\nu} - \stg{Q}_{\mu\alpha\beta}\dut{\stg{Q}}{\nu}{\alpha\beta} \right)\notag\\
    &- \frac{1}{4}\left(2\dudt{\stg{Q}}{\alpha}{\beta}{\beta}\udt{\stg{Q}}{\alpha}{\mu\nu} - \dudt{\stg{Q}}{\mu}{\beta}{\beta}\dudt{\stg{Q}}{\nu}{\beta}{\beta}\right)\notag\\ 
    &- \frac{1}{2}\left( \stg{Q}_{\alpha\beta\mu}\udt{\stg{Q}}{\beta\alpha}{\nu}-\dudt{\stg{Q}}{\beta}{\beta}{\alpha}\udt{\stg{Q}}{\alpha}{\mu\nu}\right).
\end{align}
We have now all the elements to write the STEGR field equations (obtained varying the STEGR action with respect to the metric tensor), which reads as \cite{Capozziello2021,DAmbrosio2022}
\begin{align}\label{eq:STEGR_FEs}
    \stg{G}_{\mu\nu}&:=\frac{2}{\sqrt{-g}} \nabla_{\alpha}\left(\sqrt{-g} \stg{P}_{\mu \nu}^{\alpha}\right)\notag\\
    &-\frac{1}{\sqrt{-g}}\stg{q}_{\mu \nu}+\frac{1}{2} g_{\mu \nu}\stg{Q}= \frac{8\pi G}{c^4}\mathfrak{T}_{\mu\nu}, 
\end{align}
where $\stg{G}_{\mu\nu}$ is the STEGR Einstein tensor. The variation of STEGR action with respect to the connection produces the \emph{connection field equations} \cite{Capozziello2021,DAmbrosio2022}
\begin{equation}\label{eq:AFE}
    \nabla_{\mu}\nabla_{\nu}\left(\sqrt{-g} \stg{P}_{\alpha}^{\ \mu \nu}\right)=0, 
\end{equation}
representing a set of first order differential equations for the affine connection.

Using the general results of Sec. \ref{sec:Lorentz_connection_mathematical}, it is possible to recast the STEGR connection via the tetrads $e^\alpha_{\ \beta}\in {\rm GL}(4,\mathbb{R})$ and the curvatureless hypothesis in the following form (cf. Eq. \eqref{eq:gamma_lorentz})\footnote{In Eq. \eqref{eq:GL_STEGR}, we have used a different notation with respect to those employed previously. Here it is important to underline the inverse tetrad matrix for the implication in \eqref{eq:SUBS_STEGR}.}
\begin{equation} \label{eq:GL_STEGR}
\Gamma^\alpha_{\ \mu\nu}:=(e^{-1})^\alpha_{\ \beta}\partial_\mu e^\beta_{\ \nu}.  
\end{equation}
Since STEGR is torsionless, we have (cf. Eq. \eqref{eq:TORSION_TETRAD})
\begin{equation}
T^\alpha_{\ \mu\nu}:=(e^{-1})^\alpha_{\ \beta}\partial_{[\nu} e^\beta_{\ \mu]}=0,
\end{equation}
which implies
\begin{equation}
\partial_{\mu} e^\beta_{\ \nu}=\partial_{\nu} e^\beta_{\ \mu}\quad \Leftrightarrow\quad e^\alpha_{\ \beta}\equiv e'^\alpha_{\ \beta}:=\partial_\beta \xi^\alpha,
\end{equation}
where  the tetrad is holonomic and, in addition, it can be parameterized by $\xi^\alpha=\xi^\alpha(x^\mu)$. Therefore, Eq. \eqref{eq:GL_STEGR} becomes \cite{Capozziello2021,DAmbrosio2022}
\begin{equation} \label{eq:SUBS_STEGR}
\Gamma^\alpha_{\ \mu\nu}=\frac{\partial x^\alpha}{\partial \xi^\lambda}\partial_\mu\partial_\nu \xi^\lambda.
\end{equation}
This connection can be set globally to zero, by considering the following affine (gauge) roto-translational transformation of coordinates \cite{DAmbrosio2022}
\begin{equation}
\xi^\alpha:= M^\alpha_{\ \beta}x^\beta+\xi^\alpha_0,    
\end{equation}
where $M^\alpha_{\ \beta}\in O(1,3)$ is an orthogonal matrix and $\xi^\alpha_0$ is a constant translational vector, which permits to have $\Gamma^\alpha_{\ \mu\nu}=0$, which is the so-called \emph{coincident gauge}. Physically, this means that the origin of the tangent space (expressed by $\xi^\alpha$) is coincident with the spacetime origin (given by $x^\mu$). This gauge is defined up to a linear affine transformation $a x^\mu+b$ with $a,b$ real constant values. 

It is important to note that this residual global symmetry does not vanish at infinity, ensuing significant properties at the infrared structure of the theory \cite{Jimenez2019}. In addition, recalling that the strong Equivalence Principle states that gravitation is indistinguishable from acceleration, its effects can be locally neglected via a diffeomorphic change of coordinates (i.e., LIFs). In this perspective, we understand that the affine connection is an integrable  translation. Therefore, the coincident gauge embodies and saves the strong Equivalence Principle of GR \cite{Koivisto2018}. 

It is worth noticing that the STEGR affine connection is purely inertial and it does not contain any information about gravitation. Another important implication of the coincident gauge is the explicit breaking of diffeomorphism invariance due to the particular choice of coordinates, which does not occur in other frames \cite{Hohmann2021}. The use or not of the coincident gauge affects only the boundary term \eqref{eq:scalr_Q}, which has no influence on the ensuing dynamics and therefore neither on the evolution of the metric tensor.

This particular gauge form permits to considerably simplify the calculations. In addition, the affine field equations \eqref{eq:AFE} are trivially satisfied. In TG the local Lorentz transformations are gauged through the spin connection and the calculations are simplified via the Weitzenb\"ock choice, whereas, in STG, the diffeomorphism of coordinates become the new gauge and the calculations are easily carried out through the coincident gauge. This concept is summarised in the following scheme
\begin{equation}
\begin{tikzcd}
                        & \text{loc. Lorentz trans.} \arrow[r] & \text{Weitzenb\"ock} \\
\text{gauge} \arrow[ru,"\text{TEGR}"] \arrow[rd,"\text{STEGR}"] &             &   \\
                        & \text{diff. of coord.} \arrow[r] & \text{coincident}
\end{tikzcd}
\end{equation}

There are also other two beneficial effects considering Eq. \eqref{eq:ACG}, which are
\begin{equation}
\stg{\nabla}_\mu=\partial_\mu,\qquad \stg{L}^\lambda_{\ \mu\nu}=-\lc{\Gamma}^\lambda_{\ \mu\nu}.     
\end{equation}
It is worth noticing that, in generic STG theories, it is not possible to require that the coincident gauge holds \emph{a priori}. More specifically, it is not possible, in general, to use a coordinate system which simultaneously simplifies metric and connection. When this can be achieved, it holds for a restricted set of geometries or it reduces the class of solutions (see Refs. \cite{Hohmann2019,DAmbrosio2022}, for further details).

In STEGR we have that the total DoFs are enconded in the metric tensor, having 10 components from which we have to subtract 8 diffeomorphisms as in GR, having therefore again 2 DoFs as in GR. Here we have that the 4 diffeomorphisms of coordinates become the gauge diffeomorphism symmetries. While in TEGR metric and connection are related, in STEGR the connection becomes essentially a pure gauge and all the dynamics is enclosed in the metric, which is trivially connected \cite{Jimenez2019}. It is possible to introduce a close analogy between the fields $\xi^\alpha$, parameterizing the connection, and the \emph{St\"uckelberg fields}, related to the invariance of coordinates' transformation, and also between the coincident gauge and the \emph{unitary gauge} (see Ref. \cite{Jimenez2019,Ruegg2004} for  details).

\subsection{A discussion on Trinity Gravity at Lagrangian level}
\label{sec:summary_results}
GR, TEGR, and STEGR constitute  the so-called Geometric Trinity of Gravity, but, from the above discussion, it is clear that they are nothing else but particular cases of  wide classes of geometries. According to their formulation, they are three non-communicating theories, because they start from distinct hypotheses and different dynamical-geometric objects. GR is usually conceived as the geometric formulation of gravity, whereas TEGR and STEGR as the gauge approaches to gravity, albeit also GR can be formulated in a gauge way via the use of tetrads and spin connection. 

Covariance and (strong) Equivalence Principles are at the foundations of GR. The former postulate has a more general character, which can be easily recognized  also in TEGR and STEGR, whereas the latter hides some subtleties, which are sources of confusion in  literature. For example, some papers state that such a principle does not hold in TG. More properly, it is not strictly required at the foundation of TG  but it must  hold  to provide mathematical and physical coherence for TEGR and STEGR theories. In this framework, we have discovered that the Equivalence Principle emerges naturally from TEGR and STEGR guaranteeing thus the {\it equivalence} among these three theories. This last point can be summarised as follows
\begin{equation}
\begin{tikzcd}
   &[2em] \text{foundation} \\[5ex]
\text{strong EP} \arrow[ru, "\text{GR}"] \arrow[r, "\text{TEGR}"] \arrow[rd, "\text{STEGR}"] &[6em]  \overset{\fullcirc}{\omega}^C_{\ B\mu}-\tg{K}^C_{\ B\mu}=\overset{\circ}{\omega}^C_{\ B\mu} \\[5ex]
 &[2em] \text{coincident gauge}
\end{tikzcd}
\end{equation}
Up to now,  we  underlined only the equivalence among the three theories  at  Lagrangian level, pointing out the  difference for a boundary term, namely
\begin{equation}
 \begin{tikzcd}
&\overbrace{\mathscr{L}_{\rm GR}}^{\lc{R}}\arrow[dl]  \arrow[dr] & \\
  \underbrace{\mathscr{L}_{\rm TEGR}}_{-\tg{T}-\frac{2}{e} \partial_{\mu}\left(e \tg{T}^{\mu}\right)} \arrow[rr] \arrow[ur] &                         & \underbrace{\mathscr{L}_{\rm STEGR}}_{\stg{Q}-\lc{\nabla}_\mu(\stg{Q}^\mu-\stg{\bar{Q}}^\mu)}\arrow[ul]\arrow[ll]
\end{tikzcd}
\end{equation}
As stated above, the equivalence does not hold for extensions like $f(R)$, $f(T)$, and $f(Q)$ because, in general, these aforementioned theories differ for the DoFs (see Ref. \cite{Bamba}, for a straightforward example). However, equivalence can be restored also in extensions considering appropriate boundary terms.  For example, $f(R)$, $f(T,B)$, and $f(Q,B)$ can be compared as fourth order theories when an appropriate boundary term $B$ is defined in each gravity framework. In $f(T,B)$ this equivalence has been explicitly proved considering the boundary term as in Eq. \eqref{eq:TGR_scalar} (see e.g., \cite{Bahamonde2015, Bahamonde:2016grb,Capozziello:2018qcp}). An analogue procedure should show that also $f(Q,B)$ theory can be dynamically reduced to $f(R)$ defining a suitable boundary term as in Eq. \eqref{eq:scalr_Q}.  

\section{Field equations in Trinity Gravity}
\label{sec:dyn_field_equat}
In the above discussion, equivalent representations of gravity have been compared at the level of actions and Lagrangians. Here we want to develop the same comparison at the level of field equations.

Let us start from the  Bianchi  identities, having the pivotal role to link the field equations of a theory with the conservation laws of the  gravity tensor invariants  and with the energy-momentum tensor \cite{Misner1973}. We start from the second Bianchi identity \eqref{eq:Bianchi_identity_general}, whose more explicit expression is \cite{Bahamonde2021}
\begin{equation}
\begin{aligned} \label{eq:second_Bianchi_id}
    &\nabla_{\lambda}R^{\alpha}_{\ \beta\mu\nu} + \nabla_{\mu}R^{\alpha}_{\ \beta\nu\lambda} + \nabla_{\nu}R^{\alpha}_{\ \beta\lambda\mu} \\ 
   &= T^{\rho}_{\ \mu\lambda}R^{\alpha}_{\ \beta\nu\rho} + T^{\rho}_{\ \nu\lambda}R^{\alpha}_{\ \beta\mu\rho} + T^{\rho}_{\ \nu\mu}R^{\alpha}_{\ \beta\lambda\rho},
\end{aligned}
\end{equation}
and we prove the equivalence among GR (see Sec. \ref{sec:GR_field_equations}), TEGR (see Sec. \ref{sec:TEGR_field_equations}), and STEGR (see Sec. \ref{sec:STEGR_field_equations}) in terms of their field equations, which we show to be equal to those already presented in Sec. \ref{sec:trinity_lagrangian}. 

\subsection{GR field equations}
\label{sec:GR_field_equations}
Since in GR we have $R^{\alpha}_{\ \beta\mu\nu}=\lc{R}^{\alpha}_{\ \beta\mu\nu}$ and $T^\alpha_{\ \beta\gamma}=Q_{\alpha\beta\gamma}=0$, the second Bianchi identity \eqref{eq:second_Bianchi_id} reduces to
\begin{equation}\label{eq:GR_Bianchi_id}
    \lc{\nabla}_{\lambda}\lc{R}^{\alpha}_{\ \beta\mu\nu} + \lc{\nabla}_{\mu}\lc{R}^{\alpha}_{\ \beta\nu\lambda} + \lc{\nabla}_{\nu}\lc{R}^{\alpha}_{\ \beta\lambda\mu} = 0.
\end{equation}
To simplify the calculations, thanks to the Covariance Principle, we can exploit the LIF's coordinates (cf. Eq. \eqref{eq:LIF}), where second derivatives of the metric are not null. Contracting $\alpha$ and $\lambda$, Eq. \eqref{eq:GR_Bianchi_id} becomes
\begin{equation}
    \partial_{\lambda}\lc{R}^{\lambda}_{\ \beta\mu\nu} + \partial_{\mu}\lc{R}^{\lambda}_{\ \beta\nu\lambda} + \partial_{\nu}\lc{R}^{\lambda}_{\ \beta\lambda\mu} = 0. 
\end{equation}
Using the antysimmetry in the last two indices of the Riemann tensor (cf. Eq. \eqref{eq:sym_Riemann}), we obtain
\begin{equation}
    \partial_{\lambda}\lc{R}^{\lambda}_{\ \beta\mu\nu}-\partial_{\mu}\lc{R}^{\lambda}_{\ \beta\lambda\nu} + \partial_{\nu}\lc{R}^{\lambda}_{\ \beta\lambda\mu}=0.
\end{equation}
Applying the metric to first raise up the index $\beta$ and then contracting $\beta$ and $\mu$, we have
\begin{equation}
    - \partial_\lambda\lc{R}^{\lambda}_{\ \nu} - \partial_\beta\lc{R}^{\beta}_{\ \nu}+\partial_\nu\lc{R} = 0, 
\end{equation}
from which we immediately obtain
\begin{equation}\label{eq:GRbianchicontracted}
    \partial_\mu\lc{R}^{\mu}_{\ \nu} - \frac{1}{2}\partial_\nu\lc{R}=0.
\end{equation}
Using again the metric tensor, Eq. \eqref{eq:GRbianchicontracted} becomes
\begin{equation} \label{eq:GR_Bianchi_id_final}
   \partial_\mu(\lc{R}^{\mu\nu} - \frac{1}{2} g^{\mu\nu}\hat{R}) = 0\ \Rightarrow\ \lc{\nabla}_\mu(\lc{R}^{\mu\nu} - \frac{1}{2} g^{\mu\nu}\hat{R}) = 0, 
\end{equation}
where the partial derivative is in general replaced by the covariant one. This relation leads to the Einstein field equations in  vacuum (cf. Eq. \eqref{eq:GR_equations}). The Einstein tensor $\lc{G}^{\mu\nu}$ is divergenceless, which  also implies the conservation of the energy-momentum tensor \cite{Misner1973,Boskoff}, namely
\begin{equation}
\lc{\nabla}_\mu \lc{G}^{\mu\nu}=0,\quad \Leftrightarrow\quad \lc{\nabla}_\mu \mathfrak{T}^{\mu\nu}=0.    
\end{equation}

\subsection{TEGR field equations}
\label{sec:TEGR_field_equations}
Since in TEGR curvature and non-metricity vanish, Eq. \eqref{eq:second_Bianchi_id} can be further simplified via the the Weitzenb\"ock gauge (see Sec. \ref{sec:TEGR}) as follows
\begin{equation}\label{eq:TEGR_Bianchi_id}
    \tg{\nabla}_{\lambda}R^{\alpha}_{\ \beta\mu\nu}+\tg{\nabla}_{\nu}R^{\alpha}_{\ \beta\lambda\mu}+\tg{\nabla}_{\mu}R^{\alpha}_{\ \beta\nu\lambda} = 0,
\end{equation}
where $R^{\alpha}_{\ \beta\mu\nu}\equiv\lc{R}^{\alpha}_{\ \beta\mu\nu}+\tg{\mathcal{K}}^{\alpha}_{\ \beta\mu\nu}=0$ with
\begin{align}\label{eq:tensor_Kabmn}
    \tg{\mathcal{K}}^{\alpha}_{\ \beta\mu\nu} &:= \lc{\nabla}_{\mu}\tg{K}^{\alpha}_{\ \beta\nu}-\lc{\nabla}_{\nu}\lc{K}^{\alpha}_{\ \beta\mu}\notag\\
    &+ \tg{K}^{\alpha}_{\ \sigma\mu}\tg{K}^{\sigma}_{\ \beta\nu} - \tg{K}^{\alpha}_{\ \sigma\nu}\tg{K}^{\sigma}_{\ \beta\mu},
\end{align}
including all  torsion tensor contributions and having also the following symmetry properties (cf. Eq. \eqref{eq:contortion})
\begin{align}
\tg{\mathcal{K}}^{\alpha}_{\ \beta\mu\nu} = -\tg{\mathcal{K}}_{\beta}{}^{\alpha}{}_{\mu\nu},\qquad
    \tg{\mathcal{K}}^{\alpha}_{\ \beta\mu\nu} = -\tg{\mathcal{K}}^{\alpha}_{\ \beta\nu\mu}.
\end{align}
Contracting $\alpha$ and $\lambda$, Eq. \eqref{eq:TEGR_Bianchi_id} becomes
\begin{align}\label{o}
    &\quad \tg{\nabla}_\lambda\lc{R}^{\lambda}_{\ \beta\mu\nu}+\tg{\nabla}_\mu\lc{R}^{\lambda}_{\ \beta\nu\lambda}+\tg{\nabla}_\nu\lc{R}^{\lambda}_{\ \beta\lambda\mu}\notag\\
    &+\tg{\nabla}_\lambda \tg{\mathcal{K}}^{\lambda}_{\ \beta\mu\nu}+\tg{\nabla}_\mu\tg{\mathcal{K}}^{\lambda}_{\beta\nu\lambda}+\tg{\nabla}_\nu\tg{\mathcal{K}}^{\lambda}_{\beta\lambda\mu}=0.
\end{align}
Applying the same strategy of GR (see Sec. \ref{sec:GR_field_equations}) and using the metric compatibility of TEGR, we obtain
\begin{equation} \label{eq:TEGR_FE2}
  \tg{\nabla}_{\mu}(\lc{R}^{\mu}_{\ \nu}+ \tg{\mathcal{K}}^{\mu}_{\ \nu}) - \frac{1}{2}\tg{\nabla}_{\nu}(\lc{R}+\tg{\mathcal{K}})=0,
\end{equation}
where $\tg{\mathcal{K}}_{\mu \nu}:=\tg{\mathcal{K}}^{\lambda}_{\ \mu\lambda\nu}$ and $\tg{\mathcal{K}}:=\tg{\mathcal{K}}^{\nu}_{\ \nu}$, having a formally similar definition of Ricci tensor and scalar curvature of GR. Eq. \eqref{eq:TEGR_FE2} entails twofold implications 
\begin{subequations} \label{eq:TEGR_TWO_IMPLICATIONS}
\begin{align}
   &\lc{R}_{\mu\nu}-\frac{1}{2}g_{\mu\nu}\lc{R}=-\tg{\mathcal{K}}_{\mu\nu}+\frac{1}{2}g_{\mu\nu}\tg{\mathcal{K}},\\
   &\tg{\mathcal{K}}_{\mu\nu}-\frac{1}{2}g_{\mu\nu}\tg{\mathcal{K}}=0, \label{eq:TEGR_field_equations} 
\end{align}
\end{subequations}
where the former tells that TEGR field equations are equivalent to those of GR, whereas the latter, derived using the GR vacuum field equations, gives the TEGR field equations, which are divergenceless in terms of $\tg{\nabla}$. 

Now, we prove that Eq. \eqref{eq:TEGR_field_equations} reproduces exactly Eq. \eqref{eq:TEGR_FE_final}. To this end, we first analyse $\tg{\mathcal{K}}_{\mu\nu}$, which gives
\begin{align} \label{eq:Kmunu}
    \tg{\mathcal{K}}_{\mu\nu} &= \lc{\nabla}_{\alpha}\tg{K}^{\alpha}_{\ \mu\nu} - \lc{\nabla}_{\nu}\tg{K}^{\alpha}_{\ \mu\alpha} + \tg{K}^{\sigma}_{\ \mu\nu}\tg{K}^{\alpha}_{\ \sigma\alpha} - \tg{K}^{\sigma}_{\ \mu\alpha}\tg{K}^{\alpha}_{\ \sigma\nu}\notag \\
    &=\lc{\nabla}_{\alpha}\tg{K}^{\alpha}_{\ \mu\nu}+ \lc{\nabla}_{\nu}\tg{T}_{\mu} - \tg{K}_{\sigma\mu\nu}\tg{T}^{\sigma} -\tg{K}^{\sigma}_{\ \mu\alpha}\tg{K}^{\alpha}_{\ \sigma\nu}\notag\\
    &=\lc{\nabla}_\alpha \tg{S}_\nu{}^\alpha_{\ \mu}+\lc{\nabla}_\alpha \tg{T}^\alpha g_{\mu\nu}-\tg{K}^\alpha_{\ \sigma\nu}\tg{S}_\alpha{}^\sigma_{\ \mu},
\end{align}
where we have used (cf. Eqs. \eqref{eq:contortion} and \eqref{eq:superpotential})
\begin{subequations}
\begin{align}
\tg{K}^{\alpha}{}_{\mu\alpha}&=-\tg{T}_{\mu},\\
\tg{K}^{\alpha}{}_{\alpha\mu} &= 0,\\
\tg{K}^\mu_{\ \nu\lambda}&=\tg{S}_\lambda{}^{\mu\nu}+\delta^\nu_{\lambda}\tg{T}^\mu-\delta^\mu_{\lambda}\tg{T}^\nu.
\end{align}
\end{subequations}
Now, we can analyse $\tg{\mathcal{K}}$, which gives (cf. Eq. \eqref{eq:TGR_scalar})
\begin{equation} \label{eq:K}
 \tg{\mathcal{K}} = 2\lc{\nabla}_{\lambda}\tg{T}^{\lambda}+\tg{T}=\frac{2}{e}\partial_\lambda(e\hat{T}^\lambda)+\hat{T}.
\end{equation}
Substituting Eqs. \eqref{eq:Kmunu} and \eqref{eq:K} into Eq. \eqref{eq:TEGR_field_equations}, we have
\begin{equation} \label{eq:ALTER_TEGR}
\lc{\nabla}_\alpha \tg{S}_{\nu\mu}{}^\alpha+\tg{K}^\alpha_{\ \sigma\nu}\tg{S}_\alpha{}^\sigma_{\ \mu}+\frac{1}{2}g_{\mu\nu}\tg{T}=0,
\end{equation}
which can be shown to be equal to Eq. \eqref{eq:TEGR_FE_final} by exploiting metric compatibility, and tetrad postulate (see Appendix \ref{sec:TEGR_alternative}, for more details).
 
\subsection{STEGR field equations} 
\label{sec:STEGR_field_equations}
Since STEGR is curvatureless and torsionless, Eq. \eqref{eq:second_Bianchi_id} can be further simplified via the coincident gauge (see Sec. \ref{sec:STEGR}), leading to the following expression
\begin{equation} \label{eq:STEGR_Bianchi_id}
    \partial_{\lambda}R^{\alpha}_{\ \beta\mu\nu}+\partial_{\nu}R^{\alpha}_{\ \beta\lambda\mu}+\partial_{\mu}R^{\alpha}_{\ \beta\nu\lambda} = 0,
\end{equation}
where, in this case, $R^{\alpha}_{\ \beta\mu\nu} \equiv \lc{R}^{\alpha}_{\ \beta\mu\nu} + \stg{\mathcal{L}}^{\alpha}_{\ \beta\mu\nu} = 0$. $\stg{\mathcal{L}}^{\alpha}_{\ \beta\mu\nu}$ is a function of the disformation tensor, namely
\begin{equation}\label{eq:Labmn}
\lc{\mathcal{L}}^{\alpha}_{\ \beta\mu\nu} = \lc{\nabla}_{\mu}\stg{L}^{\alpha}_{\ \beta\nu} - \lc{\nabla}_{\nu}\stg{L}^{\alpha}_{\ \beta\mu} + \stg{L}^{\alpha}_{\ \sigma\mu}\stg{L}^{\sigma}_{\ \beta\nu} - \stg{L}^{\alpha}_{\ \sigma\nu}\stg{L}^{\sigma}_{\ \beta\mu},
\end{equation}
endowed with the following symmetry properties
\begin{subequations}
    \begin{align}
        \stg{\mathcal{L}}^{\alpha}_{\ \beta\mu\nu} = - \stg{\mathcal{L}}^{\beta}_{\ \alpha\mu\nu},\qquad 
        \stg{\mathcal{L}}^{\alpha}_{\ \beta\mu\nu} = - \stg{\mathcal{L}}^{\alpha}_{\ \beta\nu\mu},
    \end{align}
\end{subequations}
where we have used Eqs. \eqref{eq:nonm_ten} and \eqref{eq:sym_nonmetricity}. 

Contracting $\alpha$ and $\lambda$ and giving the explicit expression of the Riemann tensor, Eq. \eqref{eq:STEGR_Bianchi_id} becomes
\begin{align}
   &\partial_\lambda \lc{R}^{\lambda}_{\ \beta\mu\nu}+\partial_\mu\lc{R}^{\lambda}_{\ \beta\nu\lambda}+\partial_\nu\lc{R}^{\lambda}_{\ \beta\lambda\mu}\notag\\
   &+ \partial_\lambda \stg{\mathcal{L}}^{\lambda}_{\ \beta\mu\nu}+\partial_\mu \stg{\mathcal{L}}^{\lambda}_{\ \beta\nu\lambda}+\partial_\nu \stg{\mathcal{L}}^{\lambda}_{\ \beta\lambda\mu}=0.
\end{align}
Following the same strategy adopted in GR (see Sec. \ref{sec:GR_field_equations}), we finally obtain (cf. Eqs. \eqref{eq:TEGR_TWO_IMPLICATIONS})
\begin{subequations} \label{eq:STEGR_TWO_IMPLICATIONS}
\begin{align}
   &\lc{R}_{\mu\nu}-\frac{1}{2}g_{\mu\nu}\lc{R}=-\stg{\mathcal{L}}_{\mu\nu}+\frac{1}{2}g_{\mu\nu}\stg{\mathcal{L}},\label{eq:STEGR_GR_equivalent}\\
   &\stg{\mathcal{L}}_{\mu\nu}-\frac{1}{2}g_{\mu\nu}\stg{\mathcal{L}}=0, \label{eq:STEGR_field_equations} 
\end{align}
\end{subequations}
where $\stg{\mathcal{L}}_{\mu\nu}:=\stg{\mathcal{L}}^\alpha_{\ \mu\alpha\nu}$ and $\stg{\mathcal{L}}:=\stg{\mathcal{L}}^\mu_{\ \mu}$, resembling formally the expression of Ricci tensor and scalar curvature, respectively. Let us note that, in the coincident gauge, $\stg{L}^\alpha_{\ \mu\nu}=-\lc{\Gamma}^\alpha_{\ \mu\nu}$, which soon reveals that Eq. \eqref{eq:STEGR_field_equations} is equivalent to the GR field equations. Equation \eqref{eq:STEGR_GR_equivalent} proves the equivalence between GR and STEGR field equations, whereas Eq. \eqref{eq:STEGR_field_equations} represents the STEGR field equations, which we will demonstrate to be equal to Eq. \eqref{eq:STEGR_FEs}. Let us first analyse $\stg{\mathcal{L}}_{\mu\nu}$, which yields
\begin{align}\label{eq:Lmunu}
\stg{\mathcal{L}}_{\mu\nu}& = \lc{\nabla}_{\alpha}\stg{L}^{\alpha}{}_{\mu\nu} - \lc{\nabla}_{\nu}\stg{L}^{\alpha}{}_{\mu\alpha} + \stg{L}^{\sigma}{}_{\mu\nu}\stg{L}^{\alpha}{}_{\sigma\alpha} - \stg{L}^{\sigma}{}_{\mu\alpha}\stg{L}^{\alpha}{}_{\sigma\nu}\notag\\
&=\lc{\nabla}_{\alpha} \stg{L}^{\alpha}{}_{\mu \nu}+\frac{1}{2} \lc{\nabla}_{\nu} \stg{Q}_{\mu}-\frac{1}{2} \stg{Q}_{\alpha} \stg{L}^{\alpha}{}_{\mu \nu}\notag\\
& - \frac{1}{4} \left[\stg{Q}_{\mu}{}^{\sigma}{}_{\alpha} \stg{Q}_{\nu}{}^{\alpha}{}_{\sigma} + 2 \stg{Q}^{\alpha}{}_{\sigma\nu}(\stg{Q}^{\sigma}{}_{\alpha\mu}-\stg{Q}_{\alpha}{}^{\sigma}{}_{\mu})\right],
\end{align}
where we have used
\begin{subequations}
\begin{align}
\stg{L}^{\alpha}_{\ \mu\alpha}&= - \frac{1}{2}\stg{Q}_{\mu},\label{eq:Qm}\\
\stg{L}^{\alpha}_{\ \mu\nu}&=2\stg{P}_{\ \mu \nu}^{\alpha}+ \frac{1}{2}g_{\mu\nu}(\stg{Q}^{\alpha} - \stg{\bar{Q}}^{\alpha})\notag\\
&- \frac{1}{4}(\delta^{\alpha}_{\mu}\stg{Q}_{\nu} + \delta^{\alpha}_{\nu}\stg{Q}_{\mu}).
\end{align}
\end{subequations}
Therefore, the scalar $\stg{\mathcal{L}}$ is expressed by
\begin{align} \label{eq:L}
\stg{\mathcal{L}} &= \lc{\nabla}_{\alpha}(\stg{Q}^{\alpha}-\stg{\bar{Q}}^{\alpha}) + \frac{1}{4} \stg{Q}_{\alpha \beta \gamma} \stg{Q}^{\alpha \beta \gamma} - \frac{1}{2} \stg{Q}_{\alpha \beta \gamma} \stg{Q}^{\gamma \beta \alpha}\notag \\
&\hspace{2.55cm} - \frac{1}{4} \stg{Q}_{\alpha} \stg{Q}^{\alpha} + \frac{1}{2} \stg{Q}_{\alpha} \stg{\bar{Q}}^{\alpha}\notag\\
&= \lc{\nabla}_{\alpha}(\stg{Q}^{\alpha} - \stg{\bar{Q}}^{\alpha}) - \stg{Q}.
\end{align}
Gathering together Eqs. \eqref{eq:Lmunu} and \eqref{eq:L}, using the following identity (cf. Eq. \eqref{eq:Qm})
\begin{equation}
\partial_{\alpha}\stg{Q}^{\alpha}=\lc{\nabla}_{\alpha}\stg{Q}^{\alpha}+\stg{L}^{\alpha}{}_{\sigma\alpha} \stg{Q}^{\sigma}=\lc{\nabla}_{\alpha}\stg{Q}^{\alpha}-\frac{1}{2}\stg{Q}_{\alpha}\stg{Q}^{\alpha},
\end{equation}
we then obtain
\begin{align} \label{eq:STEGR_intermediate}
    &2\partial_{\alpha}P^{\alpha}{}_{\mu\nu} + \frac{1}{2}\stg{Q}_{\alpha\mu\nu}(\stg{Q}^{\alpha} - \stg{\tilde{Q}}^{\alpha}) + \frac{1}{2}g_{\mu\nu}\partial_{\alpha}(\stg{Q}^{\alpha} - \stg{\tilde{Q}}^{\alpha})\notag  \\ 
    &+ \frac{1}{2} \stg{L}^{\sigma}{}_{\mu\nu}\stg{Q}_{\sigma} + \frac{1}{4} \stg{Q}_{\mu}{}^{\alpha}{}_{\sigma}\stg{Q}_{\nu}{}^{\sigma}{}_{\alpha} + \frac{1}{2} \stg{Q}^{\alpha}{}_{\sigma\mu}(\stg{Q}^{\sigma}{}_{\nu\alpha} - \stg{Q}_{\alpha}{}^{\sigma}{}_{\nu})\notag \\ 
    &-\frac{1}{2} g_{\mu\nu}\lc{\nabla}_{\alpha}(\stg{Q}^{\alpha} - \stg{\tilde{Q}}^{\alpha}) + \frac{1}{2} g_{\mu\nu}\stg{Q} = 0,
\end{align}
which is equal to Eq. \eqref{eq:STEGR_FEs} in an empty spacetime (see Appendix \ref{sec:STEGR_alternative}, for more details), i.e.,
\begin{align} \label{eq:STEGR_FE_simple}
\frac{2}{\sqrt{-g}}\partial_{\alpha}(\sqrt{-g} \stg{P}^{\alpha}_{\ \mu\nu}) -\frac{1}{\sqrt{-g}}\stg{q}_{\mu\nu}+ \frac{1}{2} g_{\mu\nu}\stg{Q} = 0.
\end{align}
An important remark is in order at this point. As already discussed above in the case of Lagrangians, the equivalence holds only for the theories stemming out from the scalar invariants $R$, $T$, and $Q$. In these specific cases, we obtain second order equations. This is not true for extensions, implying, in general, non-linear functions of these invariants. This fact points out again that GR, and its equivalent representations, are very peculiar cases among the theories of gravity.


\section{Solutions in Trinity Gravity}
\label{sec:trinity_solutions}
Clearly the equivalence of GR, TEGR, and STEGR has to be proven also at the solution level. In Sec. \ref{sec:dyn_field_equat},  the same field equations have been obtained, and then the same exact solutions, under the same symmetries and boundary conditions, have to be achieved.

In this perspective, performing the calculations to settle the solutions in the three gravity scenarios is useful also in view of extensions of the theories.   Recently, it has been proposed a \emph{3+1 splitting formalism in the Geometric Trinity of Gravity} \cite{Capozziello2021} entailing the following advantages: (1) simplicity in carrying out numerical analyses; (2) solving some theoretical issues existing in the various formulations of GR at the fundamental level (e.g., canonical quantization); (3) broadening this methodology also in extended and alternative gravity frameworks. 

Here, we focus the attention on one of the simplest GR solutions, represented by the \emph{Schwarazschild spacetime}. Soon after the publication of GR theory by Einstein, Schwarzschild determined the  solution, describing the  spacetime metric outside a spherically symmetric mass-energy distribution. This result is in perfect agreement with the weak field approximation  \cite{Einstein1915}.   

Jebsen, in 1921, and Birkhoff, in 1923, independently proved that the Schwarzschild solution holds outside a spherically symmetric mass distribution, even if this  varies over time. This is now known as the \emph{(Jebsen) Birkhoff theorem}, and it can be stated as follows \cite{Birkhoff1923, Johansen2006}:\\
\emph{any spherically symmetric solution of the GR field equations in vacuum has to be  static and asymptotically flat. In addition, the Schwarzschild solution is the unique solution satisfying these hypotheses.} \\

This claim entails several significant implications: (1) the uniqueness of the Schwarzschild solution in GR by imposing the spherical symmetry as starting hypothesis; (2) no emission of gravitational waves, which can be interpreted, similarly as in electromagnetism, that there exists no monopole (spherically symmetric) radiation; (3) the outcome of the Birkhoff theorem in GR gravity theory can be compared with the Gauss theorem implications in electromagnetism and in classical Newtonian gravity. 

Let us start our considerations taking into account a generic spherically symmetric metric, whose line element, written in spherical coordinates $\left\{t,r,\theta,\varphi\right\}$, in the equatorial plane $\theta=\pi/2$, and in geometric units ($G=c=1$), reads as
\cite{Misner1973,Romano2019,Boskoff}
\begin{equation} \label{eq:metric_SS2}
{\rm d}s^2=-e^{\nu(t,r)}{\rm d}t^2+e^{\lambda(t,r)}{\rm d}r^2+r^2{\rm d}\varphi^2,
\end{equation}
where $\nu(t,r),\lambda(t,r)$ are the two unknown  functions to be determined. We supplement this general metric with the well-known weak field limit on the metric time component
\begin{equation} \label{eq:WFL}
-e^{\nu(t,r)}\approx -1+\frac{2M}{r},
\end{equation}
where $M$ is the compact object mass, being the origin of the gravitational field, and $\frac{2M}{r}$ is the Newtonian gravitational potential. 

We want now to solve the field equations in  vacuum ($\mathfrak{T}_{\mu\nu}=0$, namely outside the gravitational source) in GR (Sec. \ref{sec:solution_GR}), TEGR (Sec. \ref{sec:solution_TEGR}), and STEGR (Sec. \ref{sec:solution_STEGR}). We will observe how the Birkhoff theorem emerges  also in TEGR and STEGR.
Finally, we will recover the \emph{Schwarzschild metric} in all three gravity theories, namely \cite{Misner1973,Romano2019}
\begin{equation}
    {\rm d}s^{2} = -\Big(1 - \frac{2M}{r}\Big){ \rm d}t^{2} + \Big(1 - \frac{2M}{r}\Big)^{-1} {\rm d}r^{2} + r^{2}{ \rm d}\varphi^{2},
\end{equation}
admitting $r_{\rm S}:=2M$ as event horizon (coordinate singularity) and $r=0$ as essential (physical) singularity.

Given a function $f(t,r)$, we use the following notations
\begin{equation}
\dot{f}(t,r):=\frac{{\rm d}f(t,r)}{{\rm d}t},\qquad  f'(t,r):=\frac{{\rm d}f(t,r)}{{\rm d}r}.   
\end{equation}

\subsection{ Spherically Symmetric  Solutions in GR}
\label{sec:solution_GR}
The vacuum field Eqs. \eqref{eq:GR_equations} can be recast also as
\begin{equation}\label{eq:GR_vacuum_field}
    \lc{G}_{\mu\nu}\equiv\lc{R}_{\mu\nu}=0,
\end{equation}
where $\lc{R}=0$. Analysing $\lc{G}_{tr}$ we obtain
\begin{equation}\label{eq:ind_time_B}
    \lc{G}_{tr} \equiv \frac{\dot{\lambda}(t,r)}{r} = 0,\quad\Rightarrow\quad \lambda=\lambda(r).
\end{equation}
The other independent field equations are
\begin{subequations} \label{eq:EFEs}
\begin{align} 
\lc{G}_{rr}&\equiv-e^{\lambda (r)}+r \nu '(t,r)+1=0,\label{eq:G11}\\
\lc{G}_{tt}&\equiv e^{-\lambda (r)} \left(r \lambda '(r)-1\right)+1=0.\label{eq:G44}
\end{align}
\end{subequations}
From Eq. \eqref{eq:G11}, we conclude that $\nu=\nu(r)$. All the metric components are independent of the coordinate time $t$, and this proves that the metric is \emph{static}. 

From Eq. \eqref{eq:G44} we obtain
\begin{equation}
\left[e^{-\lambda (r)} r\right]'=1\quad \Rightarrow\quad e^{-\lambda (r)}=1-\frac{C_1}{r},
\end{equation}
where $C_1$ is an integration constant.  Multiplying Eq. \eqref{eq:G44} by $e^{\lambda(r)}$ and summing it to Eq. \eqref{eq:G11}, we obtain
\begin{equation} 
\lambda'(r)+\nu'(r)=0,\quad \Rightarrow\quad \lambda(r)+\nu(r)=C_2,
\end{equation}
where the integration constant $C_2$ has to be  $C_2=0$ to achieve the \emph{asymptotic flatness}. From  the \emph{weak field limit}  consideration \eqref{eq:WFL}, we obtain
\begin{equation} \label{eq:Schwarzschild_metric}
-e^{\nu(r)}=1-\frac{2M}{r},\qquad e^{\lambda(r)}=\frac{1}{1-\frac{2M}{r}}.
\end{equation}

\subsection{Spherically Symmetric Solutions in TEGR}
\label{sec:solution_TEGR}
For solving the TEGR field equations,  we adopt the tetrad formalism. We know that each tetrad field must be associated to the related spin connection (see, for example,  Ref. \cite{Martin2019}). However, in TEGR, we can drastically simplify the calculations resorting to the Weitzenb\"ock gauge. Therefore, we can choose the diagonal tetrad
\begin{equation}\label{eq:diagonal_tetrad}
     e^{A}_{\ \mu}=\left(\begin{matrix}
      \sqrt{-e^{\nu(r)}} & 0 & 0 & 0 \\
0 & \sqrt{e^{\lambda(r)}} & 0 & 0 \\
0 & 0 & r & 0\\
0 & 0 & 0 & r\sin{\theta}
     \end{matrix}\right).
\end{equation}
Let us recall that this tetrad is related to the off-diagonal tetrad (where the spin connection is naturally vanishing \cite{Martin2019}) through a local Lorentz transformation $\Lambda^{A}_{\ B}(x)$. However, they both describe the same metric.

The non-zero torsion tensor components are
\begin{subequations}
\begin{align}
\tg{T}^t_{\ tr}&=-\frac{1}{2} \nu '(r)=-\frac{M}{r^2}\left(1-\frac{2 M}{r}\right)^{-1},\\
\tg{T}^\varphi_{\ r\varphi}&=\frac{1}{r}.
\end{align}
\end{subequations}
It is worth noticing that, physically, $\tg{T}^t_{\ tr}$ represents the redshifted radial gravitational force, because it is calculated with respect to the coordinate time $t$; whereas $\tg{T}^\varphi_{\ r\varphi}$ is the classical centrifugal force occurring in the tetrad frame.

Another important object is the contortion tensor, whose non-zero components read as
\begin{subequations}
\begin{align}
\tg{K}_{ttr}&=\frac{1}{2} e^{\nu (r)} \nu '(r)=\frac{M}{r^2},\\
\tg{K}_{\varphi r \varphi}&=r,
\end{align}
\end{subequations}
whose interpretation is closely related to that already provided for the torsion tensor (cf. Eq. \eqref{eq:eqgeo_teleparallel_TEGR}).

The superpotential components read as
\begin{subequations}
\begin{align}
\tg{S}_{\hat t}^{\ tr}&=\frac{2 e^{-\lambda (r)} \sqrt{e^{-\nu (r)}}}{r}=\frac{2}{r}\sqrt{1-\frac{2 M}{r}},\\
\tg{S}_{\hat \varphi}^{\ r\varphi}&=-\frac{e^{-\lambda (r)} \left(r \nu '(r)+2\right)}{2 r^2}=\frac{M-r}{r^3}.
\end{align}
\end{subequations}
Finally the torsion scalar is
\begin{equation}
\tg{T}=-\frac{2 e^{-\lambda (r)} \left(r \nu '(r)+1\right)}{r^2} =-\frac{2}{r^2},
\end{equation}
which represents the \qm{dynamically active part} of the scalar curvature, whereas the remaining part is included in the \qm{dynamically passive boundary term}.  

Combining these elements, it is easy to prove that $\tg{G}_{\mu\nu}\equiv\lc{G}_{\mu\nu}$ (cf. Eq. \eqref{eq:ALTER_TEGR}). Then applying the same procedure of GR (see Sec. \ref{sec:solution_GR}),  the Birkhoff theorem holds also in TEGR.  

\subsection{Spherically Symmetric Solutions in STEGR}
\label{sec:solution_STEGR}
Regarding the STEGR field equations, we adopt the coincident gauge to ease the calculations, where $\stg{\nabla}=\partial_\mu$ and $\lc{\Gamma}^\mu_{\ \alpha\beta}=-\stg{L}^\mu_{\ \alpha\beta}$. In this case, it is  immediate to get $\lc{G}_{\mu\nu}\equiv\stg{G}_{\mu\nu}$. However let us calculate the fundamental terms occurring in Eq. \eqref{eq:STEGR_field_equations} for extracting the physical information.

The non-metricity tensor has the following expression 
\begin{align}
\stg{Q}_{r\mu\nu}&=\left(
\begin{array}{ccc}
 -e^{\nu (r)} \nu '(r) & 0 & 0 \\
 0 & e^{\lambda (r)} \lambda '(r) & 0 \\
 0 & 0 & 2 r 
 \end{array}
\right)\notag\\
&=\left(
\begin{array}{ccc}
 -\frac{2 M}{r^2} & 0 & 0  \\
 0 & -\frac{2 M}{r^2 \left(1-\frac{2 M}{r}\right)^2} & 0 \\
 0 & 0 & 2 r 
 \end{array}
\right),
\end{align}
where the derivative of  gravitational potential represents the gravitational force acting on the observer and producing  the disformations. For a comparison, we have that TEGR gravitational force  makes the tetrad frame rotating (see Sec. \ref{sec:solution_TEGR}), whereas STEGR gravitational force causes expansions and contractions of the observer laboratory. The conjugate potential reads as
\begin{subequations}
\begin{align}
\stg{P}^t_{\ tr}&=\frac{r \lambda '(r)-r \nu '(r)+4}{8 r}=\frac{1}{8} \left(\lambda '(r)+\nu '(r)\right),\\
\stg{P}^r_{\ rr}&=\frac{e^{\nu (r)-\lambda (r)}}{r}=\frac{1}{r}\left(1-\frac{2 M}{r}\right)^2,\\
\stg{P}^r_{\ \varphi\varphi}&=-\frac{1}{4} r e^{-\lambda (r)} \left(r \nu '(r)+2\right)=\frac{M-r}{2},\\
\stg{P}^\varphi_{\ r\varphi}&=\frac{1}{8} \left(\lambda '(r)+\nu '(r)\right)=0,
\end{align}
\end{subequations}
while the other components are null. The last quantity, represented by the above $q_{\mu\nu}$, reads as
\begin{align}
\frac{\stg{q}_{\mu\nu}}{\sqrt{-g}}&=\left(
\begin{array}{ccc}
 \frac{2 e^{\nu (r)-\lambda (r)} \nu '(r)}{r} & 0 & 0 \\
 0 & \frac{2 r \nu '(r)+2}{r^2} & 0  \\
 0 & 0 & -\frac{r \nu '(r)+2}{e^{\lambda (r)}}
 \end{array}
\right)\notag\\
&=\left(
\begin{array}{ccc}
 \frac{4 M}{r^3}\left(1-\frac{2 M}{r}\right) & 0 & 0 \\
 0 & \frac{2}{r^2\left(1-\frac{2 M}{r}\right)} & 0  \\
 0 & 0 & \frac{2 M}{r}-2 
\end{array}
\right).
\end{align}
Substituting the above expressions in Eq. \eqref{eq:STEGR_field_equations}, we recover the same differential equations of GR (cf. Eq. \eqref{eq:EFEs}). Also in this case, we obtain  the Schwazrschild solution and the validity of the Birkhoff theorem. 

It is worth stressing that, also at this level, we cannot expect the same solutions for $f(R)$, $f(T)$, and $f(Q)$ extensions.

\section{Conclusions and  perspectives}
\label{sec:end}
We have gathered together  basic concepts    of Geometric Trinity of Gravity and derived the related dynamics pointing out analogies and differences of metric, affine, and non-metric approaches. We tried  to give a self-consistent picture of the three representations of  gravitational field.  The main statement is that equivalence is strictly achieved  for GR, TEGR, and STEGR and not for any extension of these theories.

Firstly, we introduced the  geometric arena of metric-affine gravity, where metric tensor and affine connection are two separate and independent structures. After, we provided the fundamental geometric objects, that are tetrads and spin connection. The former represents the observer laboratory, which solders the tangent space to the spacetime manifold. This procedure gives rise to anholonomic frames. They become holonomic when we are dealing with inertial frames, where a particular role is fulfilled by trivial tetrads of Special Relativity (see Sec. \ref{sec:tetrad}). The latter are intimately related with general tetrads, because they represent the inertial effects  and they are generated by local Lorentz transformations. They form the  Lorentz group, which, in turn, can be proved to give rise to a Lorentz algebra. This is a crucial aspect for defining the Fock-Ivanenko covariant derivative, useful to characterize the spin connection in terms of tetrads and to introduce the tetrad postulate (i.e., $\nabla_\mu e^A_{\ \nu}=0$). This theoretical treatment can be  interpreted from a  physical point of view as discussed in Sec. \ref{sec:spin_connection}. 

These mathematical tools allow to describe the Geometric Trinity of Gravity. Specifically, a metric formulation (encoded in the Riemannian geometry) and a gauge approach (encoded in Teleparallel Gravity) are possible. GR, TEGR, and STEGR are dynamically equivalent from the variation of their  Lagrangians up to a boundary term. Furthermore,   starting from the second Bianchi identity, it is possible to infer the field equations, which are identical in the three representations. Finally,  we analysed spherically symmetric solutions in the three theories deriving   the Schwarzschild spacetime and  the Birkhoff theorem. The approaches can be summarized as follows
\begin{equation}
\begin{tikzcd}
   &[3em] \text{Lagrangian}\arrow[d] \\[6ex]
\text{EQUIVALENCE} \arrow[ru, "\text{variation}"] \arrow[r, "\text{Bianchi Ids.}"] \arrow[rd, "\text{symmetries}"] &[16em]  \text{Field equations}\arrow[d] \\[6ex]
 &[3em] \text{Solutions}
\end{tikzcd}
\end{equation}

However, as pointed out above, although mathematical results are equivalent, the physical interpretation can be different depending on the considered variables and  observables. This fact opens several  questions. Some of them   can be listed as follows. 
\begin{itemize}

    \item Are there other equivalent formulations of gravity, outside of the Geometric Trinity? In other words, we can ask for the existence   of other representations of gravity equivalent to GR  within the metric-affine arena or, more in general, identifying other fundamental variables. The question implies also considering   extended  theories of gravity which can be "reduced" to GR (see, for example,  \cite{Kijowski:2016qbc, Capozziello2011R} for a discussion).
    
    \item From an observational point of view, what does it mean that these three theories are dynamically equivalent? This issue translates in extracting  observables from each gravity theory and then interpreting them, from a physical viewpoint, finding out suitable transformation laws which make equivalent the set of variables of each theory.
    
    \item How can we construct observational apparatuses to test  different theories dynamically equivalent to GR? This point is a direct consequence of the previous one. The question can be posed also in another way: Is it possible, if any, to discriminate different sets of  observables for equivalent descriptions of gravity from an experimental point of view?
    
    \item STG theories are the less analysed among the three approaches. A general tetrad formulation is necessary in view of  physical implications. In particular, the interpretation of gravity as a gauge theory could be particularly relevant to uniform gravity under the same standard of other fundamental theories.
    
    \item The Equivalence Principle (in its strong and weak formulations) is a fundamental aspect of GR \cite{Tino2020}.  It can be recovered in TEGR and STEGR, even if it is not at the foundation of these theories. If it were violated at  some level (e.g. at quantum level), would it be possible to state that TEGR and STEGR are more fundamental theories than GR because they do not require it as a basic principle?
    
\end{itemize}
The above ones are some of the open issues related to equivalent representations of gravity and, in particular, to Gravity Trinity. Besides the mathematical aspects, it emerges that   systematic experimental and observational  protocols are necessary to establish the set of  fundamental variables. For example, questions if metric or connection are the "true" gravitational variables are still open. Non-metricity could have a main role in this discussion due to the fact that the stringent requirement of asking for Equivalence Principle  could be relaxed. Forthcoming precision experiments \cite{Ginger}, gravitational wave astronomy \cite{Abedi}, and precision cosmology observations \cite{Cai2016} could be the tools to answer these questions.

\section*{Acknowledgements}
S.C. and V.D.F. acknowledge the support of INFN {\it sez. di Napoli}, {\it iniziative specifiche}  QGSKY, TEONGRAV, and MOONLIGHT2.
S.C. and V.D.F. thanks Gruppo Nazionale di Fisica Matematica of Istituto Nazionale di Alta Matematica for the support. The Authors thank Daniel Blixt, Nicola Menadeo, Marcello Miranda, and Davide Usseglio for the useful discussions and feedbacks. 

\appendix
\section{Derivation of TEGR field equations}
\label{sec:TEGR_alternative}
Let us derive Eq. \eqref{eq:ALTER_TEGR}, from Eq. \eqref{eq:TEGR_FE_final}, in  empty spacetime. Expanding the first terms on which the partial derivative acts, we obtain
\begin{align} \label{eq:TEGRFE1}
    &e^{A}_{\ \mu} g_{\nu\rho} \partial_{\sigma}\tg{S}_{A}{}^{\ \rho\sigma} + e^{-1} e^{A}_{\ \mu}g_{\nu\rho}\partial_{\sigma}e\notag\\
    &- \tg{S}_{B}{}^{\sigma}{}_{\nu} T^{B}{}_{\sigma\mu} + \frac{1}{2}g_{\mu\nu}\tg{T}= 0.
\end{align}
Exploiting the identity $e^{-1}\partial_{\sigma}e \equiv \partial_{\sigma} \ln(\sqrt{-g}) = \lc{\Gamma}^{\alpha}_{\ \alpha\sigma}$, the metric compatibility \eqref{eq:compatibility}, and the tetrad postulate \eqref{eq:tetrad_postulate} in Eq. \eqref{eq:TEGRFE1}, we have
\begin{align}
&\partial_{\sigma} \tg{S}_{\mu\nu}{}^{\sigma} - \tg{S}_{\alpha\nu}{}^{\sigma}\Gamma^{\alpha}{}_{\nu\sigma} - \tg{S}_{\alpha\mu}{}^{\sigma} \Gamma^{\alpha}{}_{\nu\sigma}-\tg{S}_{\mu}{}^{\rho\sigma}\Gamma^{\alpha}{}_{\rho\sigma}g_{\nu\alpha}\notag \\ 
& - \tg{S}_{\alpha}{}^{\sigma}{}_{\nu} \tg{T}^{\alpha}{}_{\sigma\mu} + \Gamma^{\alpha}{}_{\alpha\sigma}\tg{S}_{\mu\nu}{}^{\sigma} + \frac{1}{2}\tg{T} g_{\mu\nu} = 0.
\end{align}
In the above equation, we can substitute the first terms with the covariant derivative with respect to the general affine connection \eqref{eq:ACG}, splitting it in GR covariant derivative (with respect to the Levi-Civita connection) and terms involving the contorsion tensor, namely
\begin{align} \label{eq:TEGRfe2}
&\lc{\nabla}_{\sigma}\tg{S}_{\mu\nu}{}^{\sigma}-\tg{K}^{\alpha}{}_{\mu\sigma}\tg{S}_{\alpha\nu}{}^{\sigma} - \tg{K}^{\alpha}{}_{\nu\sigma}\tg{S}_{\mu\alpha}{}^{\sigma} \notag\\  
&+ \tg{K}^{\sigma}{}_{\alpha\sigma} \tg{S}_{\mu\nu}{}^{\alpha}-\tg{K}_{\nu\rho\sigma}\tg{S}_{\mu}{}^{\rho\sigma} + \tg{T}_{\sigma}\tg{S}_{\mu\nu}{}^{\sigma}\notag\\
&- \tg{S}_{\alpha}{}^{\sigma}{}_{\nu} \tg{T}^{\alpha}{}_{\sigma\mu} + \frac{1}{2}g_{\mu\nu} \tg{T} = 0.
\end{align}
Considering the antisymmetry of $S_\alpha^{\ \mu\nu}$ and $K^\alpha_{\ \mu\nu}$, we have 
\begin{subequations}
\begin{align}
&- \tg{K}^{\alpha}{}_{\nu\sigma}\tg{S}_{\mu\alpha}{}^{\sigma}-\tg{K}_{\nu\rho\sigma}\tg{S}_{\mu}{}^{\rho\sigma}=0,\notag\\
&\tg{K}^{\sigma}{}_{\alpha\sigma} \tg{S}_{\mu\nu}{}^{\alpha}+ \tg{T}_{\sigma}\tg{S}_{\mu\nu}{}^{\sigma}=0.    
\end{align}
\end{subequations}
Therefore, Eq. \eqref{eq:TEGRfe2} can be further simplified as
\begin{align}
&\lc{\nabla}\tg{S}_{\mu\nu}{}^{\sigma} + \tg{K}^{\alpha}{}_{\mu\sigma} \tg{S}_{\alpha}{}^{\sigma}{}_{\nu} - \tg{T}^{\alpha}{}_{\sigma\mu} \tg{S}_{\alpha}{}^{\sigma}{}_{\nu}  + \frac{1}{2}\tg{T} g_{\mu\nu}\notag \\
&=\lc{\nabla}_{\sigma}\tg{S}_{\mu\nu}{}^{\sigma} + \tg{K}^{\alpha}{}_{\sigma\mu}\tg{S}_{\alpha}{}^{\sigma}{}_{\nu} + \frac{1}{2} g_{\mu\nu}\tg{T} = 0, 
\end{align}
where, in the last equation, we have exploited the definition of the contortion tensor (cf. Eq. \eqref{eq:contortion}).

\section{Derivation of STEGR field equations}
\label{sec:STEGR_alternative}
We want  to derive now  Eq. \eqref{eq:STEGR_FE_simple} from Eq. \eqref{eq:STEGR_FEs} in an empty spacetime. Starting from the coincident gauge, we use the following identity (cf. Eq. \eqref{eq:Qm})
\begin{equation}\label{eq:der_met}
\frac{\partial_\alpha \sqrt{-g}}{\sqrt{-g}}=\lc{\Gamma}^\sigma_{\ \alpha\sigma}=-\stg{L}^\sigma_{\ \alpha\sigma}=\frac{1}{2}\stg{Q}_\alpha.    
\end{equation}

After  expanding $\partial_\alpha(\sqrt{-g}\stg{P}^\alpha_{\ \mu\nu})=\partial_\alpha(\sqrt{-g})\stg{P}^\alpha_{\ \mu\nu}+\partial_\alpha(\stg{P}^\alpha_{\ \mu\nu})\sqrt{-g}$, we use Eqs. \eqref{eq:der_met}, \eqref{eq:Pamn}, and \eqref{eq:qmn}. With some algebra, Eq. \eqref{eq:STEGR_FEs} finally becomes (cf. Eq. \eqref{eq:STEGR_intermediate})
\begin{align}
    &2\partial_{\alpha}P^{\alpha}{}_{\mu\nu} + \frac{1}{2}\stg{Q}_{\alpha\mu\nu}(\stg{Q}^{\alpha} - \stg{\tilde{Q}}^{\alpha}) + \frac{1}{2}g_{\mu\nu}\partial_{\alpha}(\stg{Q}^{\alpha} - \stg{\tilde{Q}}^{\alpha})\notag \\ 
    &+ \frac{1}{2} \stg{L}^{\sigma}{}_{\mu\nu}\stg{Q}_{\sigma} + \frac{1}{4} \stg{Q}_{\mu}{}^{\alpha}{}_{\sigma}\stg{Q}_{\nu}{}^{\sigma}{}_{\alpha} + \frac{1}{2} \stg{Q}^{\alpha}{}_{\sigma\mu}(\stg{Q}^{\sigma}{}_{\nu\alpha} - Q_{\alpha}{}^{\sigma}{}_{\nu})\notag\\ 
    &-\frac{1}{2} g_{\mu\nu}\lc{\nabla}_{\alpha}(\stg{Q}^{\alpha} - \stg{\tilde{Q}}^{\alpha}) + \frac{1}{2} g_{\mu\nu}\stg{Q} = 0.
\end{align}

\bibliography{references}

\end{document}